\documentclass[aps,pra,floatfix,superscriptaddress,groupedaddress,longbibliography,twocolumn,amsmath,amssymb,10pt]{revtex4-2}  

\usepackage{graphicx}
\usepackage{dcolumn}
\usepackage{bm}
\usepackage{color}
\usepackage{multirow}
\usepackage{braket}

\usepackage{float}
\usepackage{comment}

\newcommand{\Hbar}[1]{\bar{H}^{(#1)}}
\DeclareMathOperator{\Tr}{\mathrm{Tr}}

\usepackage{hyperref}

\begin{document}

\title{Decoupling Dipolar Interactions in Dense Spin Ensembles}

\author{Linta Joseph}
\email{Linta.Joseph.GR@dartmouth.edu}
\affiliation{Department of Physics and Astronomy, Dartmouth College}
\author{Wynter Alford}
\altaffiliation{Current address: Department of Physics and Astronomy, University of Rochester}
\affiliation{Department of Physics and Astronomy, Dartmouth College}
\author{Chandrasekhar Ramanathan}
\email{Chandrasekhar.Ramanathan@dartmouth.edu}
\affiliation{Department of Physics and Astronomy, Dartmouth College}

\date{\today}

\begin{abstract}
Dense spin ensembles in solids present a natural platform for studying quantum many-body dynamics. Multiple-pulse coherent control can be used to manipulate the magnetic dipolar interaction between the spins to engineer their dynamics. Here, we investigate the performance of a series of well-known pulse sequences that aim to suppress inter-spin dipolar couplings. We use a combination of numerical simulations and solid-state nuclear magnetic resonance (NMR) experiments on adamantane to evaluate and compare sequence performance. We study the role of sequence parameters like inter-pulse delays and resonance offsets. Disagreements between experiments and theory are typically explained by the presence of control errors and experimental non-idealities. The simulations allow us to explore the influence of factors such as finite pulse widths, rotation errors, and phase transient errors. We also investigate the role of local disorder and establish that it is, perhaps unsurprisingly, a distinguishing factor in the decoupling efficiency of spectroscopic sequences (that preserve Hamiltonian terms proportional to $S_z$) and time-suspension sequences (which refocus all terms in the internal Hamiltonian). We discuss our findings in the context of previously known analytical results from Average Hamiltonian Theory. Finally, we explore the ability of time-suspension sequences to protect multi-spin correlations in the system.
\end{abstract}
                             
\maketitle

\section{\label{sec:Intro} Introduction}
Interacting electronic and nuclear spins in solids can provide excellent platforms for simulating quantum many-body physics.  Recently, non-equilibrium  phenomena such as many body localization~\cite{wei_exploring_2018, alvarez_localization-delocalization_2015},  prethermalization~\cite{beatrez_floquet_2021,peng_floquet_2021}, novel dynamical phases of matter such as time crystals~\cite{choi_observation_2017,stasiuk_observation_2023,beatrez_critical_2023}, and the emergence of spin and energy hydrodynamics~\cite{peng_exploiting_2023,zu_emergent_2021} have all been successfully demonstrated in such systems. Interacting spin systems also hold promise for quantum-enhanced magnetometry via both entanglement-assisted metrology and the use of spin-squeezing techniques~\cite{cappellaro_entanglement_2005,cappellaro_quantum_2009,sahin_high_2022}. Controlling and manipulating the native system Hamiltonian is vital to these applications.

\begin{figure*}[htb]
\begin{centering}
    \includegraphics[width=0.8\textwidth]{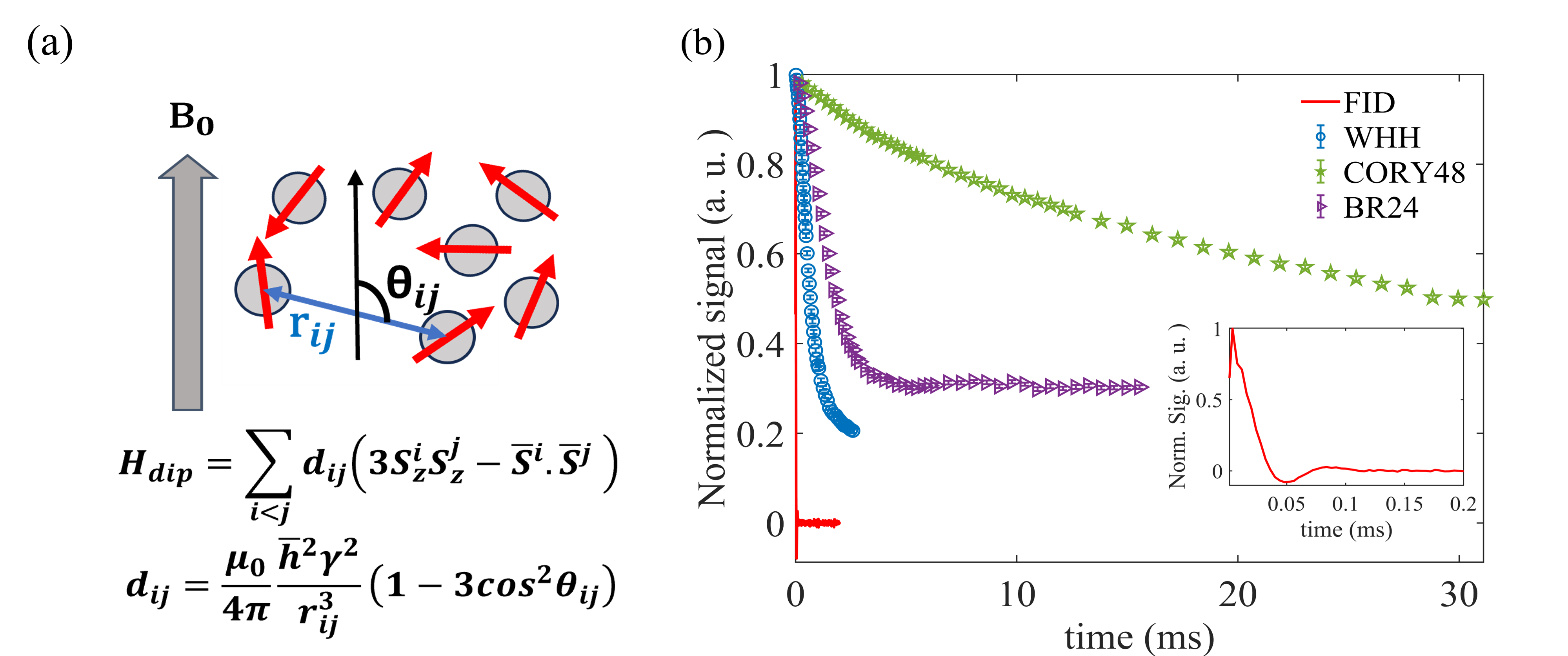}   
    \caption{(a) Dipolar coupled spins (b) Comparison of the decay of the normalized $C_{\text{avg}}$ (geometric mean of $X$, $Y$, and $Z$ autocorrelation experiments) signal over the experiment time for the BR24 (spectroscopic) and CORY48 (time-suspension) sequences at an inter-pulse delay of 4 $\mu$s. The red line shows the free induction decay (FID) signal without dipolar decoupling for reference. Inset shows a zoomed-in version of the FID signal.}
    \label{fig:cartoon_and_T2_comparison} 
	\end{centering}
\end{figure*}

The magnetic dipole interaction between spins forms a dominant part of the internal Hamiltonian in dense systems.
The spins can also experience variations in their local Zeeman interactions due to magnetic field inhomogeneities, as well as local g-factor variations for electronic spins or chemical shifts for nuclear spins.  The above interactions are always present and cannot be switched on or off with external control fields.   Controlling the systems dynamics thus requires averaging out the effect of these interactions on some experimental timescale.  Uncontrolled evolution under these interactions can lead to a rapid loss of coherence, as shown in Figure~\ref{fig:cartoon_and_T2_comparison}(b) for a solid-state nuclear magnetic resonance (NMR) experiment. The figure also shows how refocusing dipolar couplings can significantly extend observed coherence times.  

Decoupling magnetic dipolar interactions has long been a staple of solid-state NMR.  Solid-state NMR techniques to average local spin interactions perform averaging either in spin space using radiofrequency fields or in real space via magic angle spinning (MAS)~\cite{mehring_high_2012}.  Coherent averaging in spin space is accomplished using continuous-wave techniques
based on the Lee Goldburg scheme\cite{lee_nuclear-magnetic-resonance_1965,mehring_magic-angle_1972,bielecki_frequency-switched_1989,vinogradov_high-resolution_1999} or periodic multiple-pulse sequences~\cite{waugh_approach_1968,mansfield_symmetrized_1971,rhim_analysis_1973,rhim_analysis_1974} developed using the theoretical framework of Average Hamiltonian Theory (AHT)~\cite{haeberlen_coherent_1968}.  Furthermore, numerically optimized decoupling techniques have been designed using brute-force approaches~\cite{sakellariou_homonuclear_2000} and optimal control theory~\cite{tabuchi_design_2017,haas_engineering_2019,rose_high-resolution_2018}.  

Many theoretical and experimental schemes in quantum information science (QIS) have borrowed from this rich history of NMR sequences and applied them directly to average out magnetic dipolar interactions~\cite{geru_narrowing_2011,aslam_nanoscale_2017, mohammady_low-control_2018, farfurnik_identifying_2018,balasubramanian_dc_2019,khazali_scalable_2023}. Others have developed new pulse sequences that are adapted to specific application settings such as quantum dots~\cite{waeber_pulse_2019}, Rydberg atoms~\cite{cohen_quantum_2021,geier_floquet_2021} and electronic spins in diamond~\cite{choi_robust_2020, tyler_higher-order_2023-1, zhou_quantum_2020, zhou_robust_2023}.  The latter papers have attempted to simplify the conceptual framework using frame matrices and to derive general algebraic conditions for sequence design that can be tailored to the dominant dephasing mechanisms in the system.   Techniques such as reinforcement learning have also been used to discover new time-suspension sequences~\cite{peng_deep_2022} trained to be robust against a chosen subset of plausible experimental errors. Real-space averaging by magic angle spinning (MAS) is typically not considered in quantum applications. However, sample spinning has been used recently in NV diamond experiments to sense static fields~\cite{hoang_electron_2016,jin_quantum_2024}.

The challenge of averaging out magnetic dipolar interactions is complicated by the presence of local Zeeman variations.  If the two interactions are well separated in scale, it is possible to focus on decoupling the stronger interaction first on a shorter timescale, and then decouple the weaker interactions on a longer timescale.  An example of this is the interspersing of $\pi$ pulses between blocks of dipolar decoupling sequences like the four-pulse WHH (also called WaHuHa)~\cite{waugh_approach_1968} or the 8-pulse MREV8~\cite{mansfield_symmetrized_1973} when the local fields are weak compared to the dipolar couplings~\cite{farfurnik_identifying_2018}.  In the opposite regime, local variations can be refocused first~\cite{choi_robust_2020, tyler_higher-order_2023-1, zhou_quantum_2020, zhou_robust_2023}.  However, when the interactions are of the same order --- a situation that often arises in QIS applications --- the averaging process becomes more difficult.  

Here we evaluate the performance of a variety of well-known dipolar decoupling sequences from the NMR literature using both numerical simulations and experiments and examine their use in QIS applications.   Numerical simulations of sequence fidelity allow us to compare the performance of different sequences under idealized conditions and to explore the effects of various potential experimental errors.  We compare the simulation results with experimental NMR measurements on a powder sample of the plastic solid adamantane (C$_{10}$H$_{16}$).  While adamantane is an interaction-dominant nuclear spin system, the experiments are performed on a magnet without room temperature shims, resulting in the ensemble measurements being performed in an inhomogeneous magnetic field. 

Section~\ref{sec:Decoupling} introduces the spin system and dipolar decoupling sequences, section~\ref{sec:Comparisons} describes the numerical and experimental comparisons between sequences, section~\ref{sec:Errors} discusses the dominant error sources and section~\ref{sec:MQC} discusses applications to controlling correlated spin states.

\section{Decoupling Dipolar Interactions}
\label{sec:Decoupling}

\subsection{Dipolar Coupled Spin Systems}
\label{sec:system}
In an insulating solid-state spin system, the dipolar coupling between spin magnetic moments is the predominant interaction.  We assume the spins can be treated as point dipoles here.  While this is an excellent approximation for nuclear spins, it is not always true for electronic spins and depends on the spatial localization of the electron wave function.  Figure~\ref{fig:cartoon_and_T2_comparison}(a) shows a cartoon of a set of dipolar coupled spins spaced in an external magnetic field and Figure~~\ref{fig:cartoon_and_T2_comparison}(b) shows examples of how the coherence times of nuclear spins can be extended using decoupling sequences.

The Hamiltonian for a solid-state spin system (containing a single spin-1/2 species) in a strong external magnetic field (along $\hat{z}$) is given by $H_{\text{sys}}=H_{Z}+H_{\text{int}}$ where $H_{Z}$ is the Zeeman Hamiltonian and $H_{\text{int}}  = H_D + H_{\text{CS}} + H_{\text{dis}}$ is the internal Hamiltonian of the system consisting of the secular dipolar Hamiltonian ($H_D$), the chemical shift ($H_{\text{CS}}$), and the local disorder Hamiltonian ($H_{\text{CS}}$).

\begin{eqnarray}
H_{Z}&=&\hbar\omega_{L}\sum_{i=1}^{N}S_{z}^{i}\\ \notag
H_{D}&=&\sum_{i<j}d_{ij}\left(3S_{z}^{i}S_{z}^{j}-\bar{S}^{i}.\bar{S}^{j}\right)\\ \notag
H_{\text{CS}}&=&\sum_{i}\delta_{i}S_{z}^{i} \\ \notag
H_{\text{dis}}&=&\sum_{i}h_{i}S_{z}^{i} \notag
\end{eqnarray}
where $S_{\alpha}^{i}$, $\alpha=\{x,y,z\}$ are the spin operators of the $i^{\text{th}}$ spin, $\omega_{L}$ is the Larmor frequency, $\omega_{L}=\gamma B_{0}$, $B_{0}$ is the magnitude of the static longitudinal field, $\gamma$ is the gyromagnetic ratio. $d_{ij}=(\gamma^{2}\hbar^{2}/8r_{ij}^{3})(1-3\cos^{2}\theta_{ij})$ defines the dipolar coupling strength, where $r_{ij}$ is the distance between spins $i$ and $j$, and $\theta_{ij}$ is the angle between $\mathbf{r}_{ij}$ and the B$_{0}$ field as shown in Figure~\ref{fig:cartoon_and_T2_comparison}(a). Note that the forms of 
$H_{\text{CS}}$ and $H_{\text{dis}}$ are essentially the same.  We introduce both forms here to distinguish between the physical origins --- with the chemical shifts ($\delta_{i}$) indicating the presence of nuclear spins with different chemical shifts or electron spins with different g-factors --- versus a disorder term ($h_i$) resulting from local magnetic field variations that might arise due to magnetic impurities, local susceptibility variations or other imperfections.

The spins are controlled by the application of radiofrequency or microwave fields.  The control Hamiltonian $H_{\text{rf}}$ takes the form  
\begin{equation}
H_{\text{rf}}(t)=B_{1}\cos(\omega_{\text{rf}}t)\sum_{i=1}^{N}S_{\alpha}^{i}
\end{equation}
where $\alpha = \{x,y\}$.  In an interaction frame rotating about the $z$ axis at the RF frequency, the transformed Zeeman Hamiltonian takes the form of a resonance offset Hamiltonian,  $H_{\text{O}}=\Delta\omega\sum_{i}S_{z}^{i}$, where $\Delta\omega=\omega_{\text{rf}}-\omega_{L}$.  The total system Hamiltonian in this frame is $H'_{sys} = H_{\text{int}} + H_{\text{O}}.$ 

\subsection{Average Hamiltonian Theory (AHT)}
\label{sec:aht}
In Hamiltonian Engineering approaches based on AHT,  a periodic train of pulses is applied to the system to engineer an effective time-independent Hamiltonian that describes the behavior of the system when it is observed stroboscopically at periodic intervals of the drive~\cite{haeberlen_coherent_1968,gerstein_transient_1985}.  

We typically analyze the dynamics of the spins in a ``toggling'' interaction frame defined by the applied pulses. The evolution of the spins in this toggling frame is generated by a transformed Hamiltonian  $\tilde{H}_{\text{sys}}(t)=U_{\text{rf}}^{-1}(t)H'_{\text{sys}}U_{\text{rf}}(t)$. If $H_{\text{rf}}$ is a sequence that is cyclic and periodic, i.e., if $U_{\text{rf}}(t_{c},0)=\pm \openone$ and $H_{\text{rf}}(t)=H_{\text{rf}}(t+Nt_{c})$, where $t_{c}$ is the cycle time, then considerable simplification of the dynamics is possible. Under these conditions, it can be shown that, if observed at window times $Nt_{c}$, the system evolves as if under an effective Hamiltonian defined by the following Magnus expansion~\cite{magnus_exponential_1954},
\begin{equation}
\label{eqtn:Magnus_expansion_1}
	H_{\text{eff}}=\bar{H}_{\text{sys}}^{(0)}+\bar{H}_{\text{sys}}^{(1)}+... 
\end{equation}
where,
\begin{eqnarray}
	\bar{H}_{\text{sys}}^{(0)}&&=\frac{1}{t_{c}}\int_{0}^{t_{c}}d\tau\tilde{H}_{\text{sys}}(\tau) \notag \\
	\bar{H}_{\text{sys}}^{(1)}&&=\frac{-i}{2t_{c}}\int_{0}^{t_{c}}d\tau\Big[\tilde{H}_{\text{sys}}(\tau),\int_{0}^{\tau}d\phi\tilde{H}_{\text{sys}}(\phi)\Big]	
\end{eqnarray}
etc. For sufficiently short cycle times and under the condition that $t_{c}|\tilde{H}_{\text{sys}}(\tau)|\simeq t_{c}\omega_{\text{sys}}<<1$, the dynamics is primarily governed by the low order terms of the Magnus expansion~\cite{haeberlen_line_1977, maricq_application_1982, maricq_long-time_1990}. To lowest order, we often approximate  $H_{\text{eff}}\simeq\bar{H}_{\text{sys}}^{(0)}$. The evolution of the system appears to be generated by a time-independent effective Hamiltonian given by the system Hamiltonian in the toggling frame averaged over the cycle time $t_{c}$, or the `Average Hamiltonian.' 

{\renewcommand{\arraystretch}{1.5}
\begin{table*}
  \centering
\begin{tabular}{|c|c|c|c|} 
 \hline
 \multirow{3}{*}{\textbf{Sequence}} &  \multirow{3}{*}{\textbf{Hamiltonian terms averaged to zero}} & {\textbf{Zeroth order}}   &  {\textbf{Chemical shift}} \\
  &  & \textbf{effective Hamiltonian}  & \textbf{scaling factor} \\
   &  & ($a_{i}=\delta_{i}+ h_{i} + \Delta\omega$) & ($\Delta_{\text{SF}}$) \\
 \hline
WHH & $H_{D}^{0(\delta), 1}$, $H_{O}^{1}$, $H_{DO}^{1}$, $H_{\alpha D}^{1}$ & $\frac{1}{3}\sum_{i}a_{i}(S_{xi}+S_{yi}+S_{zi})$ & $1/\sqrt{3}$ \\ 
\hline
 MREV8 & $H_{D}^{0, 1}$, $H_{DO}^{1}$, $H_{\alpha D}^{1}$, $H_{\epsilon D}^{1}$, $H_{\epsilon}^{0}$ & $\frac{1}{3}\sum_{i}a_{i}(S_{xi}+S_{zi})$ & $\sqrt{2}/3$ \\  
 \hline
 MREV16 & $H_{D}^{0, 1}$, $H_{DO}^{1}$, $H_{\alpha D}^{1}$, $H_{\epsilon D}^{1}$ & $\frac{1}{3}\sum_{i}a_{i}(S_{zi})$ & $1/3$ \\  
 \hline
 BR24 & $H_{D}^{0, 1, 2(\delta), 3(\delta)}$, $H_{\alpha D}^{1}$ & $\frac{2}{9}\sum_{i}a_{i}(S_{xi}+S_{yi}+S_{zi})$ & $2/3\sqrt{3}$\\  
 \hline
CORY48 & $H_{D}^{0,1,2(\delta),3(\delta)}$, $H_{O}^{0, 1}$, $H_{DO}^{1}$, $H_{\epsilon}^{0, 1, 2}$, $H_{\epsilon D}^{1}$  &  0  &  \\  
 \hline
 YXX24 & $H_{D}^{0, 1}$, $H_{O}^{0, 1, 2}$, $H_{DO}^{1}$, $H_{\epsilon}^{0, 1, 2}$, $H_{\epsilon D}^{1}$  & 0 &  \\  
 \hline
 YXX48 & $H_{D}^{0,1}$, $H_{O}^{0, 1, 2}$, $H_{DO}^{1}$, $H_{\epsilon}^{0, 1}$, $H_{\epsilon D}^{1}$   & 0 &  \\  
 \hline
\end{tabular}
\caption{Comparison of dipolar decoupling sequences. Notation: $D$ - dipolar, $O$ - resonance offset, $\epsilon$ - over or under rotation error, $\alpha$ - phase transient error, $\delta_{i}$ - chemical shift, $h_{i}$ - disorder field, $\Delta\omega$ - resonance offset frequency. Superscripts refer to the order of terms in the Magnus expansion that are decoupled; $(\delta)$ in the superscripts denote that the terms are zero only for $\delta$-function pulses.} 
\label{table:sequence_comparison}  
\end{table*}}

\subsection{Dipolar Decoupling Sequences}
\label{sec:sequences}
Using AHT, several multiple-pulse sequences that perform coherent averaging of dipolar interactions in the spin space have been developed. These dipolar decoupling sequences can be divided into two broad classes. 
Time-suspension sequences aim to set $H_{\text{eff}} = 0$ (in Equation~\ref{eqtn:Magnus_expansion_1}), at least to the lowest order, while spectroscopic sequences set the zeroth order effective dipolar Hamiltonian, $H_{D}^{0}=0$, but preserve the chemical shift, disorder, and resonance offset terms. In high-resolution NMR spectroscopy, efforts are made to suppress disorder and optimize offsets to allow accurate measurement of chemical shifts~\cite{haeberlen_high_2012}.  In the absence of such efforts, the results from spectroscopic sequences are strongly compromised by the local disorder.

The effective zero-order Hamiltonian for the spectroscopic sequences is $H_{\text{eff}}^{0}=\Delta_{\text{SF}}\sum_{i}a_{i}(S_{z'}^{i})$, where $\Delta_{\text{SF}}$ is a sequence-specific scaling factor, $a_{i}=\delta_{i}+ h_{i} + \Delta\omega$, and $z'$ is the quantization axis for the effective Hamiltonian.

At higher spin concentrations, decoupling sequences do not use $\pi$ pulses, as these do not refocus the dipolar interactions between similar spins.
The shortest pulse sequence that partially refocuses dipolar interactions to form an echo is the solid echo~\cite{mansfield_multiple-pulse_1981} sequence, consisting of two $\pi/2$ pulses with a 90-degree phase offset between them. However, the shortest cyclic sequence that sets the zeroth order average dipolar Hamiltonian to zero is the WHH sequence, a symmetrized combination of two solid echoes. Still, it only manages to refocus the lowest order term in the effective dipolar Hamiltonian ($\bar{H}_{\text{D}}^{0}$) for infinitesimal $\delta$-function pulses~\cite{waugh_approach_1968}. Symmetrizing sequences can improve effectiveness by setting all odd order terms in the average Hamiltonian to zero~\cite{mansfield_symmetrized_1973}. The MREV8 sequence improves on the WHH by being robust to finite pulse widths and RF inhomogeneities~\cite{rhim_analysis_1973}. MREV16 combines two MREV8 sequences~\cite{ladd_coherence_2005}. In addition to the properties of MREV8, MREV16 is particularly effective in protecting $Z$ magnetization ($Z=\sum_i S_z^i$), as it engineers an effective Hamiltonian in the Z direction. BR24~\cite{burum_analysis_1979} is the best-known spectroscopic sequence and cancels the effective dipolar Hamiltonian to the third order for ideal pulses. The time-suspension sequences still improve upon the decoupling ability by canceling out other interactions in addition to the dipolar Hamiltonian.  

To our knowledge, all spectroscopic sequences have been designed using AHT.  Several time-suspension sequences have also been designed using AHT, including the CORY48 sequence, which has long been considered the ``gold standard'' for decoupling experiments~\cite{cory_time-suspension_1990}. The YXX24 (also called Peng24 in~\cite{stasiuk_frame_2023})  and YXX48 sequences were discovered using reinforcement learning techniques and were `trained' to be robust to common experimental errors~\cite{peng_deep_2022}. The specific set of pulses and delays for all sequences are included in Appendix~\ref{sec_supp:Sequences}.

Table~\ref{table:sequence_comparison} shows the AHT analysis up to second order for these sequences. Experimentally, sequence performance depends on a number of  key factors --- higher order terms in the Magnus expansion and experimental constraints such as finite pulse widths, and pulse errors. The first is inherent in sequence design, while the latter results from imperfect experimental control. Extensive analytical results are available for the shorter sequences incorporating the effect of pulse errors into the average Hamiltonian calculation~\cite{gerstein_transient_1985, haeberlen_resonance_1971}.   Analytical calculations quickly become difficult for longer sequences.
We focus on
four spectroscopic sequences, WHH, MREV8, MREV16, and BR24 and three time-suspension sequences CORY48, YXX24 and YXX48 and compare them using both numerical simulations and experiments in the following section.

\section{Sequence Comparisons}
\label{sec:Comparisons}
\subsection{\label{sec:sims} Numerical Simulations}
Numerical simulations of the system evolution under decoupling allow us to compare the performance of the different sequences and better understand the effect of pulse errors even for longer sequences where analytical results are unavailable. Additionally they allow us to explore both the effect of individual errors and the interplay between different effects.

\begin{figure*}[]
\begin{centering}
    \includegraphics[width=0.45\textwidth]{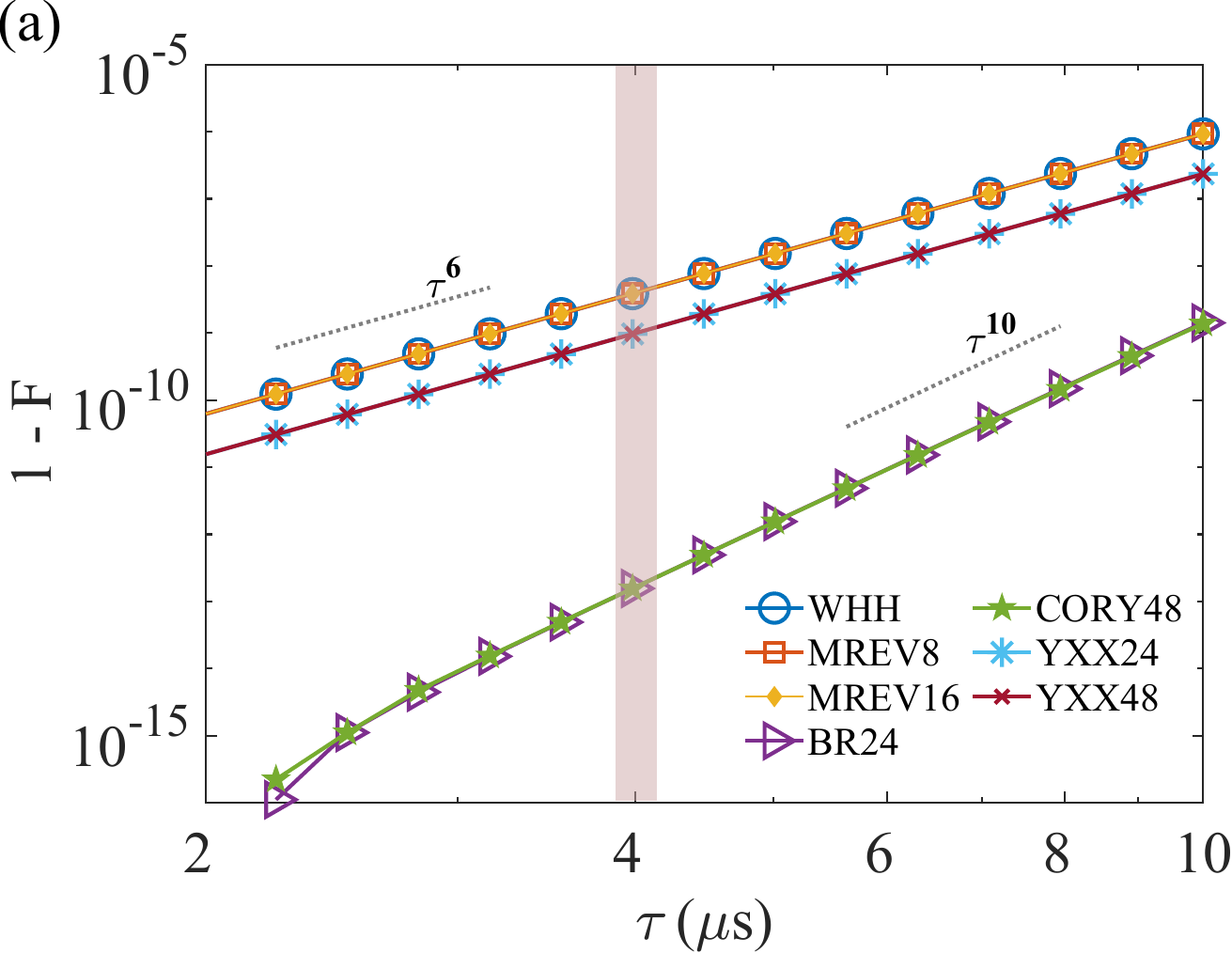} \hspace*{0.03in}
    \includegraphics[width=0.45\textwidth]{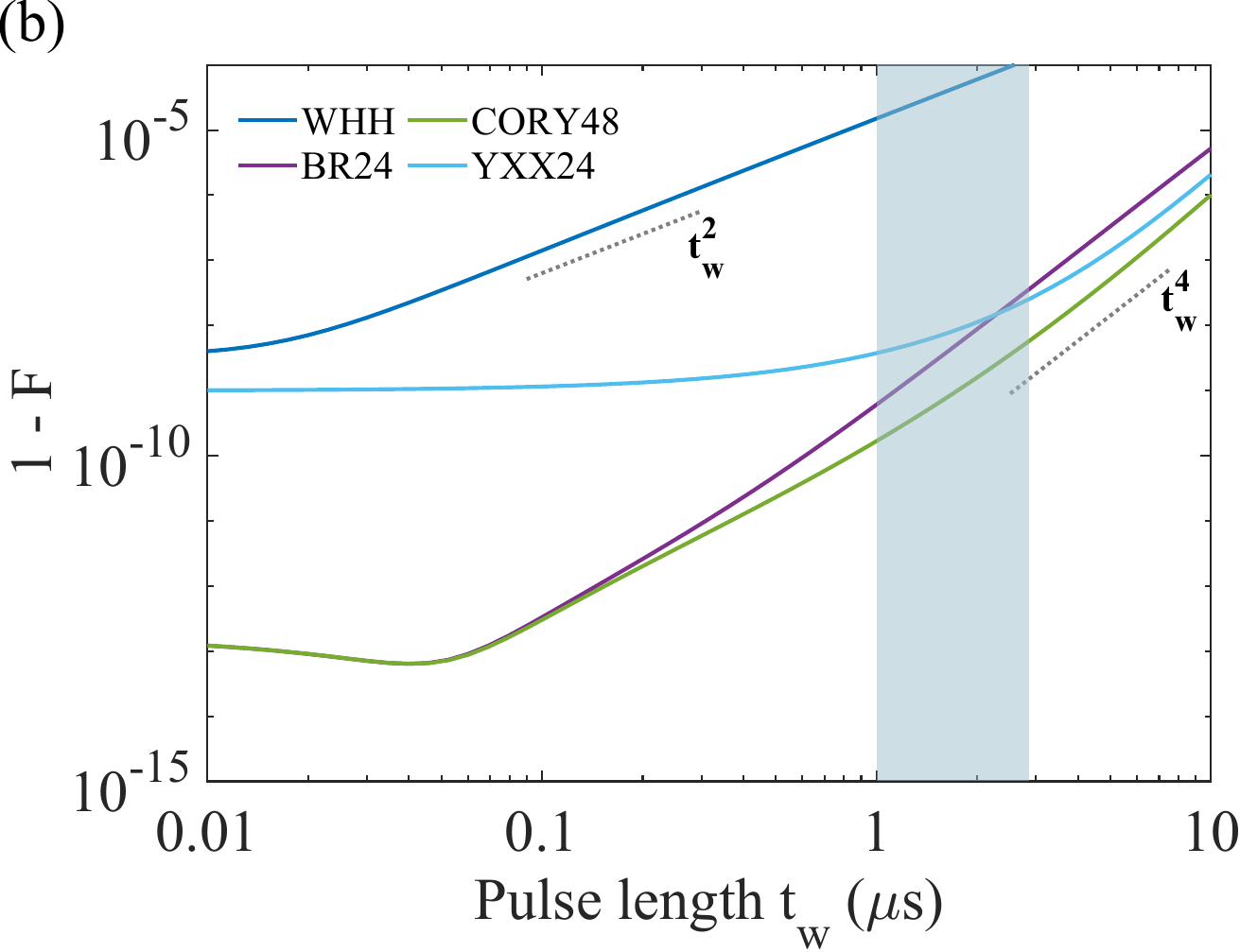} 
    \caption{(a) Dependence of sequence fidelities on the inter-pulse spacing $\tau$ in the ideal, $\delta$-pulse limit. (b) Dependence of sequence fidelities on the pulse width $t_w$ at $\tau=4$ $\mu$s (red shaded region in (a)). Blue-shaded region indicates the typical pulse widths (1-3 $\mu s$) used in our experiments. All other pulse errors are set to zero.}
	\label{fig:tau_and_tw_simulations} 
 \end{centering}
\end{figure*}

We perform numerical simulations on a small system of 8 spins, and average the results over 16 sets of dipolar couplings, each set sampled from a Gaussian distribution with $3\sigma=5000$ Hz set as the approximate maximum dipolar coupling strength. We choose a mid-range value of 5000 Hz in our simulations since the maximum dipolar coupling strength in solids that are frequently used to study NMR spin dynamics varies from less than 500 Hz ($^{1}$H-$^{1}$H coupling in Adamantane) to 10-15 kHz ($^{19}$F-$^{19}$F coupling in flourapatite and calcium flouride). Experimental unitaries ($U_{\text{exp}}$) over one cycle of each pulse sequence are calculated. In the ideal case, this can be decomposed into rotations by instantaneous $\delta$-function pulses along axes specified by the pulses and evolution under the system (dipolar) Hamiltonian during inter-pulse delays. We then compare $U_{\text{exp}}$ to the theoretical unitary operator expected from the ideal sequence, $U_{\text{th}}$, using the fidelity metric $F=\text{Tr}(U_{\text{th}}^{\dag}U_{\text{exp}}^{1/M})$, where $M$ is the length of the decoupling sequence. $U_{\text{th}}$ is set to the identity in our studies to probe the efficiency of decoupling. Since the sequences used have different cycle times $t_c = M\tau$, the experimental unitary is scaled by the inverse of the sequence length $M$ to obtain an effective unitary over $\tau$ and ensure a fair comparison between sequences. 

Figure~\ref{fig:tau_and_tw_simulations}(a) compares the infidelity ($1-F$) of the sequences with respect to the inter-pulse delay $\tau$ under `ideal' conditions of $\delta$-pulses and no pulse errors. We see three distinct ``subsets" of sequences based on infidelity scaling. Infidelities of WHH, MREV8, MREV16, YXX24, and YXX48 scale similarly with the inter-pulse spacing, with different constant pre-factors. CORY48 and BR24 both show lower infidelity, but faster scaling, indicating that the leading terms contributing to the infidelity are of higher order in the Magnus series. The first set of sequences cancels the dipolar Hamiltonian to the first order, while the second set achieves cancellation to the $3^{\text{rd}}$ order under ideal pulses. This is consistent with the scaling with $\tau$ seen in Figure~\ref{fig:tau_and_tw_simulations}(a). Note that, if $H^{(m)}$ is the leading order effective Hamiltonian, the fidelity of the unitary propogator at time $t=nt_c=nM\tau$, assuming $U_{\text{th}}$ is set to identity, becomes $F = \Tr(\exp(-iH_{\text{eff}}n\tau))$. Employing the series expansion of the exponential and truncating at the second order term for small values of $\tau$, we have $F = \Tr(\openone-iH_{\text{eff}}n\tau-(1/2!)(H_{\text{eff}}n\tau)^2)$. For infinitesimal pulses with no errors, we can write $H_{\text{eff}} \sim H^{(m)} \equiv \tau^m h_{\text{eff}}$, with $h_{\text{eff}}$ consisting only of linear or bilinear spin operators and constant factors. We verified numerically that $\Tr(h_{\text{eff}})=0$ for average Hamiltonian terms of orders up to five. The infidelity scaling for a leading order of $m$ then becomes $\propto \tau^{2(m+1)}$ from the second order term in the Taylor expansion. This is the scaling that we see in Figure~\ref{fig:tau_and_tw_simulations}(a).

For better visualization of sequence behavior, we show only WHH and YXX24 as representative examples of the first class and BR24 and CORY48 as representative examples of the second class in the subsequent figures. More numerical results on all seven sequences can be found in Appendix~\ref{sec_supp:representative_sequences}.

In practice, the pulses are never infinitesimal and have a finite width $t_{w}$. We need to account for system evolution under the internal Hamiltonian during the pulse application time, altering the unitary operator for a $\pi/2$ $X$ pulse from $U_{\text{pulse}}=e^{-iX\pi/2}$ to $U_{\text{pulse}}=e^{H_{\text{sys}}t_w+iX\pi/2}$ where $X = \sum_i S_x^i$ (assuming a perfectly calibrated pulse). This consideration also alters the average Hamiltonian obtained. For example, the WHH sequence only achieves dipolar decoupling for $\delta$-function pulses.  Figure~\ref{fig:tau_and_tw_simulations}(b) shows sequence performance with varying pulse widths at $\tau = 4$ $\mu$s. WHH suffers a significant reduction in effectiveness as soon as finite pulse widths are introduced. CORY48 and BR24 show a lower decrease in absolute fidelity and retain comparable performance in the presence of finite pulses.
WHH, for finite pulses, has a linear dependence on pulse width in the zeroth order average Hamiltonian, causing sequence infidelity to scale as $t_w^2$. For all other sequences considered here, the leading order term in $t_w$ is the 2$^{\text{nd}}$ order dipolar Hamiltonian. In this case, the infidelity scaling with respect to the pulse length also depends on the dipolar coupling strength and the inter-pulse delay that are present to the 2$^{\text{nd}}$ power in this term. Such complex interplay of parameters is difficult to tease out analytically, and numerical results provide an easier approach. Here, for the chosen dipolar coupling and inter-pulse spacing, we see that the insensitivity to pulse length is only for short pulses.  As the pulses get longer the fidelity drops very quickly, scaling by the 4$^{\text{th}}$ power of $t_w$ for all sequences except WHH.

\subsection{\label{sec:expts} Experimental Results}
We used a powder sample of the the plastic solid adamantane (C$_{10}$H$_{16}$), which is a model 3D spin system, to compare the relative performance of the decoupling sequences described in the previous sections. This solid has a diamond-like arrangement of carbon atoms and high degrees of internal motion that cancel out intramolecular dipolar interactions. The molecules are further arranged in an fcc lattice. The maximum dipolar coupling strength between protons in the system is approximately $\omega/2\pi=420$ Hz and the $^1$H NMR peak has a linewidth of $~13.8$ KHz~\cite{ernst_high-speed_1998,alvarez_localization-delocalization_2015}. 

We used a 7 T (300 MHz proton Larmor frequency) Oxford superconducting magnet and a commercial Bruker spectrometer to conduct the experiments. The lack of room-temperature shims in the NMR magnet introduces a non-negligible degree of local magnetic field inhomogeneity in these experiments, resulting in 244 Hz linewidth for the proton NMR signal for a water sample (see~\ref{fig_supp:signal_details}(c) in Appendix~\ref{sec_supp:Expts}). A water sample was used to calibrate the $\pi/2$ pulses using standard pulse train experiments~\cite{gerstein_transient_1985}. 

\begin{figure*}[]
\begin{centering}
    \includegraphics[width=0.8\textwidth]{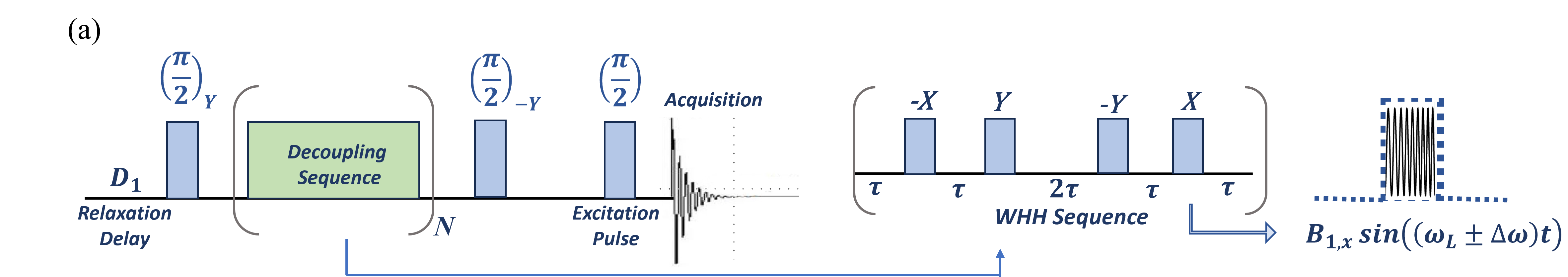} \\
    \vspace*{0.25in}
    \includegraphics[width=0.31\textwidth]{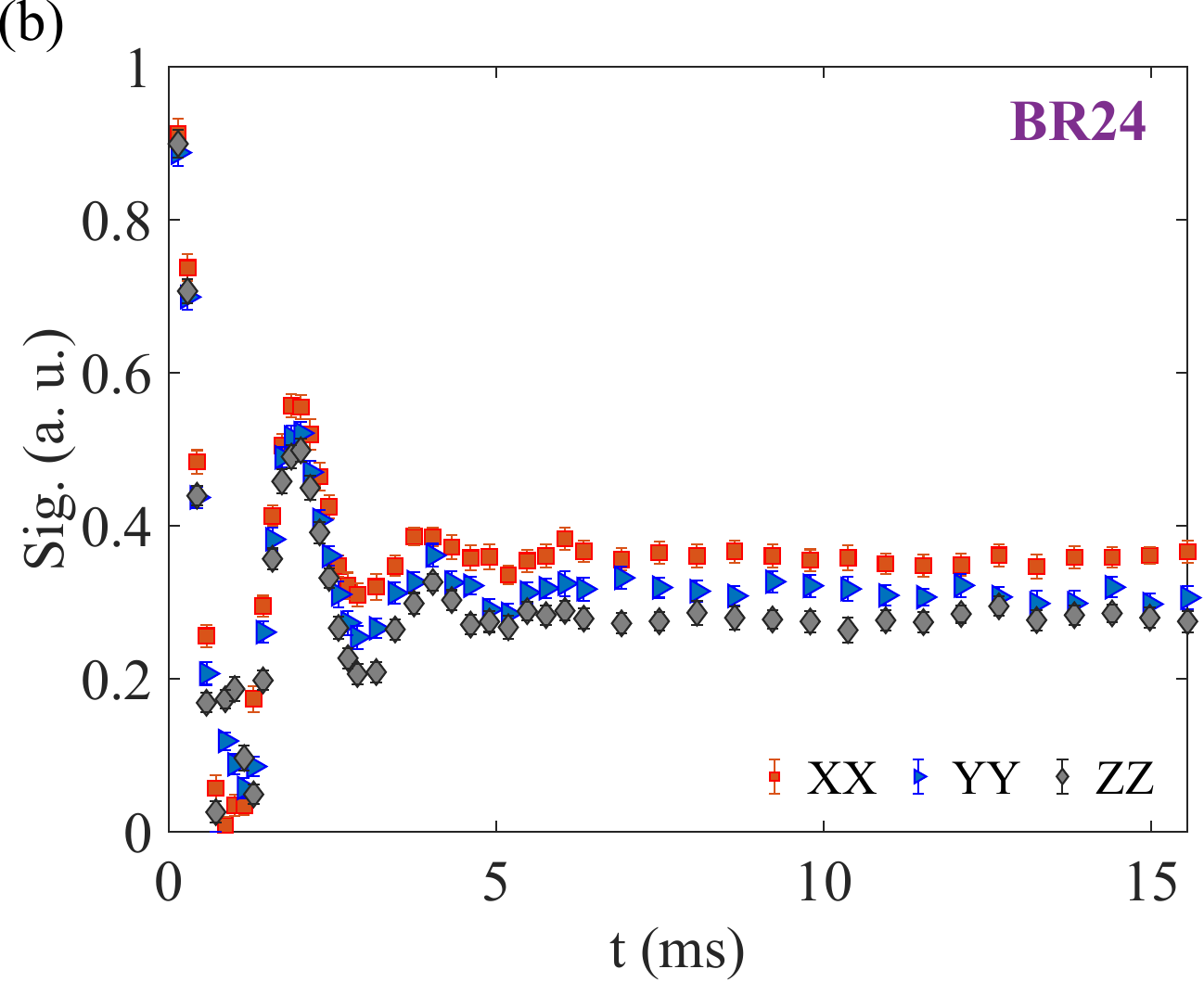} \hspace*{0.05in}
    \includegraphics[width=0.31\textwidth]{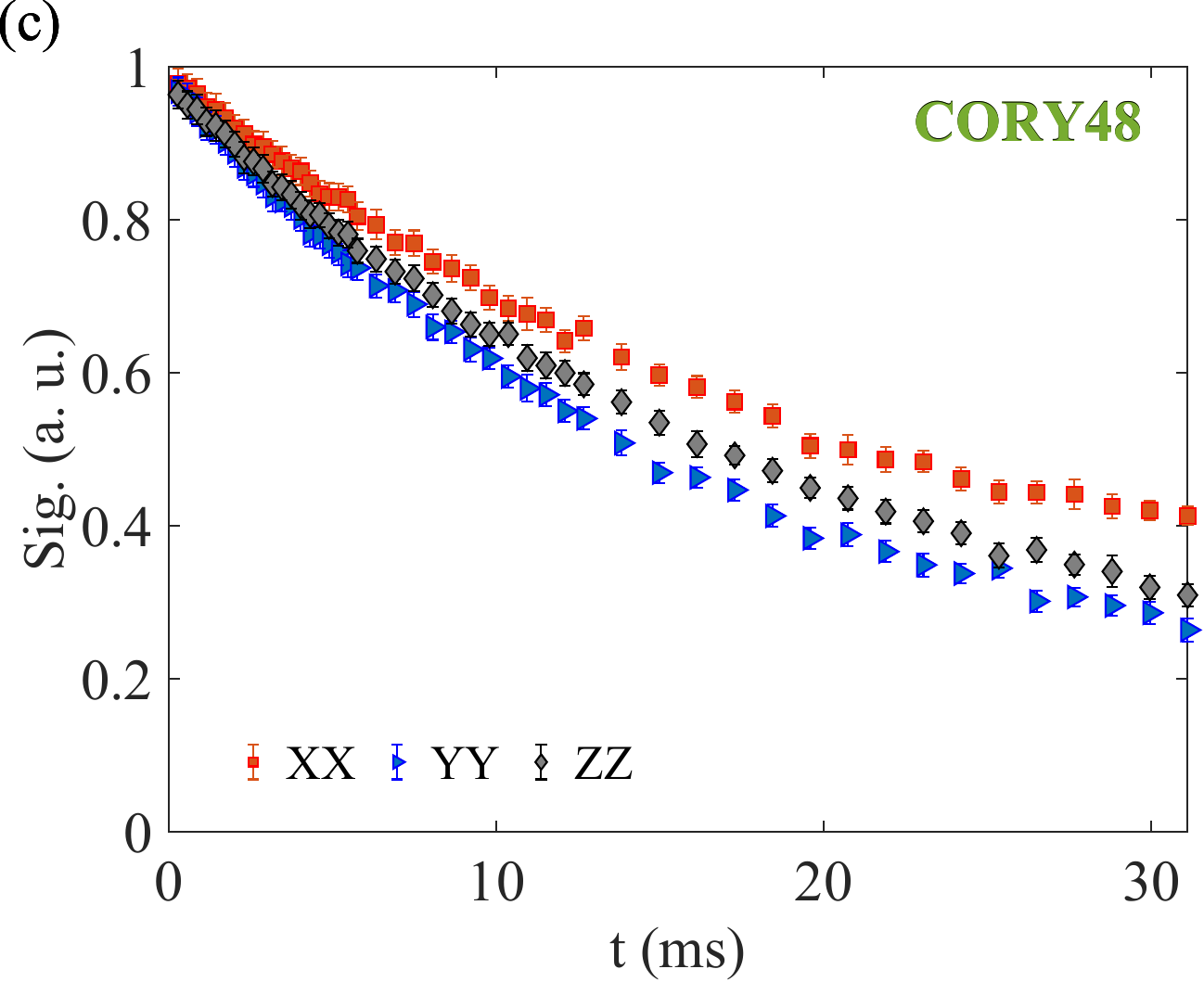}     \hspace*{0.05in}
    \includegraphics[width=0.31\textwidth]{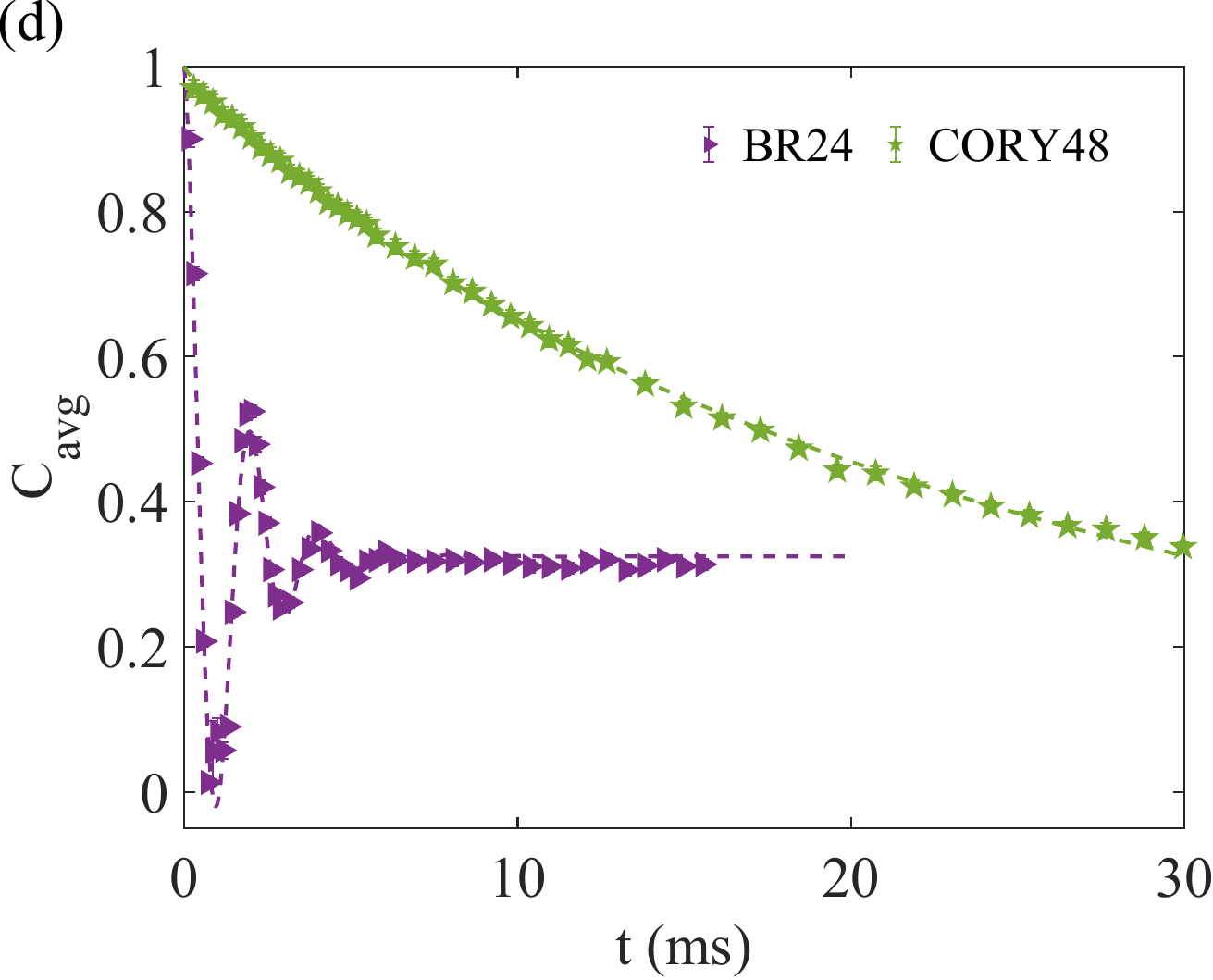} 
   	\caption{(a): Experimental scheme for the {\em XX} autocorrelation experiment. The Figure also shows the WHH sequence as an example and the important pulse parameters. Decay curves for the signal from the $X$, $Y$ and $Z$ autocorrelation experiments for (b) BR24 and (c) CORY48 sequences for an off-resonance frequency, $\Delta=-1.25$ kHz and $\tau = 4$ $\mu$s. (c) $\text{C}_{\text{avg}}$ with fits (dotted lines) for both sequences.}
	\label{fig:exp_scheme_and_autocorrelation_data} 
\end{centering}
\end{figure*} 

 Ideally, to quantify the average performance of the sequence without favoring a particular initial state or readout axis, we should measure the fidelity of the sequence, $F=\text{Tr}(U_{\text{th}}^{\dag}U_{\text{exp}}^{1/M})$. This metric would require performing process tomography on the system, which is not possible since we are limited to collective control of the spins. Instead, in this work, to experimentally estimate sequence fidelity, we use the average autocorrelation $C_{\text{avg}}$, the geometric mean of the three autocorrelations, $C_{\text{avg}}=(C_{xx}C_{yy}C_{zz})^{\frac{1}{3}}$, where $C_{xx}=\frac{1}{2^N}\left\langle S_{x}(t)S_{x}(0)\right\rangle_{\beta=0}$. This quantity has previously been shown to closely approximate the state fidelity~\cite{peng_deep_2022}. Figure~\ref{fig:exp_scheme_and_autocorrelation_data}(a) shows the experimental scheme with spin initialization and readout to measure the $C_{xx}$ autocorrelation. In this experiment, we initialize the spin system in a state with $\delta\rho(0) = X/2^N =\sum_{i}S_{x}^{i}/2^N$  (achieved using a collective rotation of the spins from their original equilibrium state, $\rho_{eq}=\frac{1}{2^N}(\openone+\epsilon Z)$, where $Z=\sum_{i}S_{z}^{i}$ and $\epsilon\sim10^{-5}$). After the initialization pulse, we apply multiple blocks of the decoupling sequences to the system. The $N^{\text{th}}$ block contains $N$ repetitions of the pulse cycle of the decoupling sequence. The inverse of the initialization pulse aligns the magnetization back along $Z$ and we wait for 100 ms for off-diagonal components to dephase, for a cleaner signal. The $X$ magnetization, $\text{Tr}(\delta\rho(t)X)/2^N$, is then measured after another excitation pulse. The NMR free induction decays are collected after each full block of the decoupling sequences. $N$ varies from 1 to 148 with an uneven sampling. We perform similar measurements for $C_{yy}$. For $C_{zz}$, no preparation pulse is used before the decoupling sequence is applied. 

To analyze the decoupling performance, we fit the normalized decay curves $C_{\text{avg}}$ as follows (see Appendix~\ref{sec_supp:normalization} for details on the normalization procedure).  For the time-suspension sequences we fit the data to a stretched exponential of the form 
\begin{equation}
    C_{\text{avg, ts}}=C_0 e^{-t^{g}/T_{2,\text{eff}}}
\end{equation} 
where $g$ is the stretch factor, while for  spectroscopic sequences we fit the data to 
\begin{equation}
C_{\text{avg, spectro}}=C_0\cos(2\pi ft)e^{-t^{g}/T_{2,\text{eff}}}+C_1,
\end{equation}
where $f$ quantifies the oscillation of the magnetization about the static effective field created by the pulse sequence. $C_1$ captures the `pedestal' from the conserved fraction of the initial magnetization, if any. 

Figures~\ref{fig:exp_scheme_and_autocorrelation_data}(b) and (c) show the $X$, $Y$, and $Z$ autocorrelation measurements for the BR24 spectroscopic sequence and the CORY48 time-suspension sequence at a finite offset frequency. The spectroscopic sequences show an oscillation of the $X$ magnetizations (or $Y$ or $Z$) for nonzero resonance offset. They may also exhibit persistent, nondecaying magnetization about some axes, even at long times, as seen in Figure~\ref{fig:exp_scheme_and_autocorrelation_data}(b). For the time-suspension sequences, on the other hand, the $X$, $Y$, and $Z$ autocorrelations exhibit very similar behavior and all three of them consistently decay to zero at long times. These contrasting behaviors can be understood by recalling that the effective Hamiltonian engineered by the spectroscopic sequences in the rotating frame is a rescaled and reoriented version of the chemical shift Hamiltonian (see Table~\ref{table:sequence_comparison}). The existence of such a preferred effective field direction would mean that the initial magnetization in this direction does not decay even at long times and that the $X$, $Y$, and $Z$ magnetizations exhibit a precessional motion about it. Since time-suspension sequences engineer the effective Hamiltonian to be zero, no such preferred directions or spin precession exist in this case.  Figure~\ref{fig:exp_scheme_and_autocorrelation_data}(d) shows the $C_{\text{avg}}$ data and corresponding fits for BR24 and CORY48.

\begin{figure*}[]
\begin{centering}
    \includegraphics[width=0.31\textwidth]{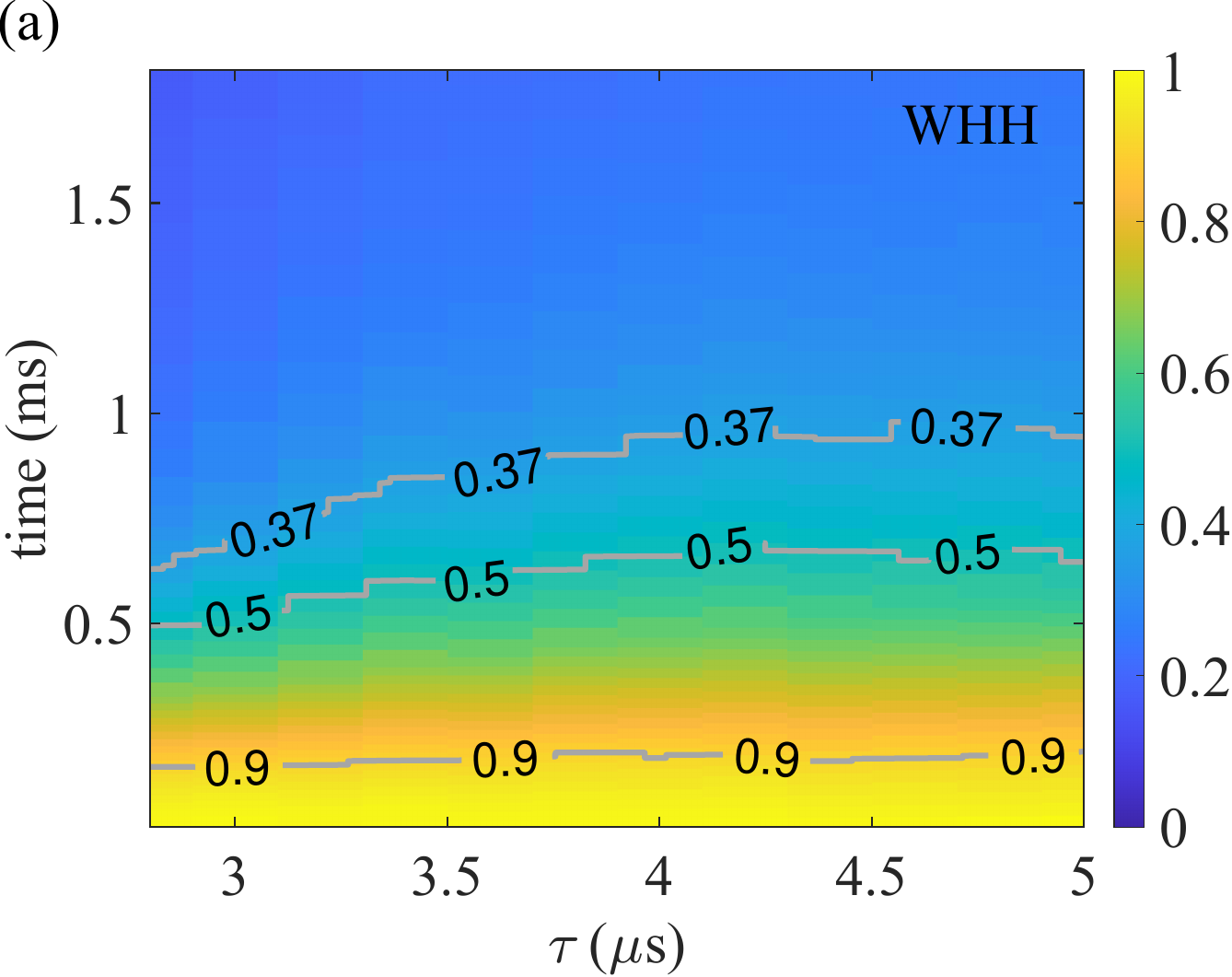} \hspace*{0.05in}
    \includegraphics[width=0.31\textwidth]{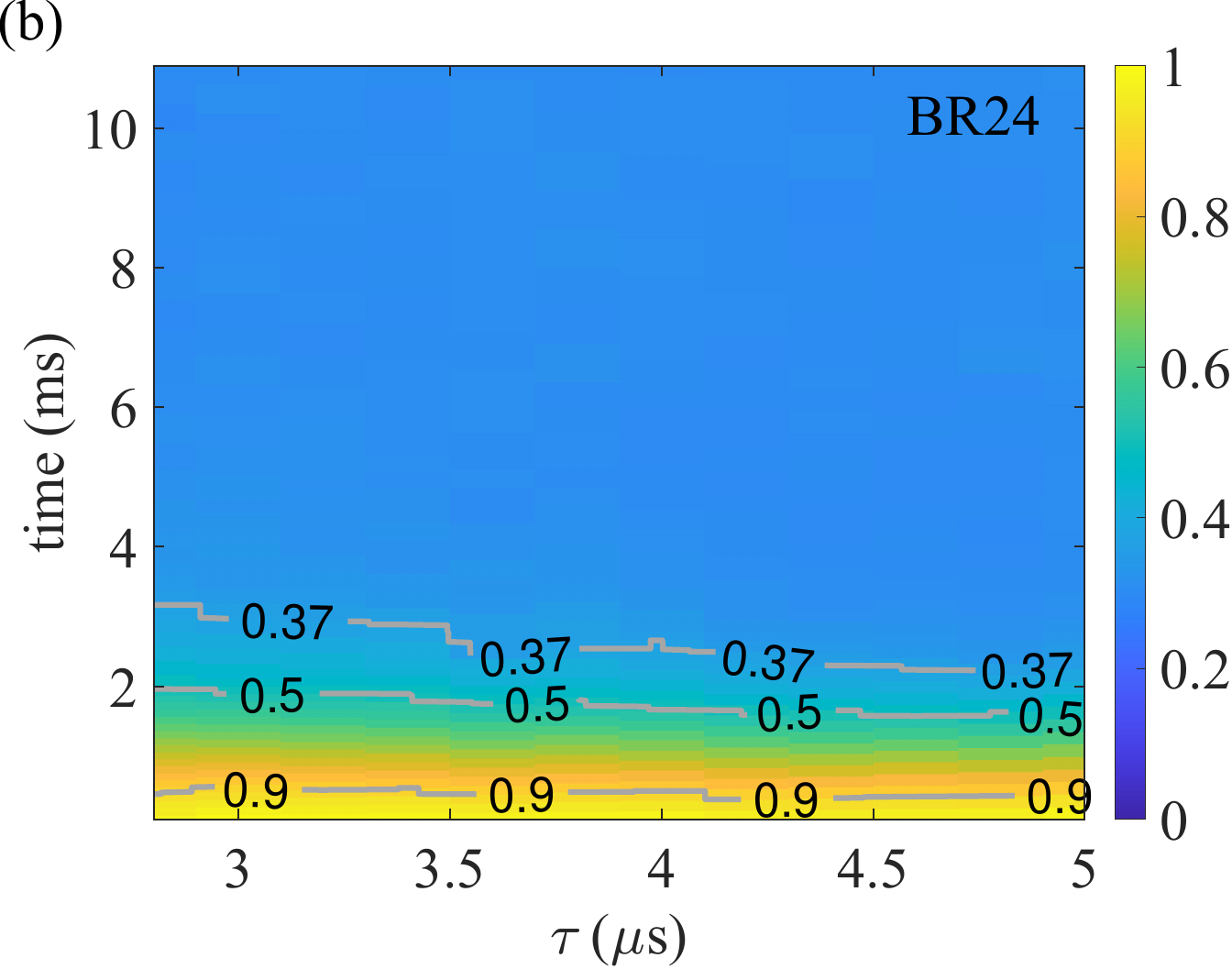} \hspace*{0.05in}
    \includegraphics[width=0.31\textwidth]{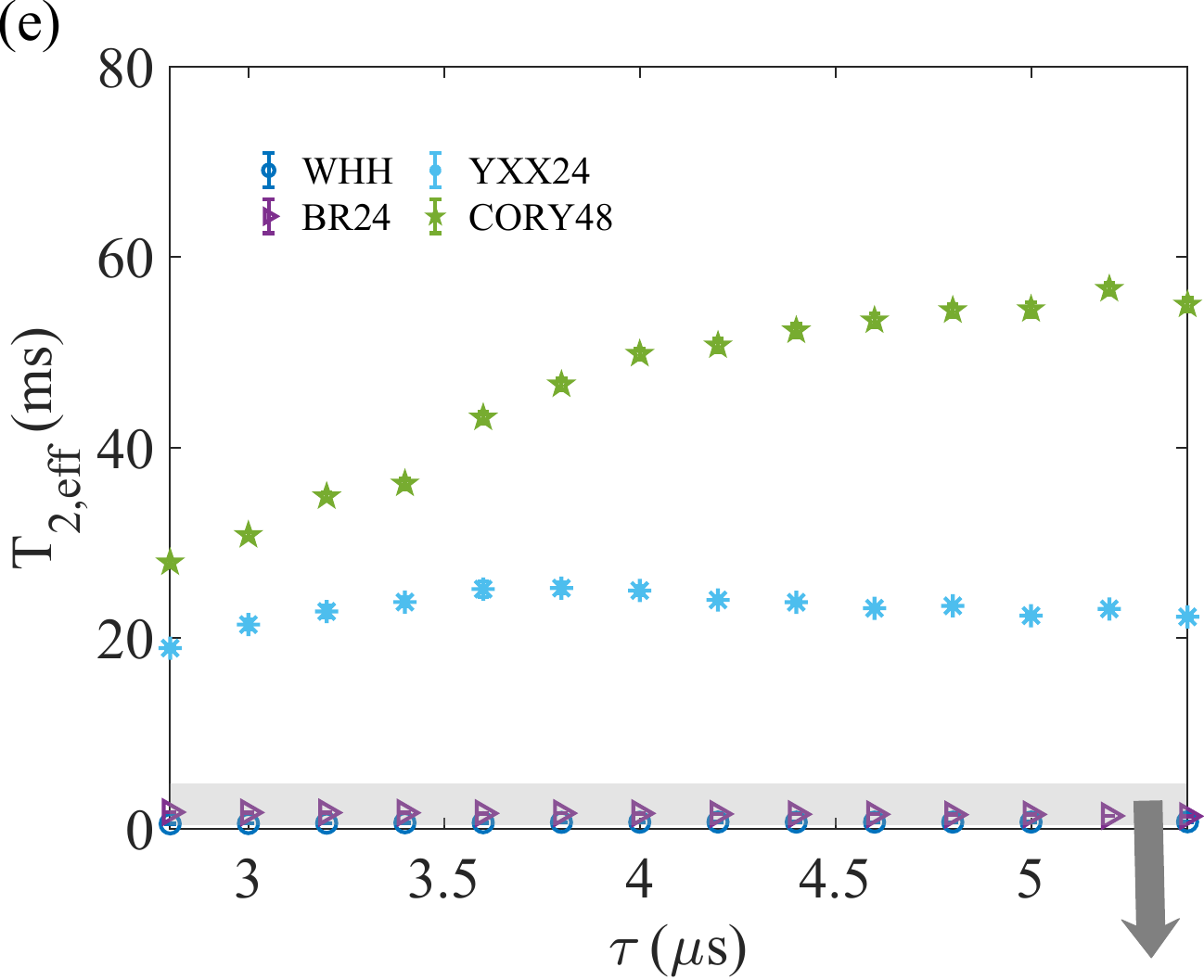}    
    \includegraphics[width=0.31\textwidth]{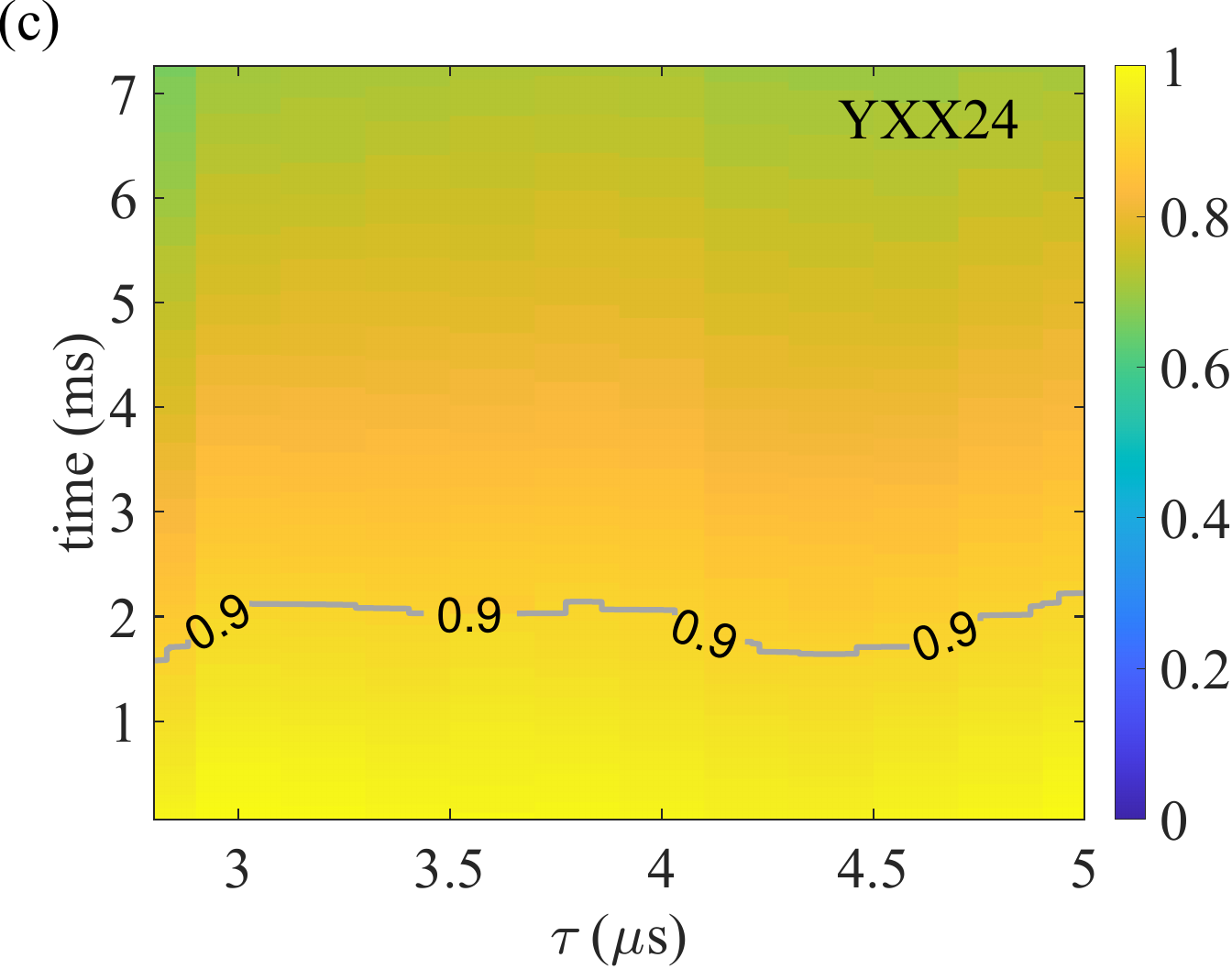} \hspace*{0.06in}
    \includegraphics[width=0.31\textwidth]{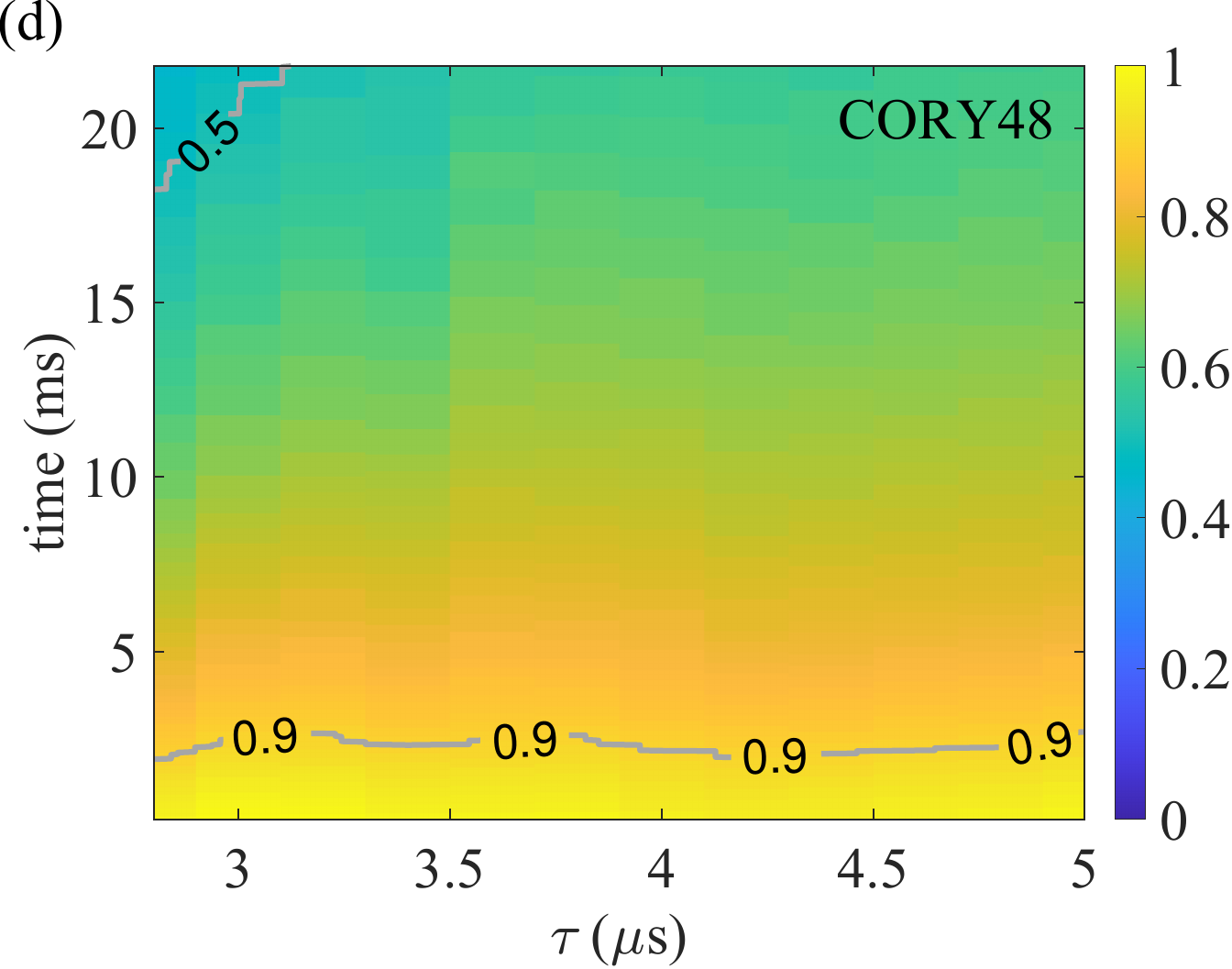}    \hspace*{0.05in}
    \includegraphics[width=0.31\textwidth]{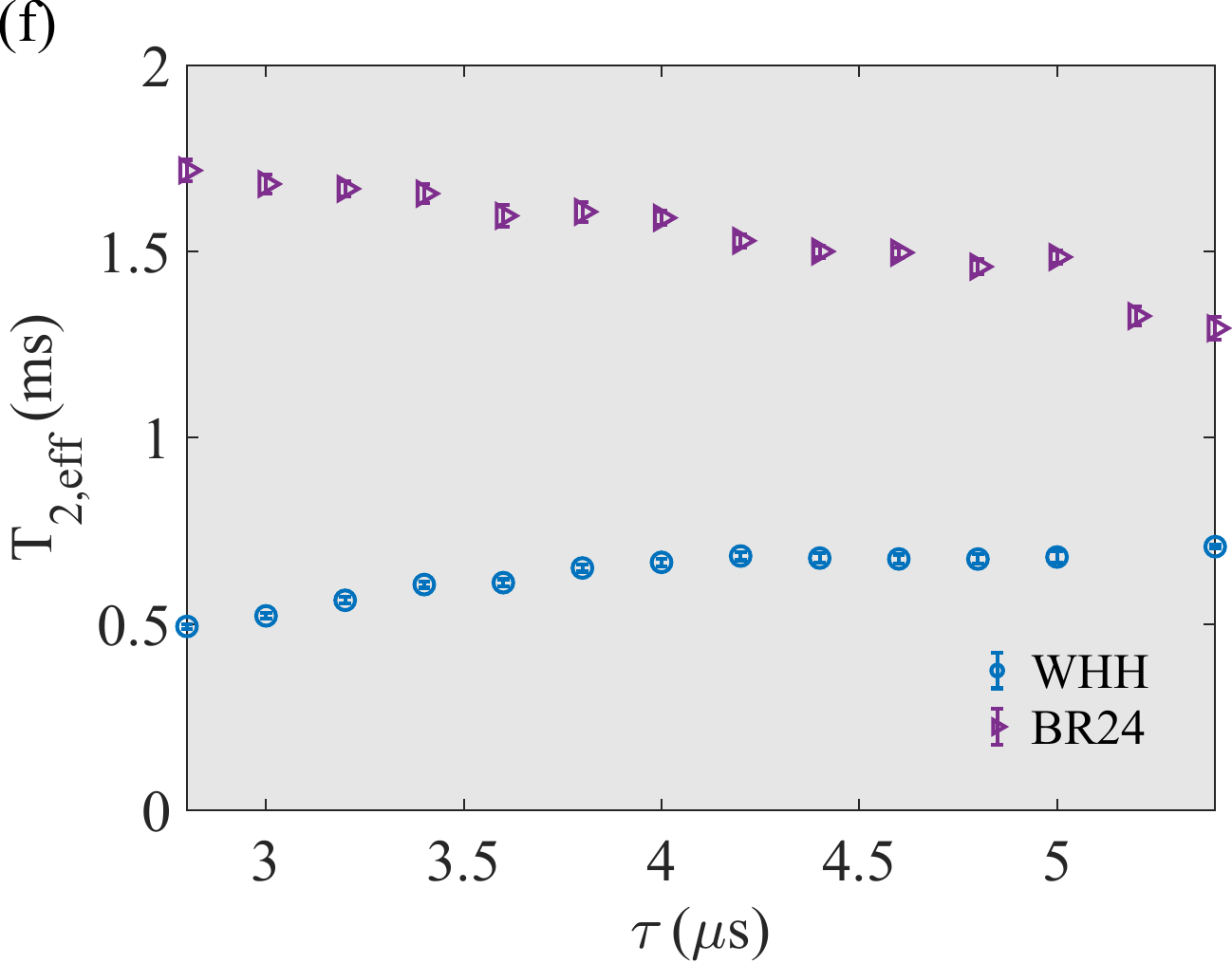} 
   	\caption{(a)-(d): Color plots showing the normalized signal intensity (geometric mean of $X$, $Y$, and $Z$ autocorrelation experiments, $C_{\text{avg}}$) for WHH, BR24, YXX24 and CORY48 sequences respectively, with experiment time on the $Y$ axis and inter-pulse spacing $\tau$ on the $X$ axis. Grey lines show the contours as a guide to the decay of the signal over time. (e): Effective coherence time $T_{2,\text{\,eff}}$ extracted from fits to the $C_{\text{avg}}$ decay curves in (a)--(d) at each $\tau$ for all sequences studied. (f): Zoomed-in version of (e) to show the spectroscopic sequences.}
	\label{fig:tau_data} 
\end{centering}
\end{figure*} 

 Figures~\ref{fig:tau_data}(a)--(d) show the experimental coherence decay curves for four representative sequences as we vary the inter-pulse delay $\tau$. The y-axis shows the total experiment time.  Figures~\ref{fig:tau_data}(e) and (f) show the dependence of the effective decay time extracted from the fits on the inter-pulse spacing. The stretch factors for these fits were generally constrained to vary between [0.5, 2.5]. They tend to cluster around 0.8 for spectroscopic sequences and 1.5 for the time-suspension sequences for low $\tau$ values and tend to diverge slightly for higher inter-pulse spacing (see Figure~\ref{fig_supp:fit_details}(b) in Appendix~\ref{sec_supp:Expts}).  The time-suspension sequences are seen to clearly perform better than the spectroscopic sequences under our experimental conditions.  Even the BR24 sequence, which is designed to refocus dipolar interactions to third order in the Magnus expansion, is seen to perform poorly.  We will return to this point later.
 
 We expect the sequences to perform their best for an intermediate value of $\tau$. At longer cycle times, the assumption that the stroboscopic evolution of the system can be approximated using the lowest order term in the Magnus expansion (`the average Hamiltonian') breaks down. The higher order terms can no longer be ignored, and sequence performance generally tends to worsen in this regime.  The question of convergence of the Magnus expansion is non-trivial and is briefly explored in Appendix~\ref{sec_supp:aht}. In the opposite limit, sequence performance generally improves as the cycle time decreases. However, in this regime, experimental errors limit sequence performance at low $\tau$ values. These limitations can crop up in two ways: (i) the electronics limit our ability to access the lowest inter-pulse spacing regimes and (ii) phase transient errors tend to become more important as pulses get closer to each other. 
 
We observe that YXX24 and BR24 show peak performance at relatively lower $\tau$ values (3.5 and 3 $\mu$s). However, for WHH and CORY48 the performance improves with $\tau$ in the regime studied, though it appears to plateau at around 5 $\mu$s. We achieve a maximum effective coherence time of 56.6 ms with the CORY48 sequence at $\tau = 5.2$ $\mu$s. 
 
\begin{figure*}[]
\begin{centering}
    \includegraphics[width=0.34\textwidth]{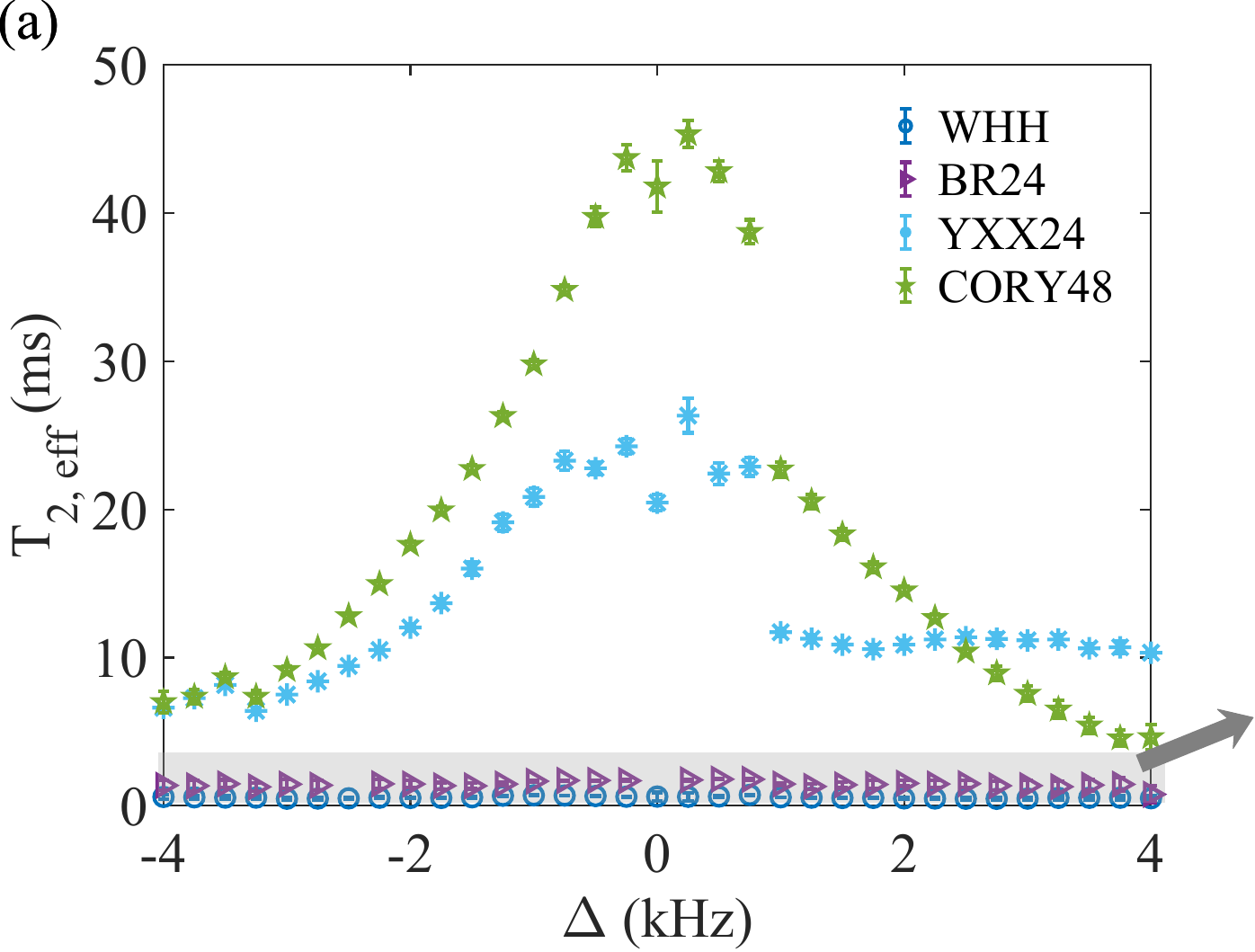} \hspace*{-0.02in}
    \includegraphics[width=0.315\textwidth]{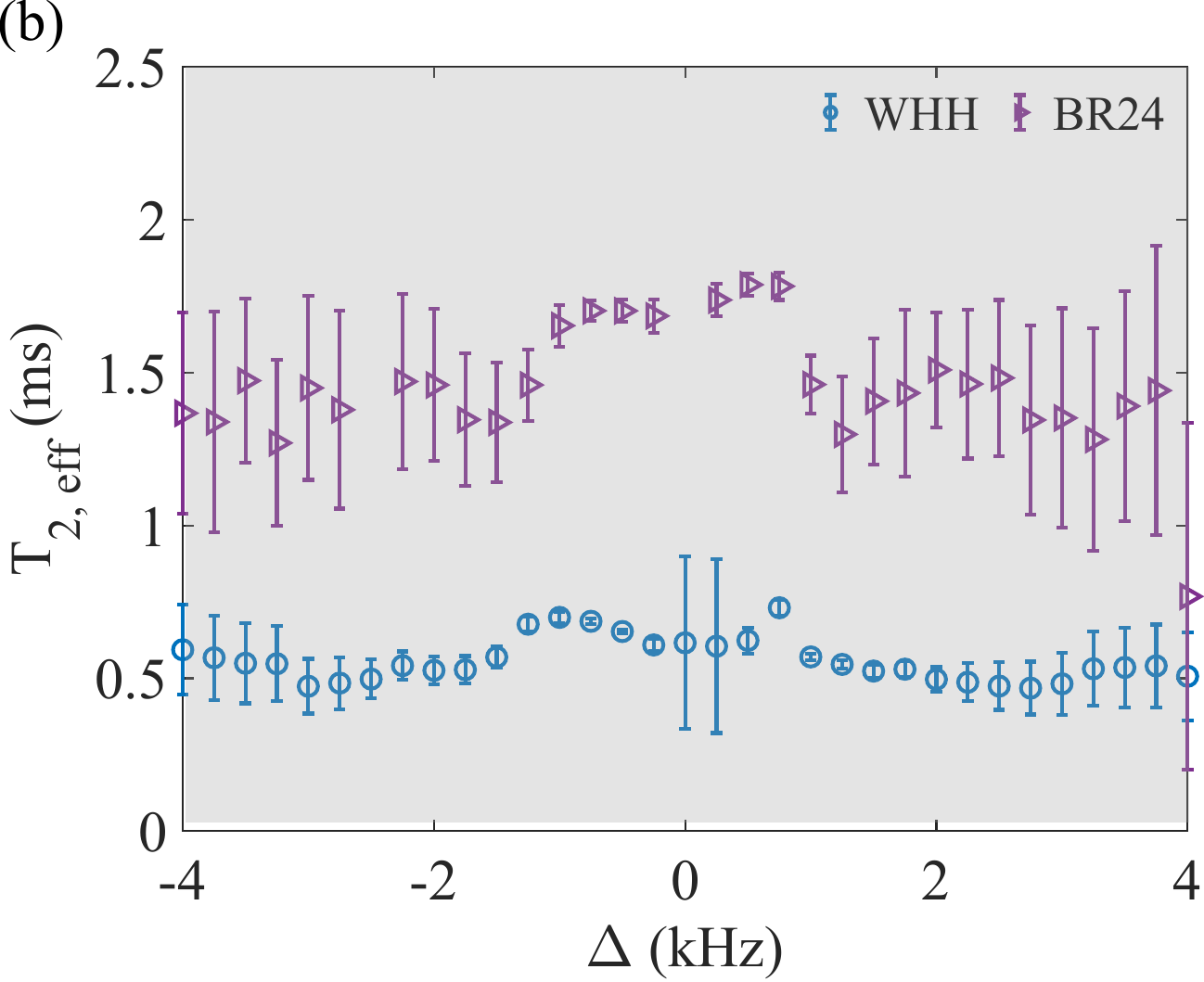} \hspace*{0.12in}
    \includegraphics[width=0.3\textwidth]{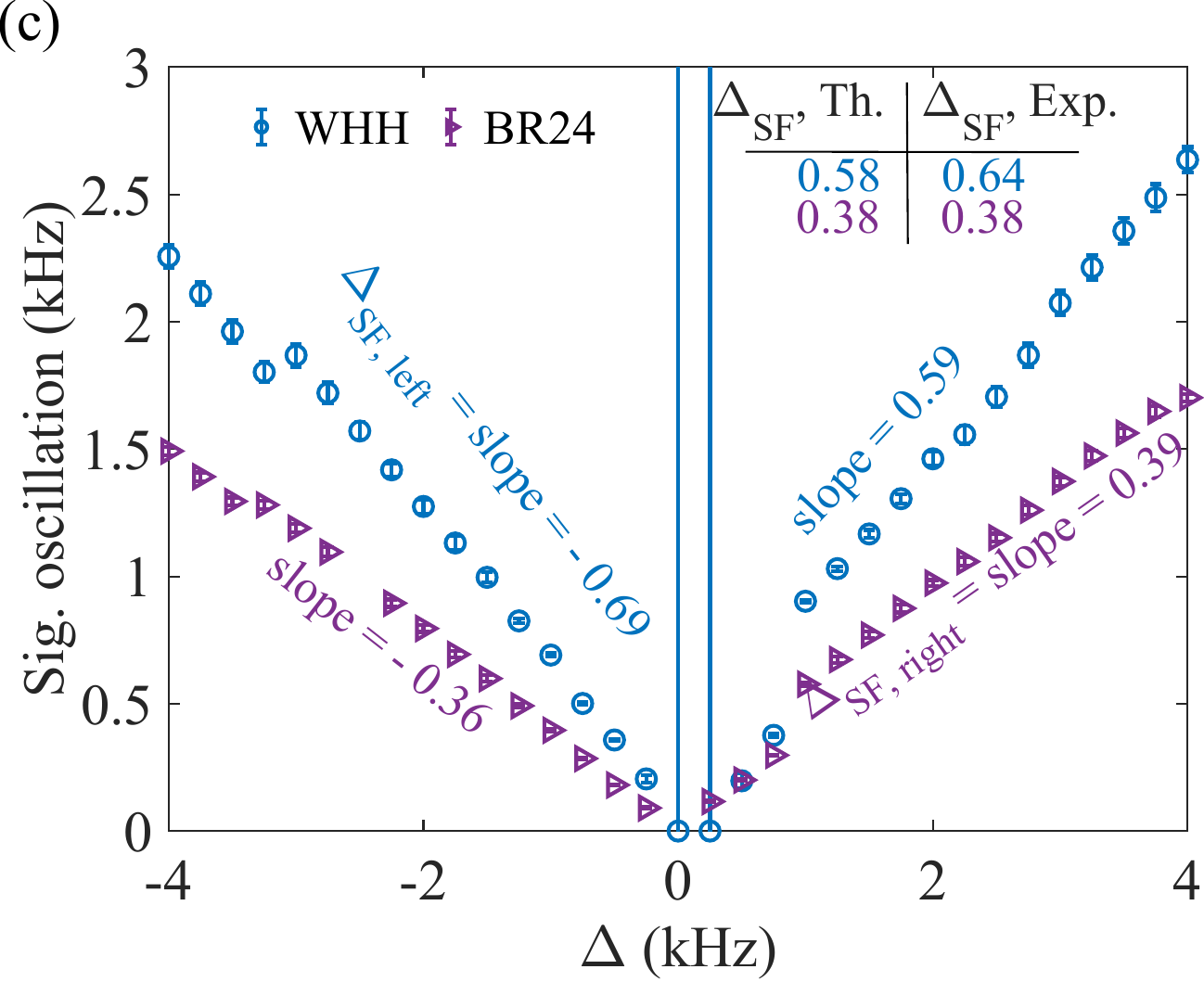}     
   	\caption{(a) $T_{2,\,\text{eff}}$ from the fits to $C_{\text{avg}}$ over a range of off-resonance frequencies for all sequences at $\tau = 4$ $\mu$s. (b) Zoomed-in version of (a) to better visualize $T_{2,\,\text{eff}}$ of spectroscopic sequences. (c) Linear scaling of the oscillation frequency of the $C_{\text{av}g}$ decay with the frequency offset $\Delta$ for the spectroscopic sequences and a comparison of the theoretically and experimentally obtained chemical shift scaling factors.}
	\label{fig:delta_data} 
\end{centering}
\end{figure*} 

\subsection{Sensing Resonance Offsets}
\label{sec:sensing_offsets}
As noted above, spectroscopic sequences aim to preserve local chemical shift ($\delta_i$) or offset ($\Delta$) information while suppressing dipolar interactions. Here we examine the ability of the different sequences to measure such frequency shifts that are key to performing spectroscopy or DC magnetic field sensing. It should be noted that while chemical shift information (or small $g$ factor variations in EPR) is desirable, the local disorder fields ($h_i$) are typically not. However, it is not possible to distinguish between the two as they have the same symmetry --- which is why DC magnetic field sensing is difficult with solid-state spin systems.  In Section~\ref{sec:Errors}A, we consider the influence of local disorder fields as an undesirable control error.

Figures~\ref{fig:delta_data}(a) and (b) show the dependence of the effective decay time on the global resonance offset frequency $\Delta$. The step size of 250 Hz we consider for these experiments is appropriate for adamantane, which has a relatively weak average dipolar coupling strength of $~420$ Hz. Stretch factors for these fits, constrained to vary within the range [0, 3], are all between 0.5 and 2 (see Figure~\ref{fig_supp:fit_details}(d) in Appendix~\ref{sec_supp:Expts}), but show a larger variation with $\Delta$, in contrast to the variation seen with $\tau$  discussed earlier. We observe the best performance for the spectrosocopic sequences slightly off-resonance, asymmetrically decaying away from the central peak. This can be explained by the second averaging effect~\cite{haeberlen_high_2012}. Time-suspension sequences perform their best on resonance and show a more or less symmetric decay around the central peak, as expected. 

To quantify the sensing efficiency of spectroscopic sequences, we examine the oscillation frequency of the $C_{\text{avg}}$ signal and compare the data with the expected scaling factors for each sequence. These data and a comparison of the scaling factors obtained experimentally and theoretically are given in Figure~\ref{fig:delta_data}(c). 
We observe, on average, an order of magnitude better coherence time with the time-suspension sequences compared to the spectroscopic sequences in all our experiments. We also note that the absolute effective decay constants we obtained for these sequences are lower than the highest values previously reported in the literature. 

\section{Control Errors}
\label{sec:Errors}
Resonance offset effects, rotation errors and phase transients comprise the significant control errors in our system.  While all errors are typically present in the system at all times, we numerically explore their effects individually in the limit of ideal $\delta$-function pulses and finite pulse widths below. 

\begin{figure*}
\begin{centering}
    \includegraphics[width=0.31\textwidth]{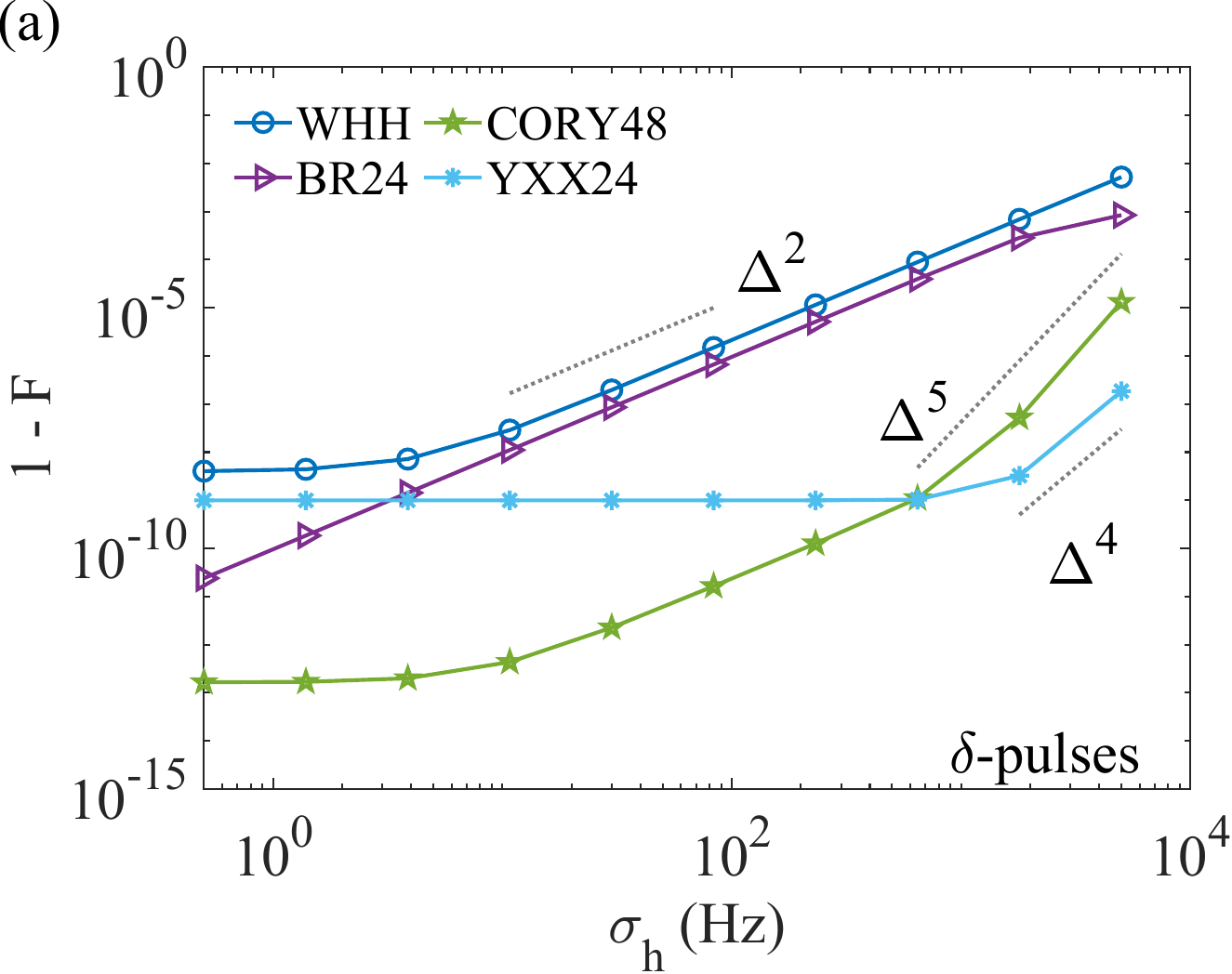}  \hspace*{0.01in}
    \includegraphics[width=0.31\textwidth]{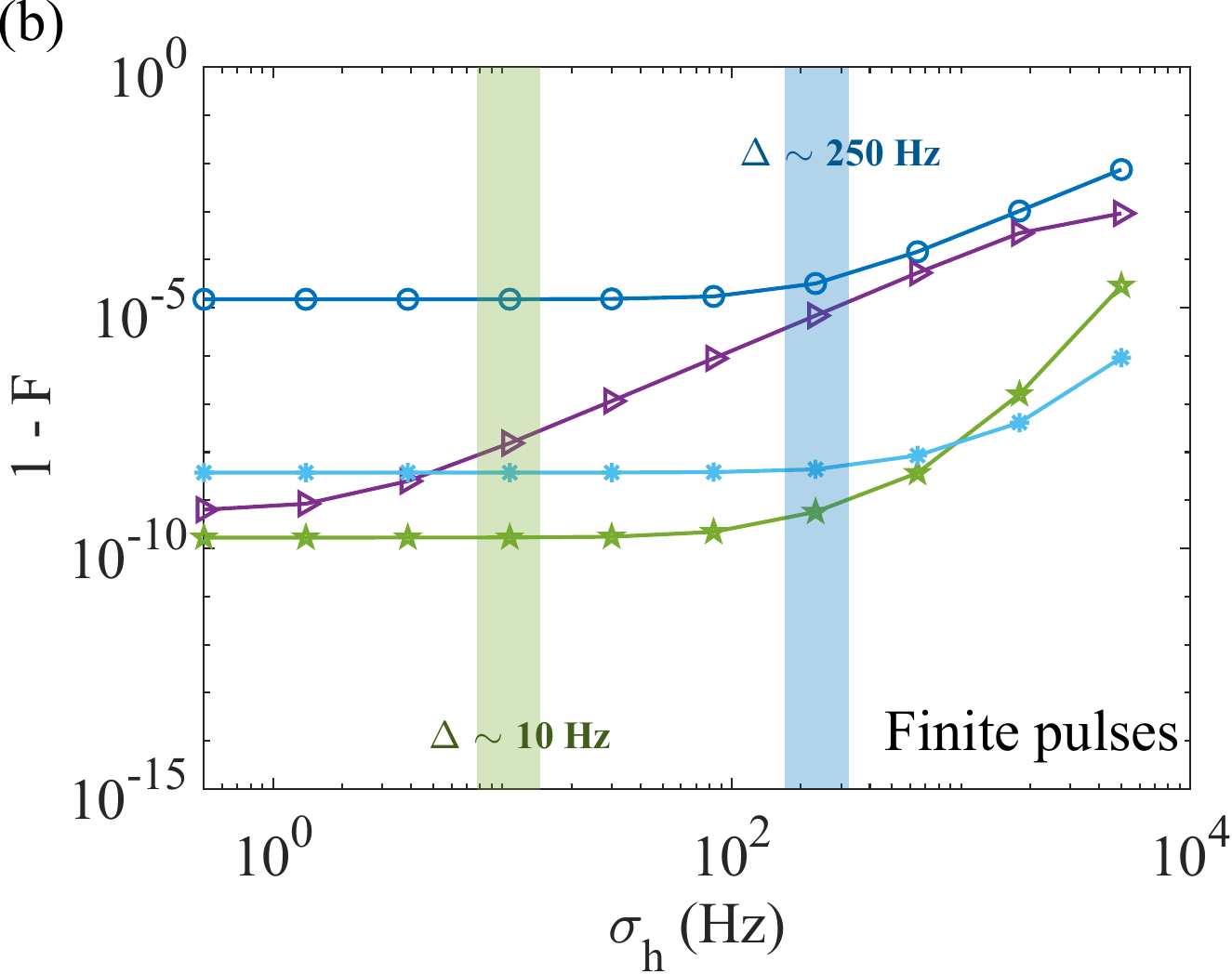}  \hspace*{0.01in}
    \includegraphics[width=0.31\textwidth]{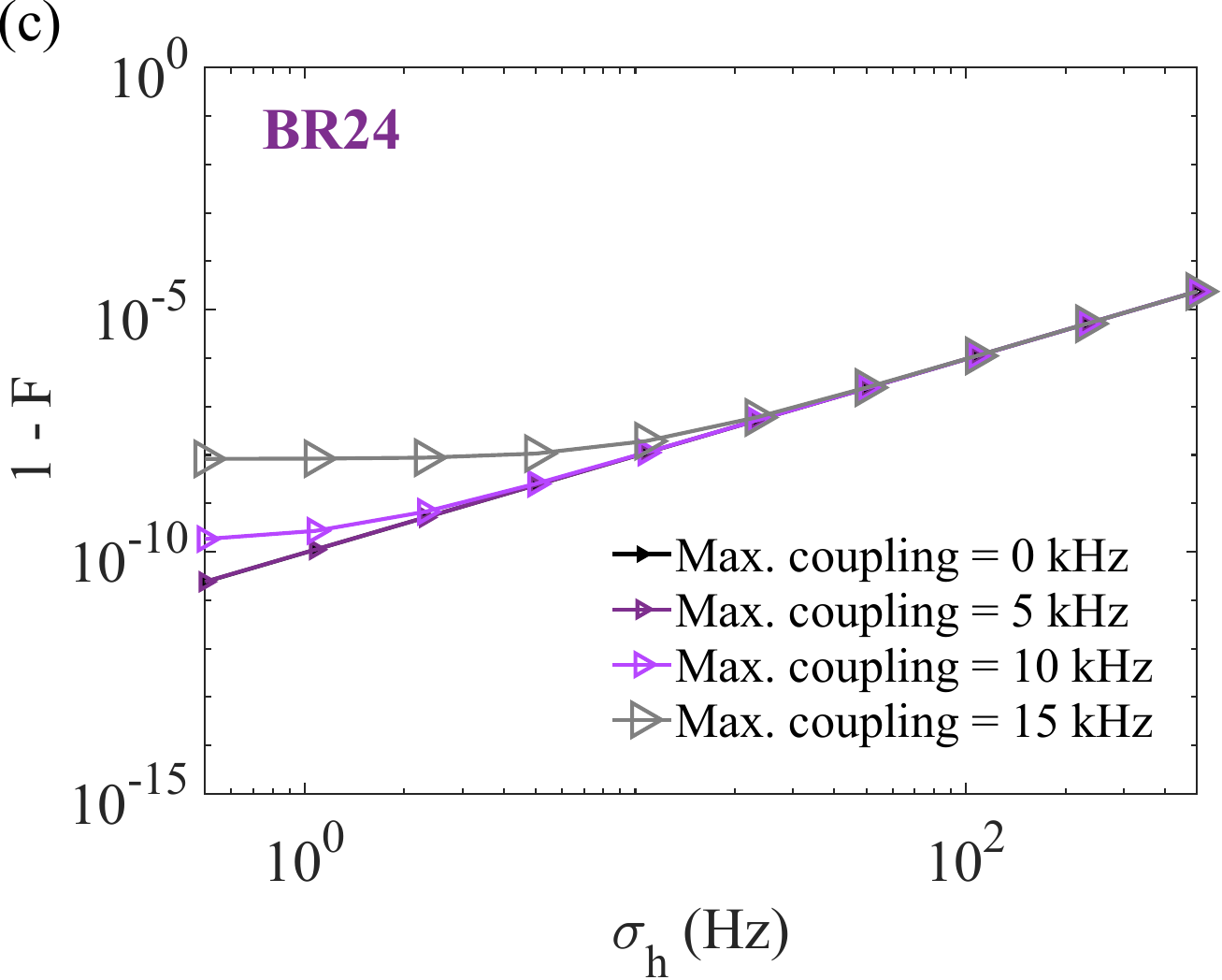}
    \caption{Effect of a distribution of resonance offset errors (disorder) on sequence fidelity for (a) ideal $\delta$-pulses, (b) finite pulses with $t_w = 1$ $\mu$s. The shaded regions are estimates of the magnetic field inhomogeneity in the two superconducting magnets we use for our experiments. The green (10 Hz) and blue (250 Hz) regions represent the shimmed, 9.4 T magnet and the unshimmed, 7 T magnet respectively. (c) Offset dependence of BR24 sequence fidelity for varying dipolar coupling strengths for finite pulses with $t_w=1$ $\mu$s. $\tau$ is fixed at 4 $\mu$s for all three cases, and all other pulse errors are set to zero.}
	\label{fig:disorder_simulations} 
\end{centering}
\end{figure*}

\subsection{Local Disorder Errors}
\label{sec:disorder_simulations}
Resonance offset errors can arise in a poorly calibrated system, or more typically in a system with a broad distribution of resonance frequencies due to local disorder or an inhomogeneous magnetic field.  Here, we  consider the effects of a distribution of locally varying off-resonance errors on sequence fidelity. We use a Gaussian distribution centered around zero ($\langle h_i \rangle \sim 0$) with a standard deviation of $\sigma_h$ and average the sequence fidelity over 100 samples of this distribution for $\sigma_h$ ranging from 0.5 Hz to 5 kHz as shown in Figure~\ref{fig:disorder_simulations}.  Our approach is informed by the experimental reality of NMR measurements in a macroscopic sample --- a local cluster of spins experiencing an offset frequency different from another cluster due to the longitudinal field being non-uniform or distorted by the sample. The signal is an average over all such clusters. The ideal unitary, $U_{\text{th}}$, is set to unity for all sequences. 

Figures~\ref{fig:disorder_simulations}(a) and (b) show the effects of resonance offset errors on sequence performance for $\delta$-function and finite pulses, respectively. 
We see $(1-F) \propto \sigma_h^2$ scaling for WHH, BR24 and CORY48. WHH and BR24, being spectroscopic sequences, have a leading order $H_O^{(0)}\propto h_i$. The $\sigma_h^2$ scaling of CORY48 infidelity is possibly from the leading order $H_{DO}^{(2)}\propto\sigma_h(J\tau)^2$ term. YXX24 and YXX48 cancel offset terms to the third order and the remaining leading order term is the cross term, $H_{DO}^{(2)}\propto\ \sigma_h^2(J\tau)$. We don't see the effect of this term for the chosen $J\tau$ and range of errors until $\sigma_h$ values of about 1 kHz, indicating considerable robustness to offset errors. For the case of finite pulses, we see similar trends, but infidelity is now dominated by the finite pulse width effect, as expected.

The case of BR24 is particularly interesting. It goes from being as good at decoupling as the time-suspension sequences at low disorder values to being only as good as the WHH sequence at significant disorder values. This is not surprising, as the spectroscopic sequences protect Hamiltonian terms $\propto S_{z}$, such as the disorder. Figure~\ref{fig:disorder_simulations}(c) further shows the disorder dependence of the fidelity of the BR24 sequence over a range of dipolar coupling strengths. The convergence of fidelity to a similar value even at low offset strengths of $\sim25$ Hz further clarifies that the loss of fidelity is due to inhomogeneous broadening effects rather than from higher order dipolar-offset cross terms. This explains the poor performance of the BR24 sequence in Figure~\ref{fig:tau_data} compared to CORY48, even though both sequences average dipolar couplings out to third order for ideal pulses.

\begin{figure}
\begin{centering}
    \includegraphics[width=0.45\textwidth]{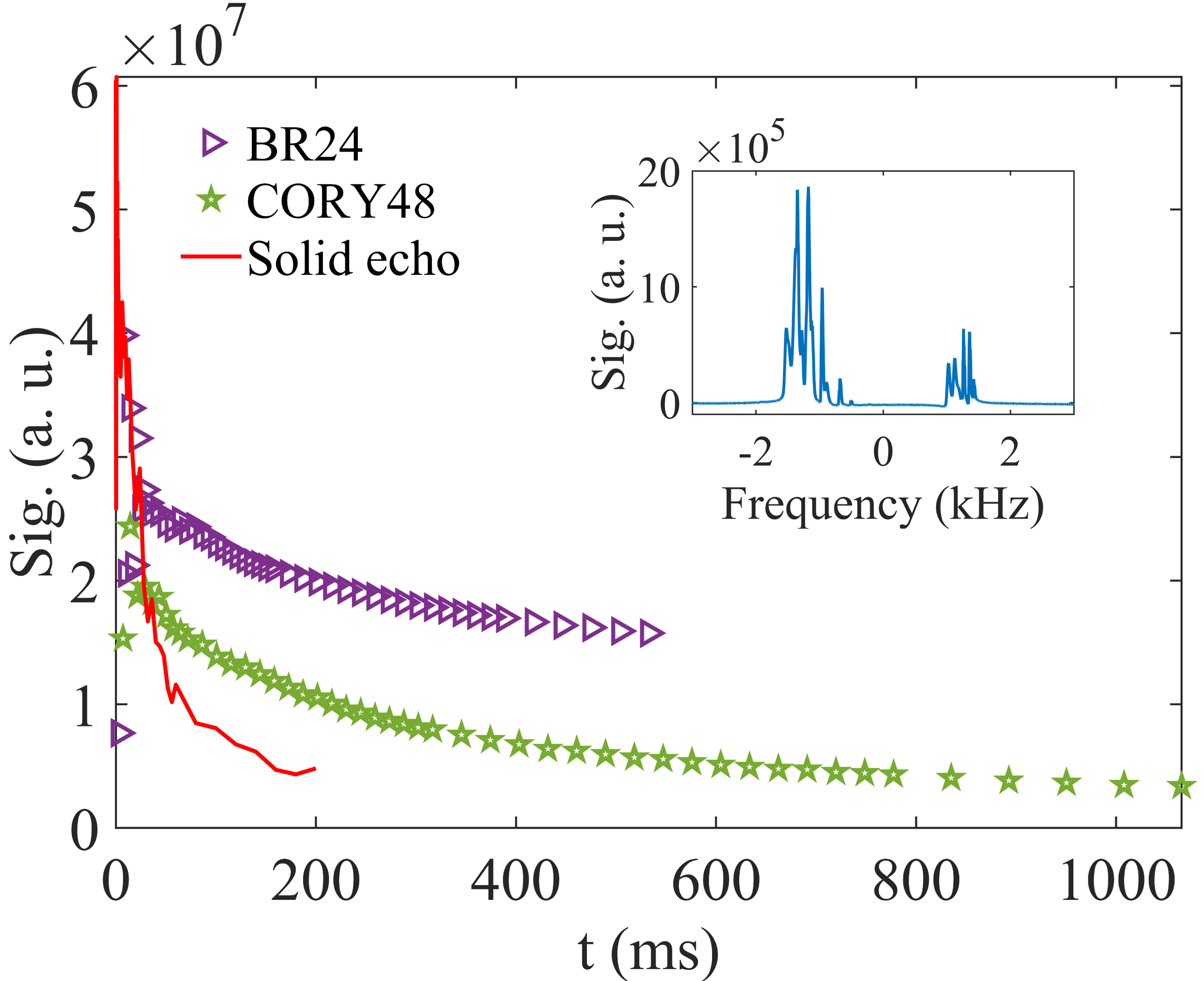}
    \caption{ Experiments on a liquid crystal in a shimmed magnet. $Z$ autocorrelation experiments  showing the comparison of coherence decay under the application of solid echo sequences (red solid line), BR24 (purple triangles), and CORY48 (green pentagrams) in a liquid crystal sample in a shimmed, 9.4 T magnet. NMR spectrum of the liquid crystal ZLI1132 is shown in the inset.}
	\label{fig:liqcrys_data} 
\end{centering}
\end{figure}

To contrast the behavior of the decoupling sequences in the presence and absence of local disorder, we show the results of experiments performed on a liquid crystal sample in a well-shimmed, 9.4 T magnet, with estimated field inhomogeneity of the order of 30 Hz (green shaded region in Figure~\ref{fig:disorder_simulations}(b) is at 10 Hz --- compared to about 250 Hz for the adamantane experiments). We used a Merck ZLI1132 liquid crystal which was subject to repeated cycles of heating (outside the magnet) and cooling (inside the magnet) to create an ordered nematic phase. In a strong magnetic field, the orientational ordering of the liquid crystal molecules prevents intramolecular dipolar interactions from averaging to zero. However, intermolecular dipolar couplings average to zero~\cite{emsley_nmr_1975} .

The NMR spectrum in the inset of Figure~\ref{fig:liqcrys_data} shows a complex spectrum indicating a cluster of strongly dipolar coupled spins. The individual peaks are about 30 Hz in width, suggesting higher longitudinal field homogeneity.  We show that BR24 and CORY48 sequences exhibit similar performance in $Z$ autocorrelation experiments under these conditions that minimize inhomogeneous offset errors. This starkly contrasts the adamantane data that shows at least an order of magnitude difference in the effective coherence time achieved by the two sequences (Figure~\ref{fig:cartoon_and_T2_comparison}(b)).

\subsection{Rotation Errors \& Phase Transients} 
\label{sec:overrot_and_phtr}
Rotation errors result from the over or under rotation of the spin magnetization by a pulse deviating slightly from the ideal $\pi/2$ rotation. To incorporate the effect of this error, we modify the RF unitary operator describing system evolution during an $X$ pulse to $e^{-iS_{x}\pi(1+\epsilon)/2}$, where $\epsilon$ denotes the fractional over or under rotation error. An ideal pulse has $\epsilon=0$. A miscalibration of the $\pi/2$ pulse or changes in pulse power after calibration can cause this type of global rotation error. Inhomogeneity of the transverse control field over the macroscopic sample can also cause rotation errors with spatially varying $\epsilon_{i}$. 

Phase transients originate from the electronics of the setup.  During the rise and fall times of a pulse, a `transient', 90-degrees out of phase component is induced in the circuit which results in rotation errors during the pulse. 

Recent work~\cite{stasiuk_frame_2023} shows that this type of unitary error can be compensated for by performing systematic measurements and a frame change technique, as long as experimental conditions and probe tuning are stable. It works exactly for a leading-edge dominated phase transient error and approximately for a transient effect balanced over the leading and trailing edges. We have not performed the frame change technique in our data in this paper. We have incorporated the phase transient errors in our simulations by adding the appropriate unitary operators at the beginning and end of each $\pi/2$ pulse. For example, the RF unitary for an $X$ pulse, after incorporating rotation error and phase transient effect looks as, $U_{\text{pulse}}=e^{-iS_{y}\alpha_{tr}\pi/2}e^{-iS_{x}\pi/2(1+\epsilon)}e^{-iS_{y}\alpha_{l}\pi/2}$, where $\alpha_{l}$ and $\alpha_{tr}$ are the strengths of the leading and trailing edge phase transients respectively, as a fraction of the $\pi/2$ pulse strength. It has been shown that symmetric phase transients ($\alpha_{l} = \alpha_{tr}$) are less harmful than asymmetric transients~\cite{haeberlen_high_2012}. Since pulse calibration techniques can be used to suppress asymmetric transients, we assume that the phase transients are symmetric here.

\begin{figure*}
\begin{centering}
    \includegraphics[width=0.45\textwidth]{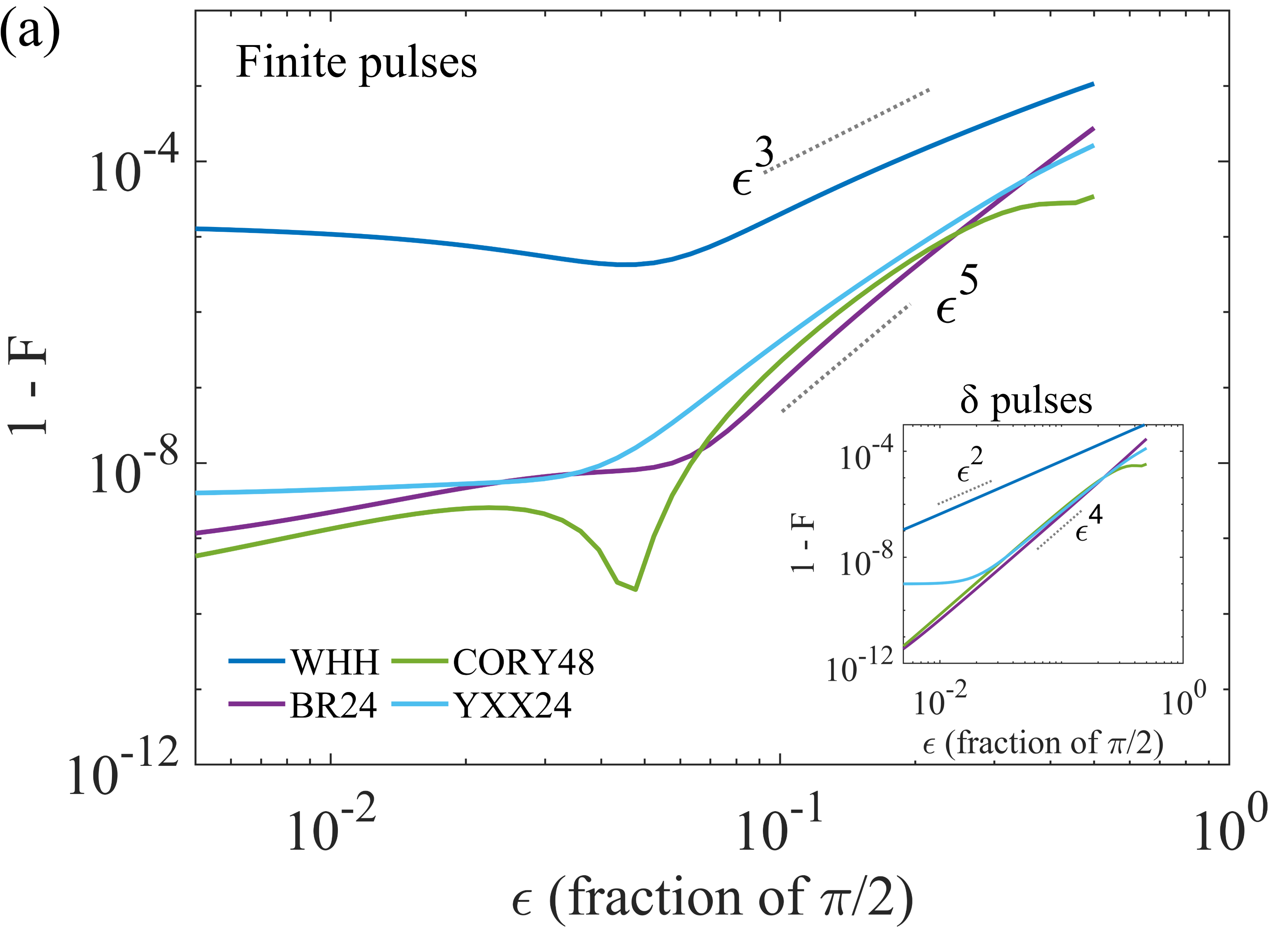} \hspace*{0.1in}
    \includegraphics[width=0.45\textwidth]{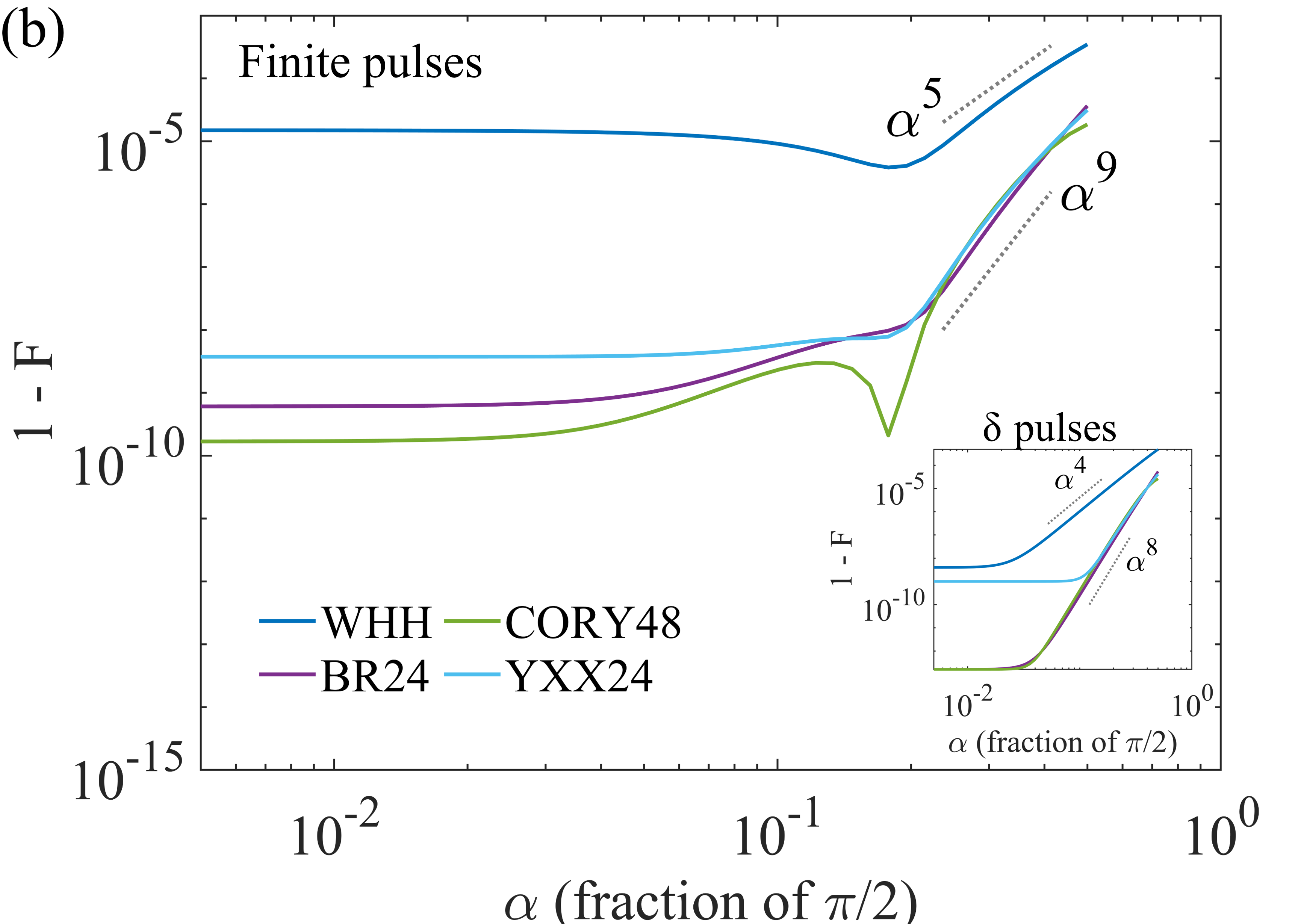}
    \caption{Effect of (a) over or under rotation errors and (b) phase transients on sequence fidelity for finite pulses with $t_w = 1$ $\mu$s. $\tau$ is fixed at 4 $\mu$s for all  cases, and all other pulse errors are set to zero. Insets show results for $\delta$-function pulses, all other conditions held the same.}
	\label{fig:overrot_and_phtr_simulations} 
\end{centering}
\end{figure*}

Figures~\ref{fig:overrot_and_phtr_simulations}(a) and (b) show the scaling of sequence fidelities with rotation errors and phase transients respectively.  The insets show the case of infinitesimal pulses and the main figures show the results for finite-width pulses. For rotation errors, the $\epsilon^2$ scaling with $\delta$-function pulses for the WHH sequence likely originates from $H_{\epsilon}^{(0)}$. YXX24 and CORY48 sequences both have second order over-rotation error-dipolar cross terms with $\epsilon^2$ dependence, leading to the $\epsilon^4$ scaling. Analytical calculations incorporating phase transient errors are not available for most sequences even for pulses of infinitesimal width. Furthermore, it is evident that for finite pulses, the infidelity scaling of the sequences with respect to both over-rotation and phase transient errors is different from the case of infinitesimal pulses. 

\section{Protecting Correlated States}
\label{sec:MQC}
While the numerical simulations used the trace unitary fidelity to characterize sequence performance, the experiments shown previously all used high-temperature initial states in which the spins were largely uncorrelated, with $\rho \sim \openone - \delta\rho$, $\delta\rho = \epsilon \sum_i S_z^i$.  In order to examine our ability to protect highly-correlated spin states (with $\delta \rho$ containing higher order spin operators strings $S_\alpha^i S_\beta^j S_\gamma^k...$) using these sequences, we examined how well the CORY48 sequence could protect multiple quantum coherences (MQCs).

For a system of interacting spin-1/2 particles, the density operator can be represented in a particular single spin basis $S_\alpha$ as $\rho=\sum_{n}\rho_{n}$ where 
$\rho_{n}=\sum_{i,j}\ket{m_{\alpha,i}}\bra{m_{\alpha,j}}$ such that $|m_{\alpha,i}-m_{\alpha,j}|=n$. An operator of the form $\ket{m_{\alpha,i}}\bra{m_{\alpha,j}}$ is a multiple quantum coherence of order $n$ in this basis where $m_{\alpha,i}$ denotes the magnetic quantum number along $S_\alpha$. The intensity of a coherence order is defined by $I_{n}= \text{Tr}(\rho_{n}^{\dag}\rho_{n})$.

\begin{figure*}[]
\begin{centering}
    \includegraphics[width=0.3\textwidth]{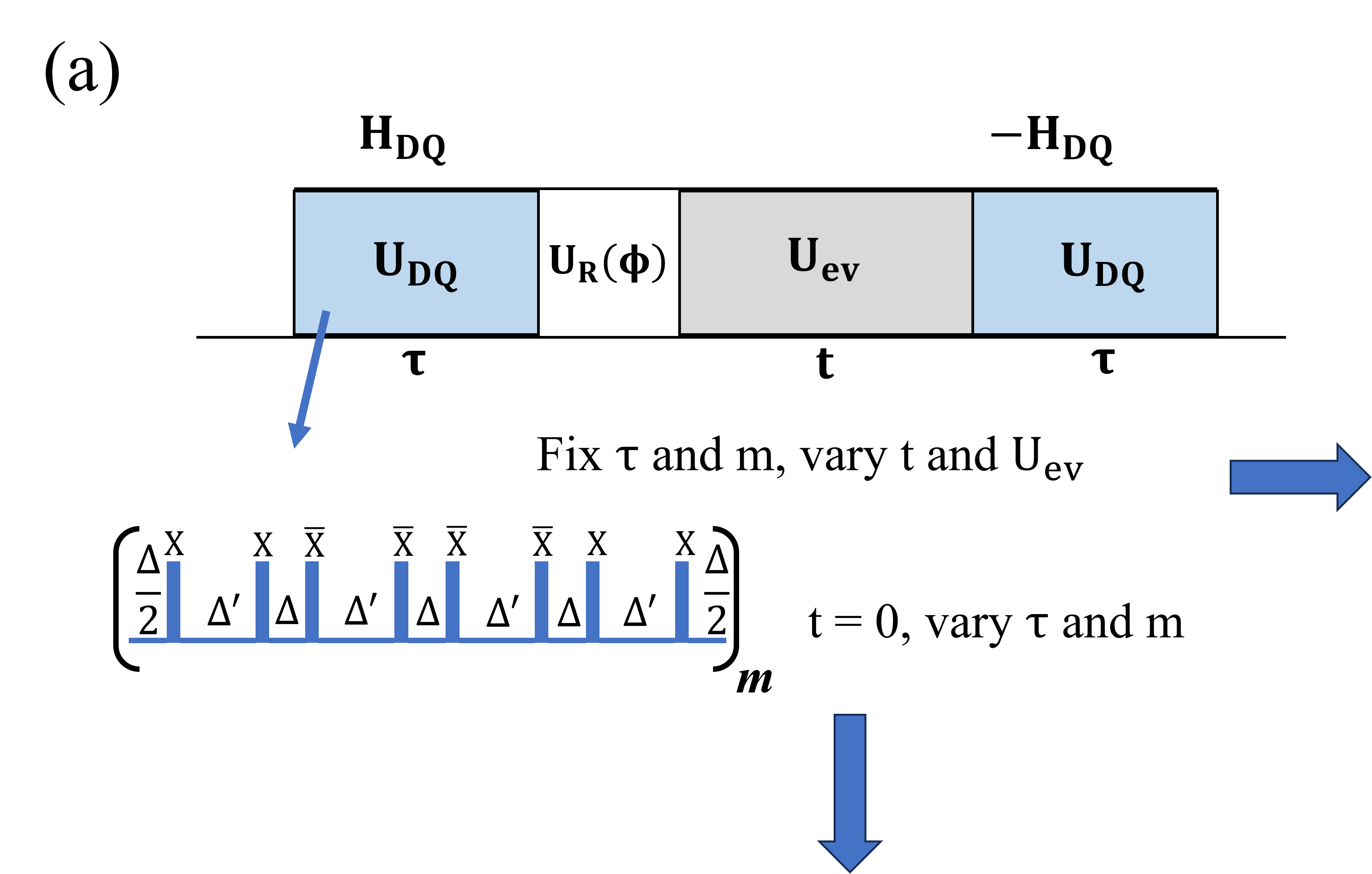}
    \includegraphics[width=0.3\textwidth]{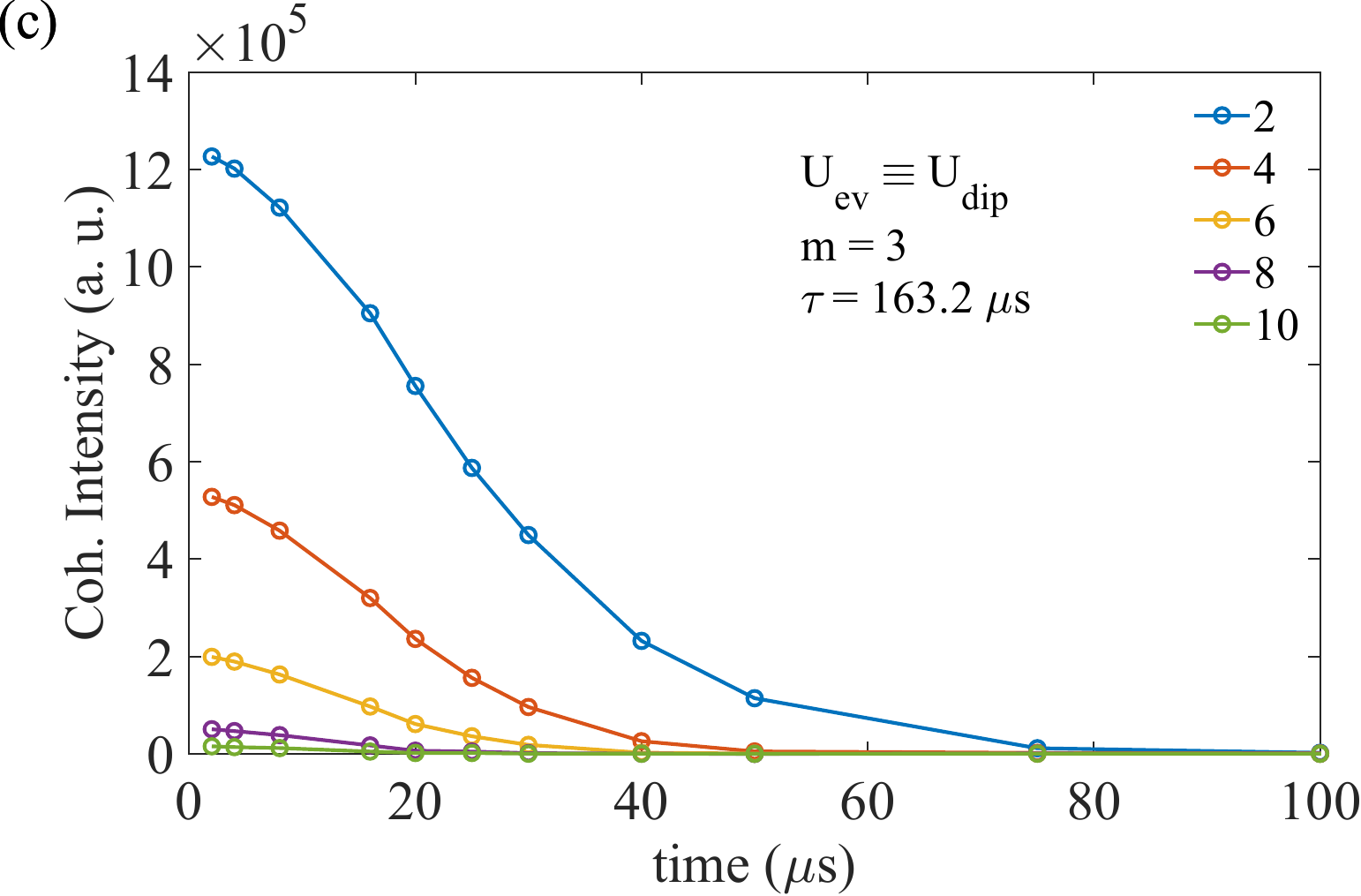} 
    \hspace*{0.03in}
    \includegraphics[width=0.3\textwidth]{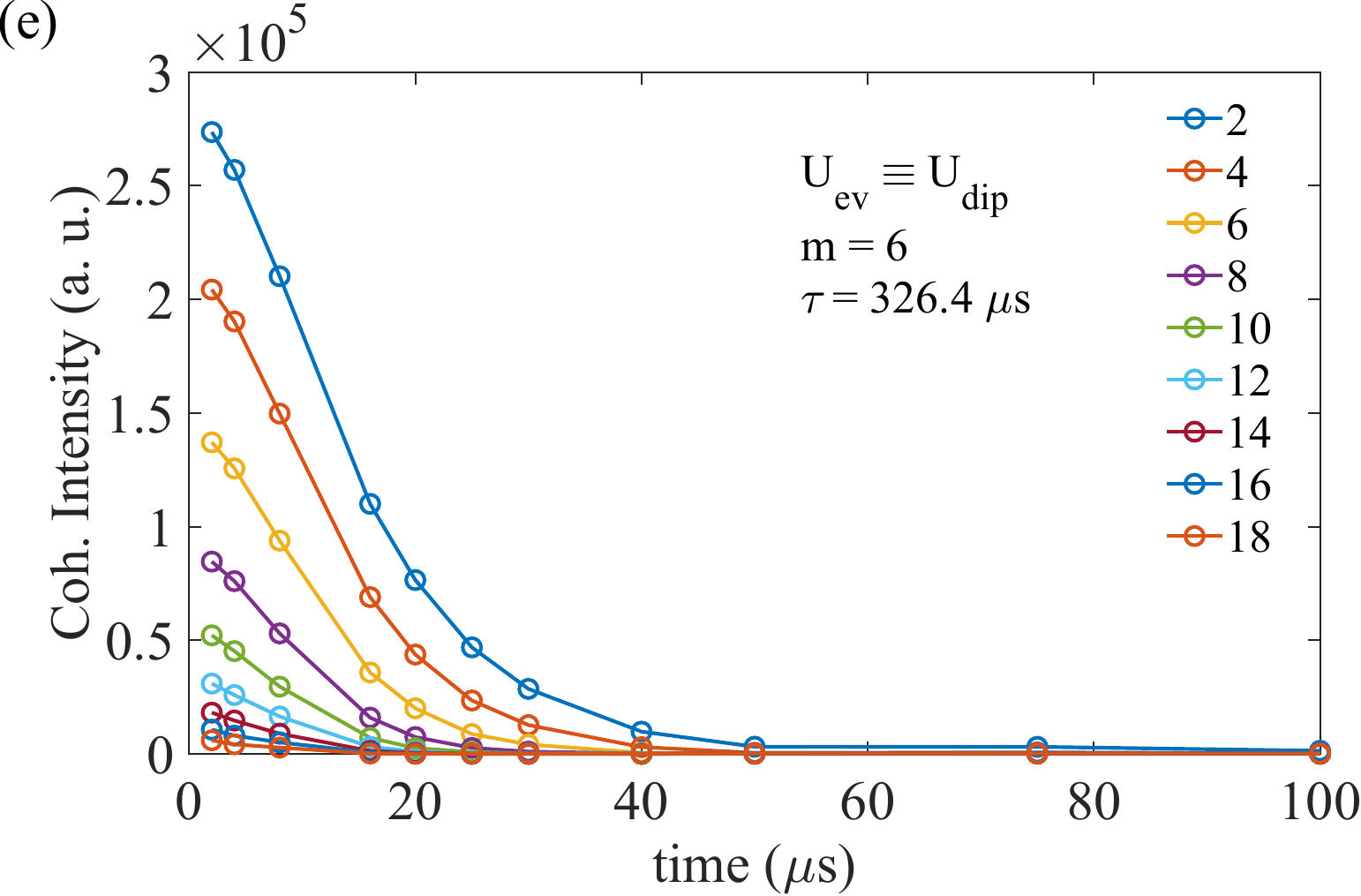} 
    \includegraphics[width=0.3\textwidth]{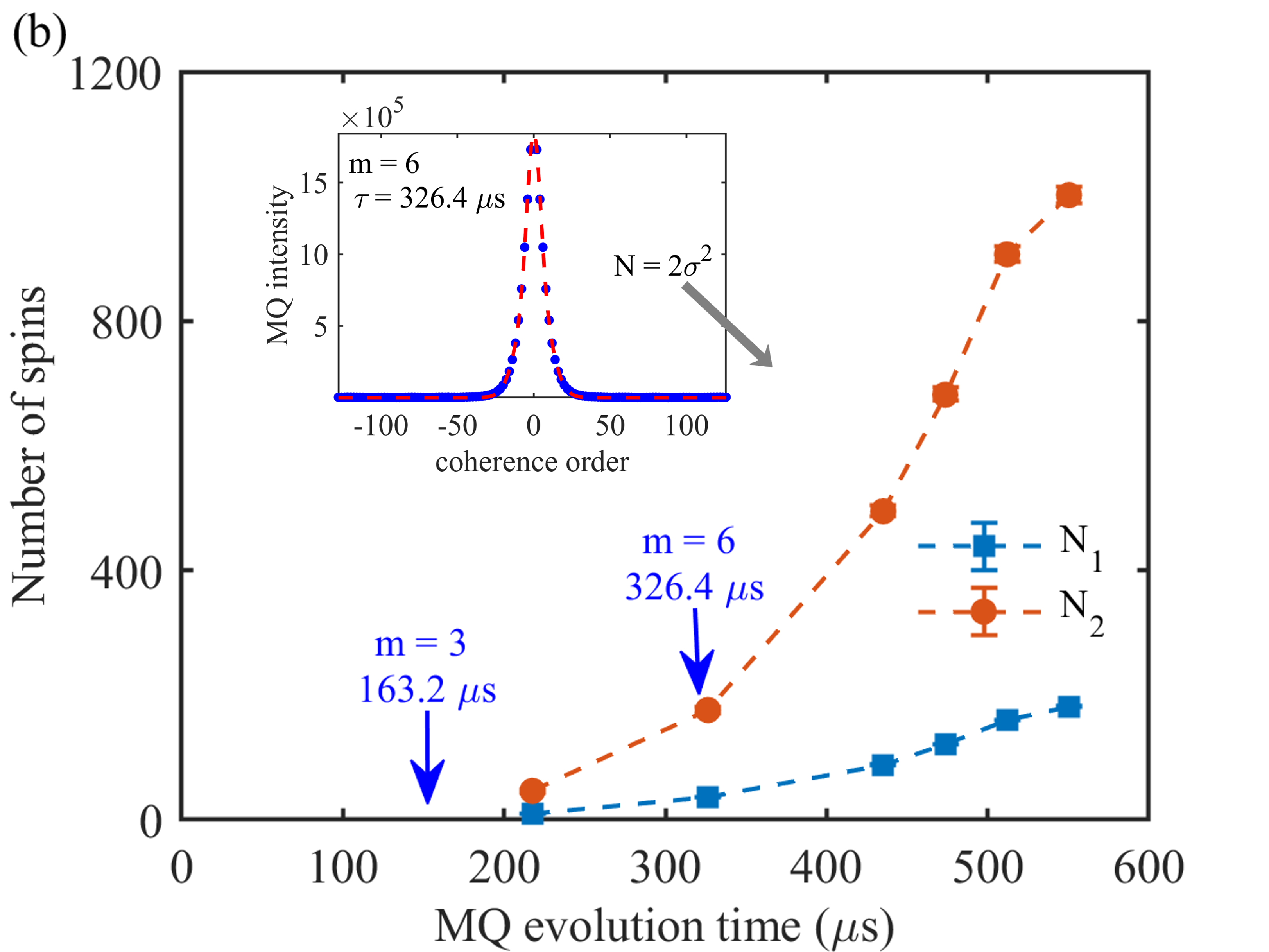} 
    \includegraphics[width=0.3\textwidth]{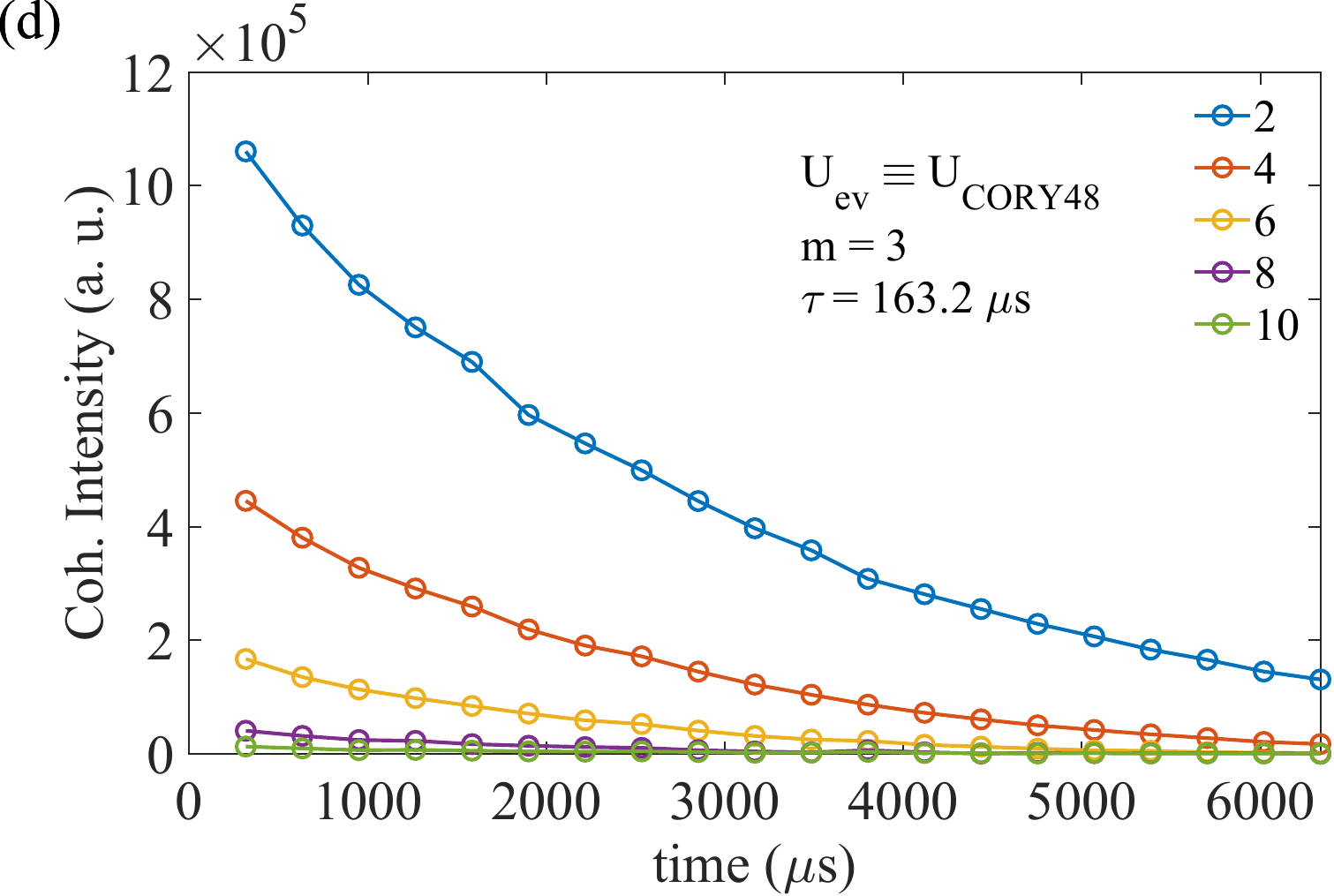}  \hspace*{0.03in}
    \includegraphics[width=0.3\textwidth]{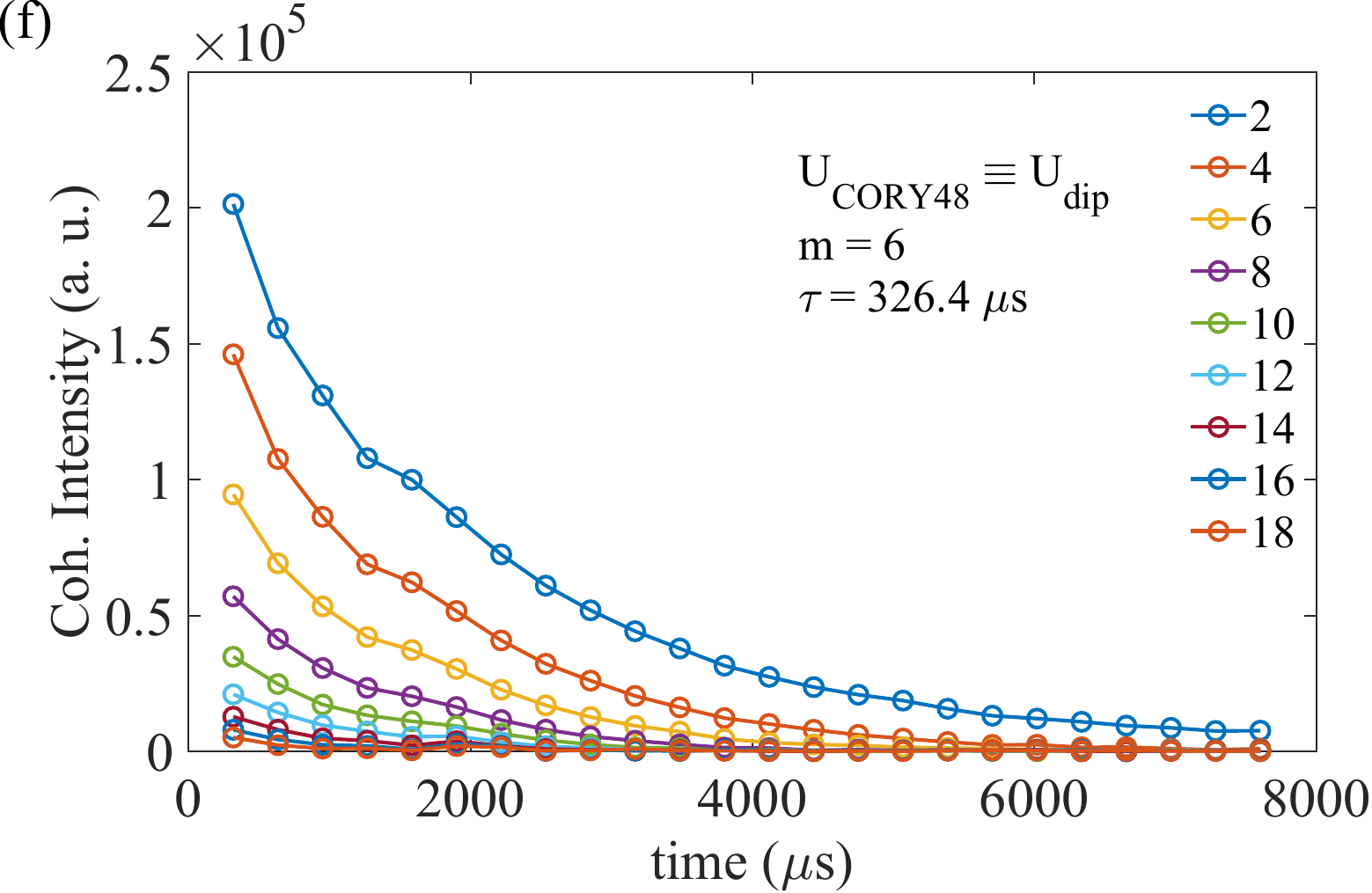} 
   	\caption{Protecting multiple-quantum correlated initial states using dipolar decoupling sequences. (a) shows the experimental scheme to study the growth and decay of coherences under an evolution Hamiltonian. (b)  shows the number of correlated spins extracted from the width of each of the two Gaussian fits to the coherence distribution, $N= 2\sigma^2$, where $\sigma$ is the standard deviation of the Gaussian. Inset shows the MQC distribution (with a double Gaussian fit) when the number of cycles of the double quantum sequence ($m$) is set to 6, resulting in the total time for MQ growth, $\tau = 326.4$ $\mu$s. (c) and (e) show the decay of the different coherence orders when they are allowed to evolve under the natural system Hamiltonian for a time $t$ after the MQC growth (forward evolution for $\tau = 163.2$ $\mu$s and $\tau = 326.4$ $\mu$ s respectively) and before the backward evolution to refocus the coherences. (d) and (f) show similar data as (c), but the natural system Hamiltonian is replaced by multiple cycles of the CORY48 sequence, thus protecting the coherences from decaying due to dipolar interactions.}
	\label{fig:mqc_data} 
\end{centering}
\end{figure*} 

MQCs are crucial in investigating many-body dynamics using NMR~\cite{cappellaro_dynamics_2007,zhang_nmr_2009,ramanathan_experimental_2011} and a form of out-of-time-ordered correlator (OTOC)~\cite{wei_exploring_2018}. We use the double-quantum Hamiltonian ($H_{\text{DQ}}=1/2\sum_{j,k}J_{jk}(S_{x}^{j}S_{x}^{k}-S_{y}^{j}S_{y}^{k})$) in our experiments. The double quantum Hamiltonian causes double spin flips, i.e., transitions between spin states with $\Delta m_{z}=\pm2$. 

As shown in the experimental scheme in Figure~\ref{fig:mqc_data}(a), we let the system evolve under $m$ cycles of the double-quantum Hamiltonian for a total time $\tau$ to allow (even numbered) coherence orders to develop in the system. A collective $z$-rotation by $\phi$ phase tags the coherence, following which the evolution is refocused using $m$ cycles of $-H_{\text{DQ}}$. The pulse sequence used in the experiments to engineer $H_{\text{DQ}}$ is also shown. $-H_{\text{DQ}}$ is created by phase shifting the pulses in the original sequence by $\pi/2$. It can be shown that the signal from this experiment is $S(T) = \sum I_n(\rho_{\tau})e^{in\phi}$, where $T$ is the total experiment time, $\rho_{\tau}$ is the density matrix at time $\tau$, and $I_n(\rho_{\tau})$ denotes the intensity of the coherence order $n$ in $\rho_{\tau}$.

We set $\tau=163.2$ $\mu$s ($m=3$) or  $\tau=326.4$ $\mu$s ($m=6$) in our experiments. The inset in Figure~\ref{fig:mqc_data}(b) shows the intensities of coherence orders $I_{n}$ derived from a discrete Fourier transform of $S(T)$ with respect to $\phi$ for $m=6$. Under certain assumptions, the approximate number of correlated spins ($N$) can be calculated from the second moment of the coherence distribution~\cite{baum_multiplequantum_1985}. Figure~\ref{fig:mqc_data}(b) shows the growth of such correlated clusters as a function of the evolution time $\tau$ under $H_{\text{DQ}}$, and the presence of clusters containing about $10^3$ spins.

We studied the decay of multiple quantum correlations due to the dipolar interactions by inserting a time window $t$ between the phase tagging step and the backward evolution of the MQC experiment sequence~\cite{krojanski_scaling_2004,cho_decay_2006} (see Figure~\ref{fig:mqc_data}(a) for the experiment scheme). In Figures~\ref{fig:mqc_data}(c) and (e), we see that free evolution under the dipolar Hamiltonian leads to significant decay in the MQC intensities. In Figures~\ref{fig:mqc_data}(d) and (f) we also show the effect of applying multiple cycles of dipolar decoupling sequences during the delay time $t$ to protect multiple quantum coherences. We see an increase of up to two orders of magnitude in the characteristic decay time of the multiple quantum coherence intensities using this method - similar to that observed for $^{19}$F spins in calcium fluoride~\cite{cho_decay_2006} .

We also observe that the longest decay times observed in this data set go up to the order of only a millisecond. This is in contrast to the refocusing experiments done on single-quantum initial states discussed earlier, where coherence times of several tens of milliseconds were observed. This comparison also serves to illustrate the sensitivity of MQCs to decoherence and errors in the implementation of the pulse sequences. 

\section{\label{sec:Conc} Conclusions}
We compared the efficiency, limitations, and susceptibility to errors of several dipolar decoupling sequences, both qualitatively and quantitatively. We find that the new `machine-learned' time-suspension sequences perform comparably to the CORY48 sequence in our simulations and experiments, demonstrating the power of using such techniques for quantum control, supplementing well-known analytical tools and physical intuition. 

We believe that re-assessing and benchmarking these sequences developed at different times and contexts from a unified quantum simulation perspective is valuable in choosing the right sequences for the right applications. System-specific pulse-sequence design, where system-specific noise and errors are included in first-principles modeling to assess the performance of multiple-pulse sequences, can become a valuable tool in future experimental investigations.

Going forward, there exists the potential to discover new and possibly improved dipolar decoupling pulse sequences by further modifying the action spaces of the reinforcement learning algorithms using physical intuition and insights from analytical techniques. There are also several open questions about the regimes of validity and convergence of the Magnus expansion and the consequences of these on pulse sequence design and performance. Another fruitful avenue of investigation might be integrating pulse shaping and optimization techniques into the already established AHT framework. Recent work has also been in unifying pulse-sequence design for various applications, distilling available analytical approaches into a simple checklist of conditions and pictorial representations of the toggling frame evolution. 

\begin{acknowledgments}
We thank Pai Peng, Paola Cappellaro, Madhumati Seetharaman and Ethan Williams for helpful discussions. This work was partially supported by the NSF under grant No. OIA-1921199. LJ acknowledges support of a QISE-NET Triplets Award (NSF award DMR-1747426). 
\end{acknowledgments}

\noindent\hrulefill 

\setcounter{figure}{0}         
\renewcommand\thefigure{A\arabic{figure}} 
\setcounter{table}{0}
\renewcommand{\thetable}{A\arabic{table}}

\appendix
\section{Numerical simulations of Small Spin Systems}
\label{sec_supp:Simulations}

\subsection{Choosing representative sequences} 
\label{sec_supp:representative_sequences}
\begin{figure}[]
\begin{centering}
    \includegraphics[width=0.22\textwidth]{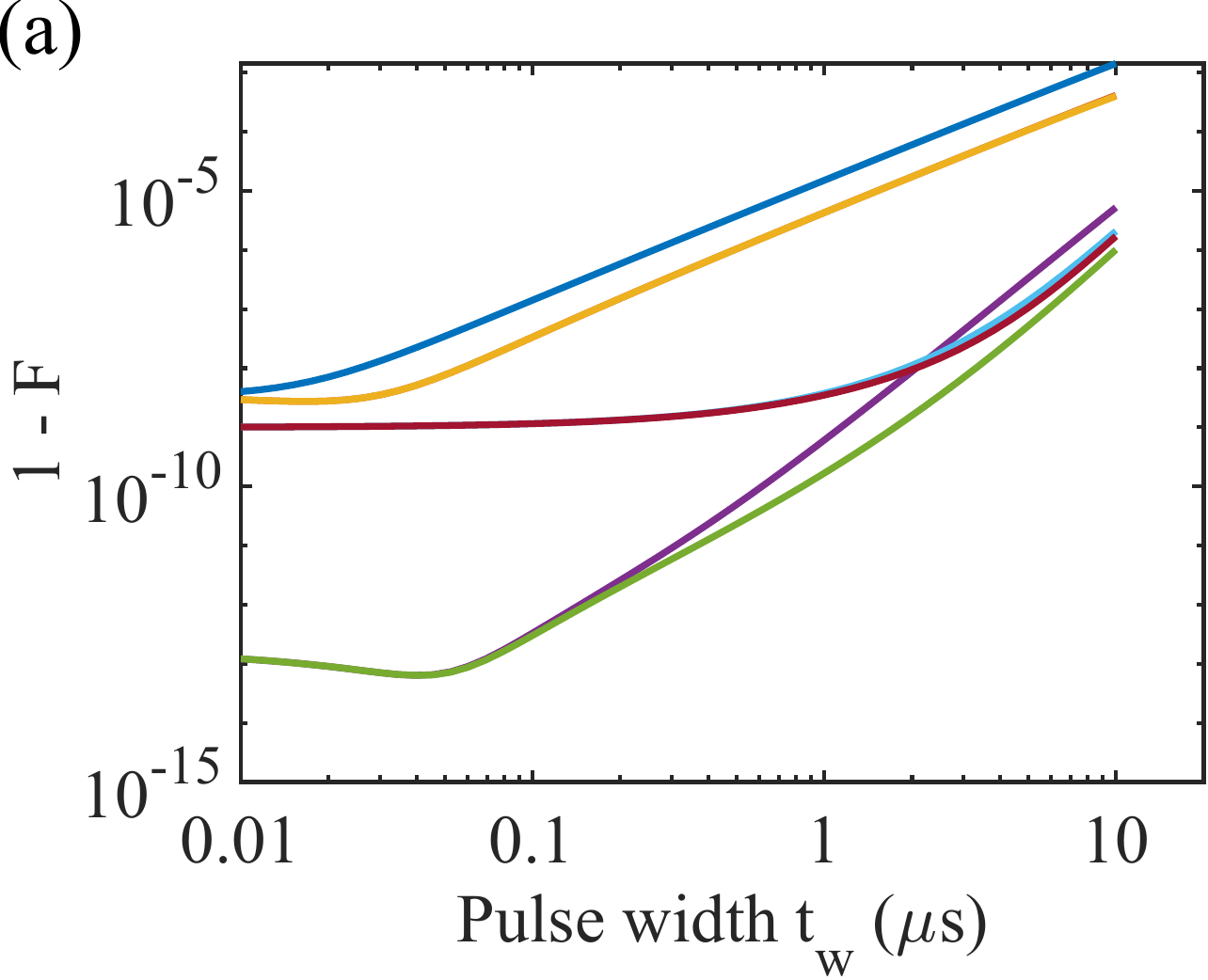} 
    \hspace{0.3cm}
    \includegraphics[width=0.22\textwidth]{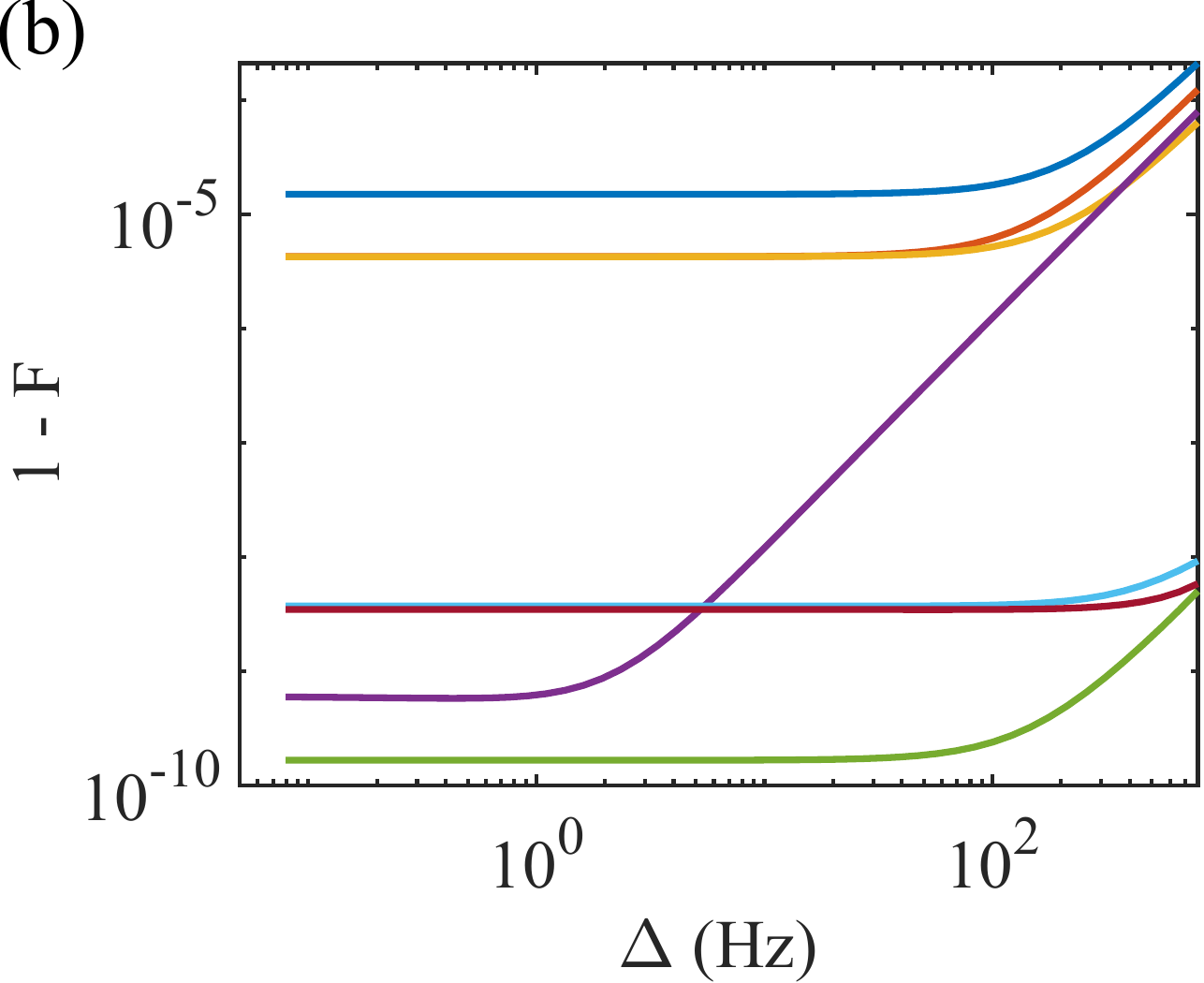} \\

    \includegraphics[width=0.22\textwidth]{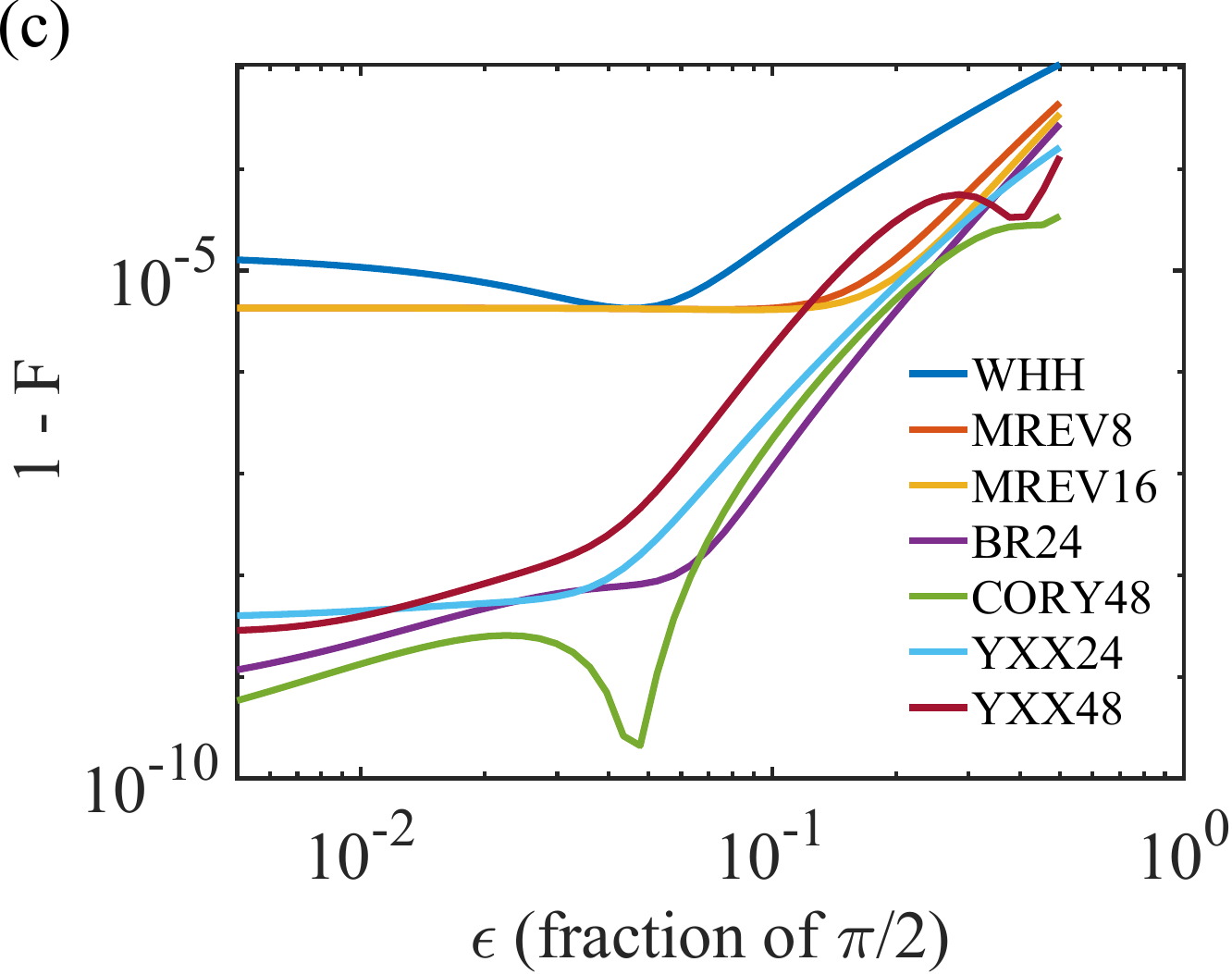}
    \hspace{0.3cm}
    \includegraphics[width=0.22\textwidth]{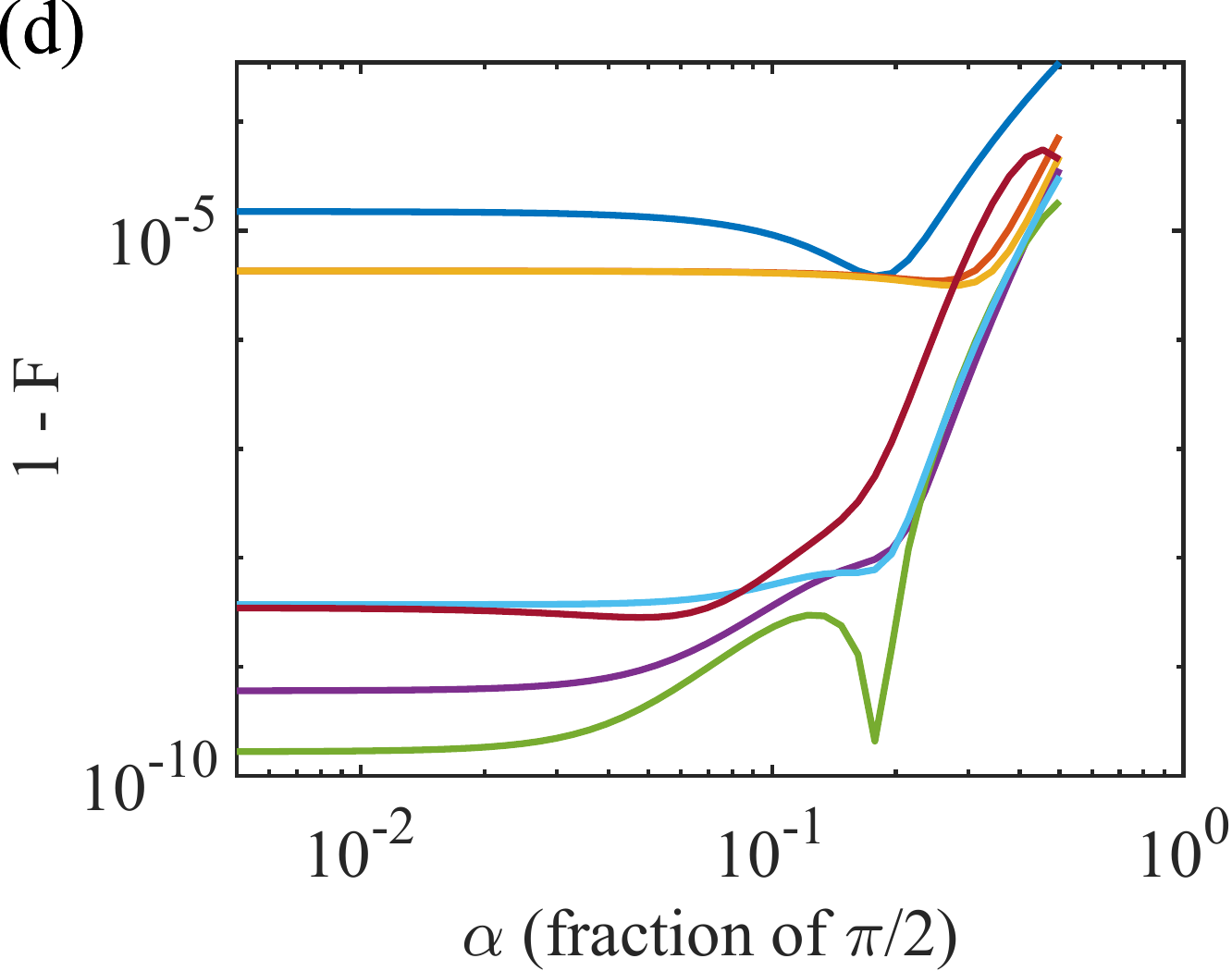} 
 
    \caption{Numerical simulation results showing the clustering of sequences into different classes based on their symmetries and the average Hamiltonian terms they average to zero. We have shown four of these seven sequences as representative sequences in the main paper. Figures show the dependence of the sequence infidelity ($1 - F$) on (a) pulse width, (b) resonance offset, (c) over or under-rotation errors, and (d) phase transients. All data are for an inter-pulse delay ($\tau$) of 4 $\mu$s. All other errors are set to zero in (a). (b), (c), and (d) are at a finite pulse-width ($t_{w}$) of 1 $\mu$s, and only one type of error is incorporated in each case.}
	\label{fig_supp:simulation_details_all_sequences} 
\end{centering}
\end{figure}

Here, we include figures showing simulation results for the fidelity-response of all seven sequences discussed in Section \ref{sec:sims}. In addition to the four sequences (WHH, BR24, YXX24, and CORY48 shown in Figures \ref{fig:tau_and_tw_simulations}, \ref{fig:disorder_simulations}, and \ref{fig:overrot_and_phtr_simulations}), we include MREV8, MREV16, and YXX24 in Figure \ref{fig_supp:simulation_details_all_sequences}. MREV8 and MREV16 exhibit responses to pulse errors that are parallel to the WHH sequence and are near-identical to each other, while YXX48 behaves very similarly to YXX24 under non-ideal conditions, exhibiting similar robustness and scaling behavior. These observations justify our choice of WHH and YXX24 as representative sequences for the above two classes. 

\subsection{Global resonance offset error} 
\label{sec_supp:global_offset_error}
\begin{figure}[]
\begin{centering}
    \includegraphics[width=0.22\textwidth]{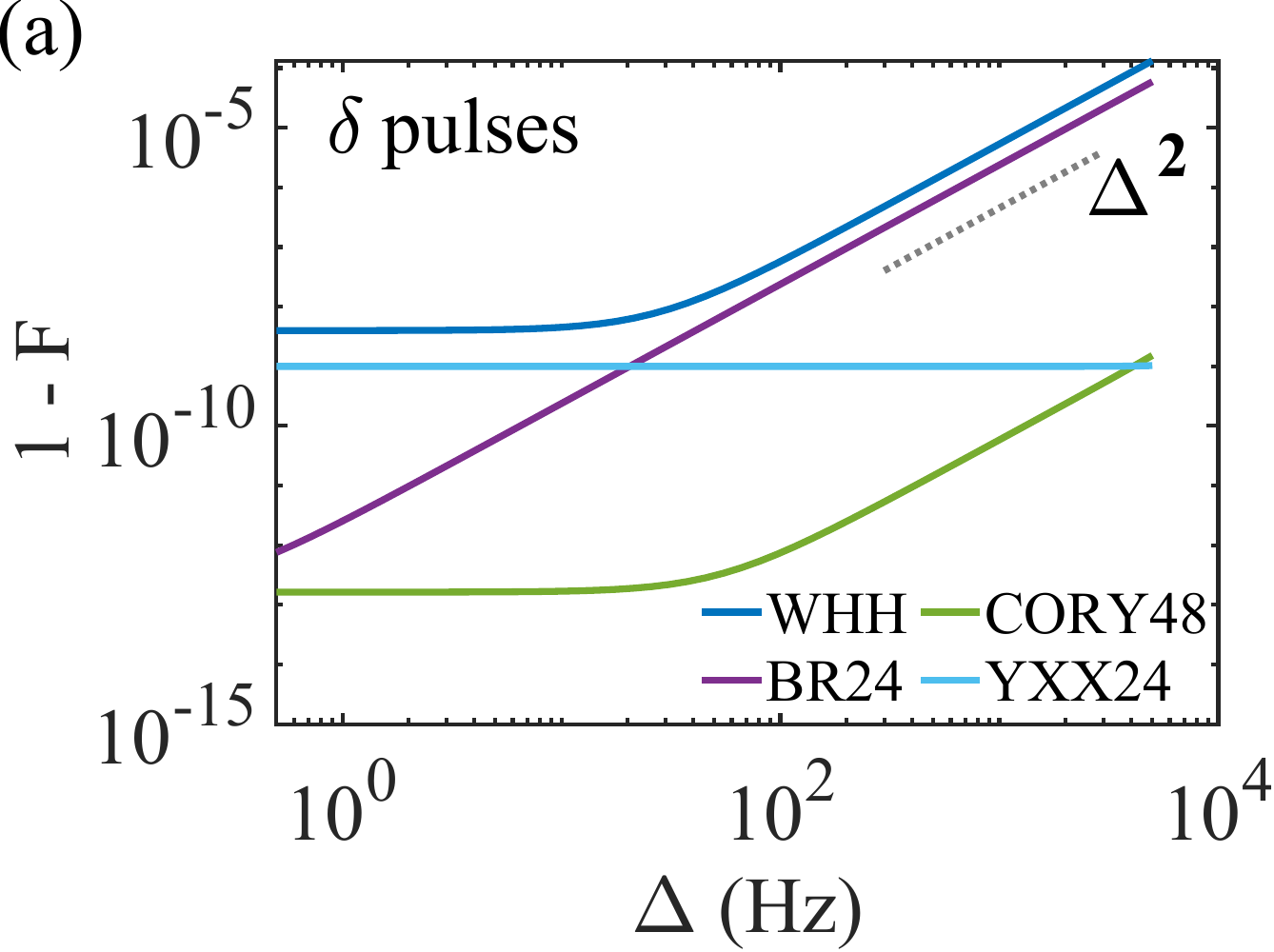} 
    \hspace{0.3cm}
    \includegraphics[width=0.22\textwidth]{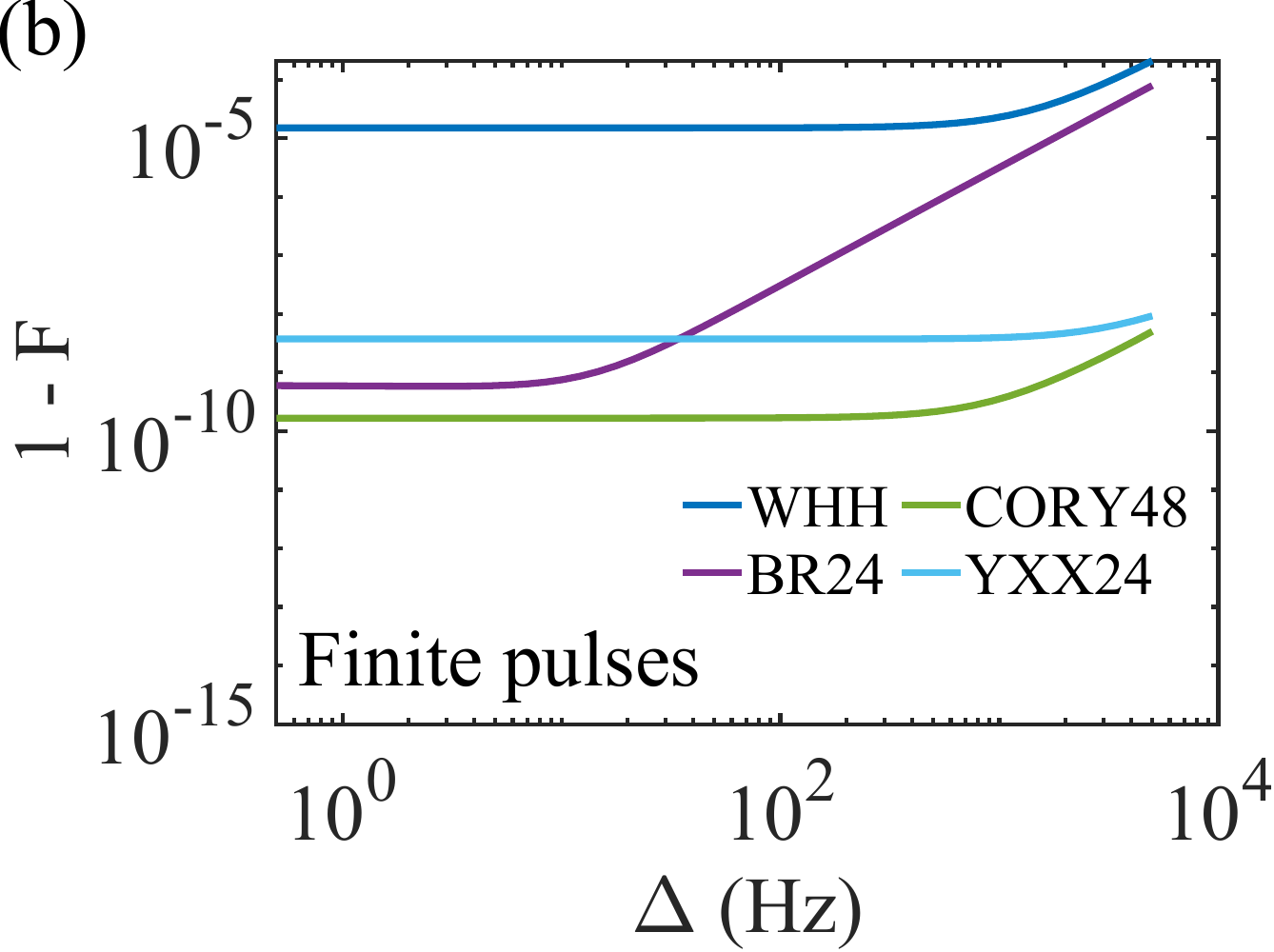} 
    \caption{Effect of global resonance offset errors on sequence fidelity for (a) ideal $\delta$-pulses and (b) finite pulses with $t_w = 1$ $\mu$s.}
	\label{fig_supp:global_offset_error} 
\end{centering}
\end{figure}

In Section \ref{sec:disorder_simulations}, we discuss the effect of local disorder on sequence fidelity. Figure \ref{fig_supp:global_offset_error} shows that numerical results for the effect of global offset error for infintesimal and finite pulses are very similar to the effect of local disorder (Figure \ref{fig:disorder_simulations}).  

\subsection{Average Hamiltonian Analysis}
\label{sec_supp:aht}
\begin{figure*}[]
\begin{centering}
    \includegraphics[width=0.22\textwidth]{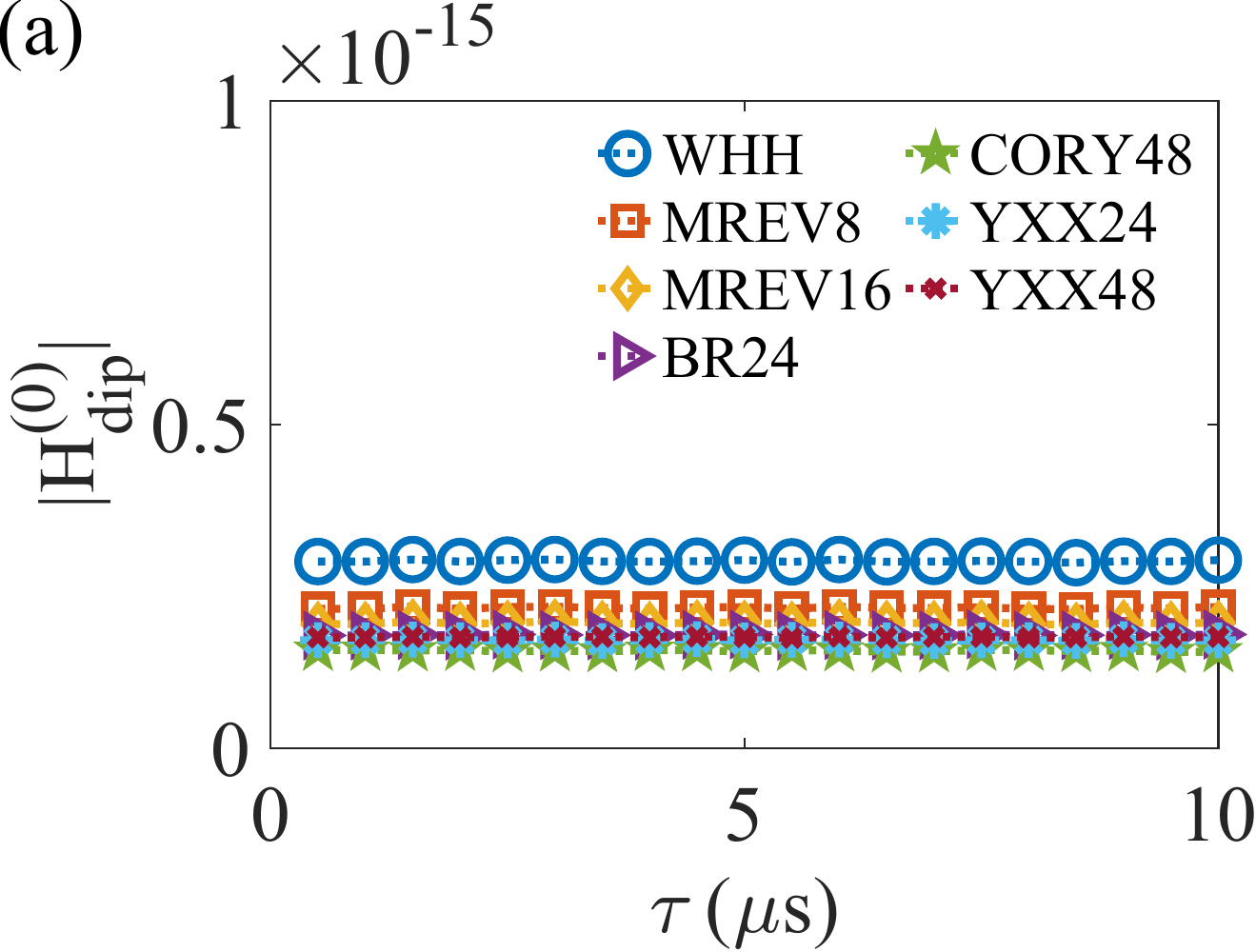} 
    \hspace{0.3cm}
    \includegraphics[width=0.22\textwidth]{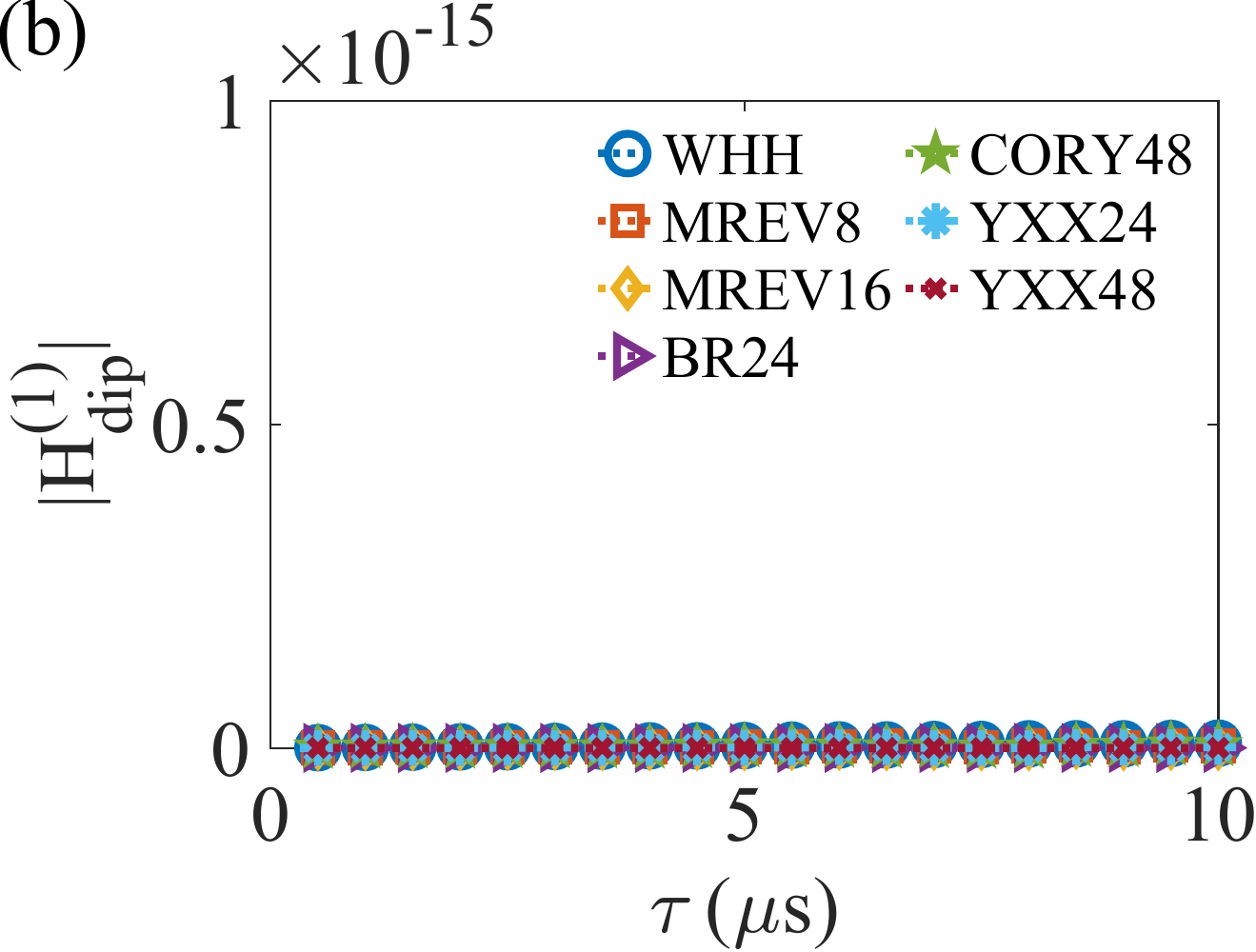}
        \hspace{0.3cm}
    \includegraphics[width=0.22\textwidth]{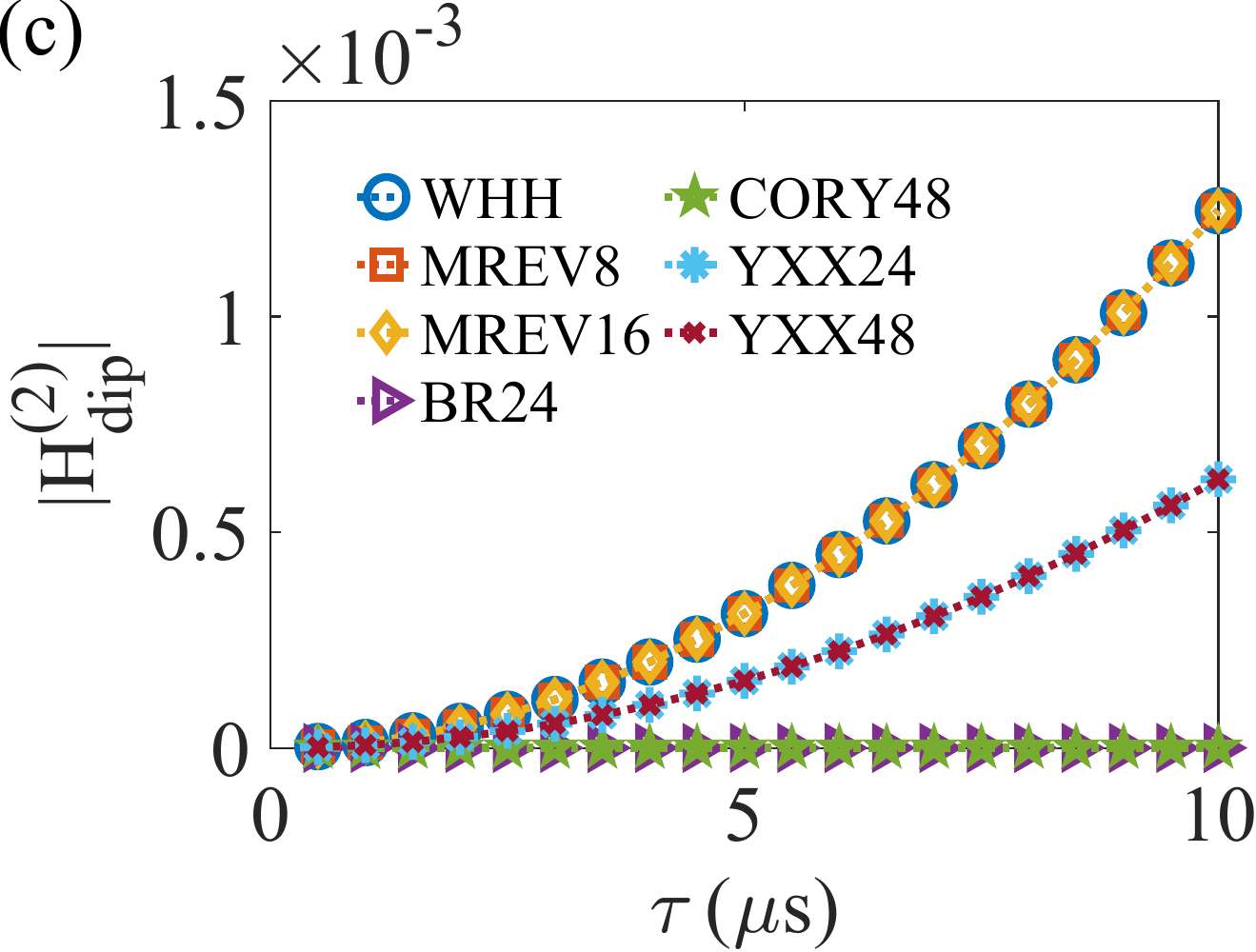} 
        \hspace{0.3cm}
    \includegraphics[width=0.22\textwidth]{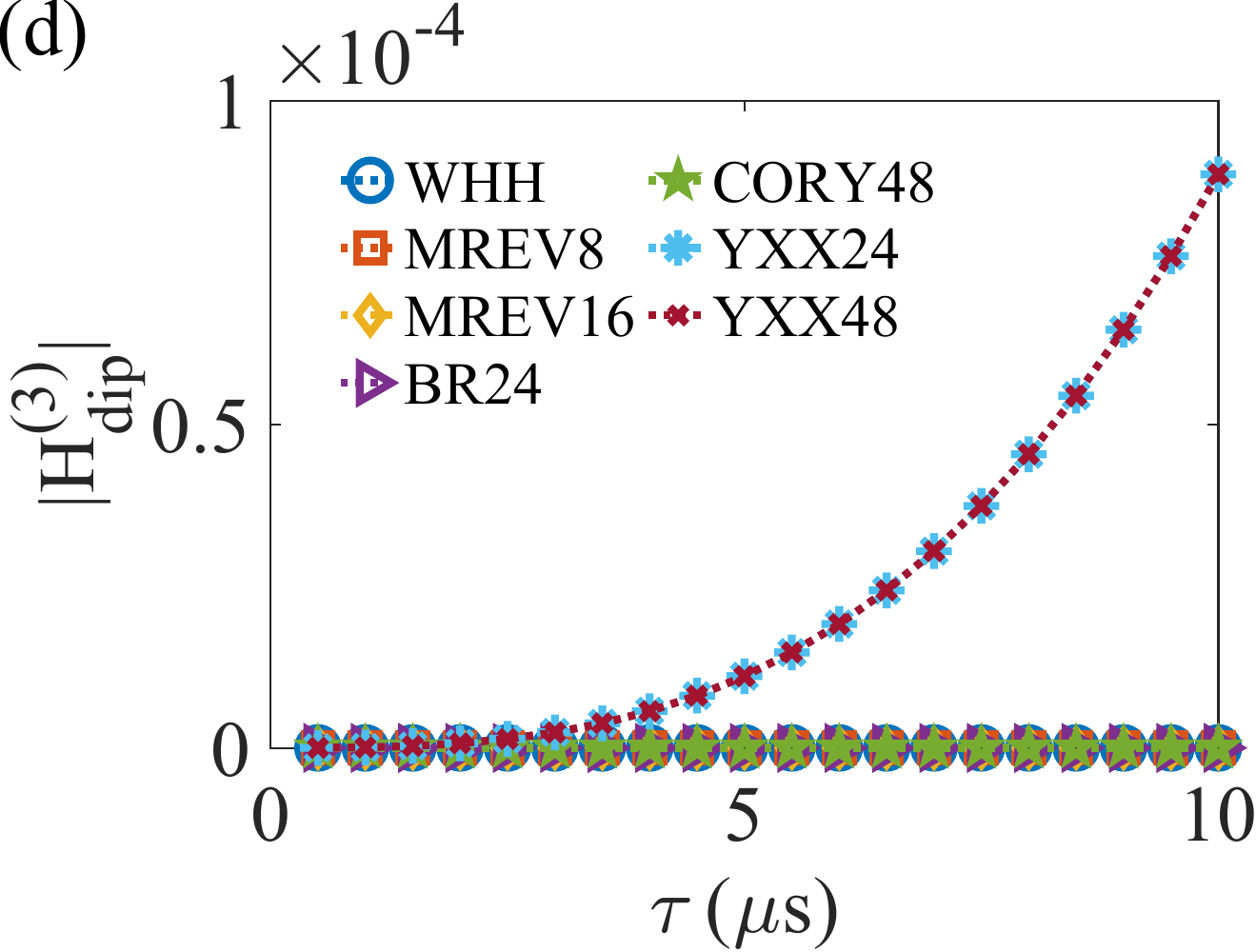} \\
    \includegraphics[width=0.22\textwidth]{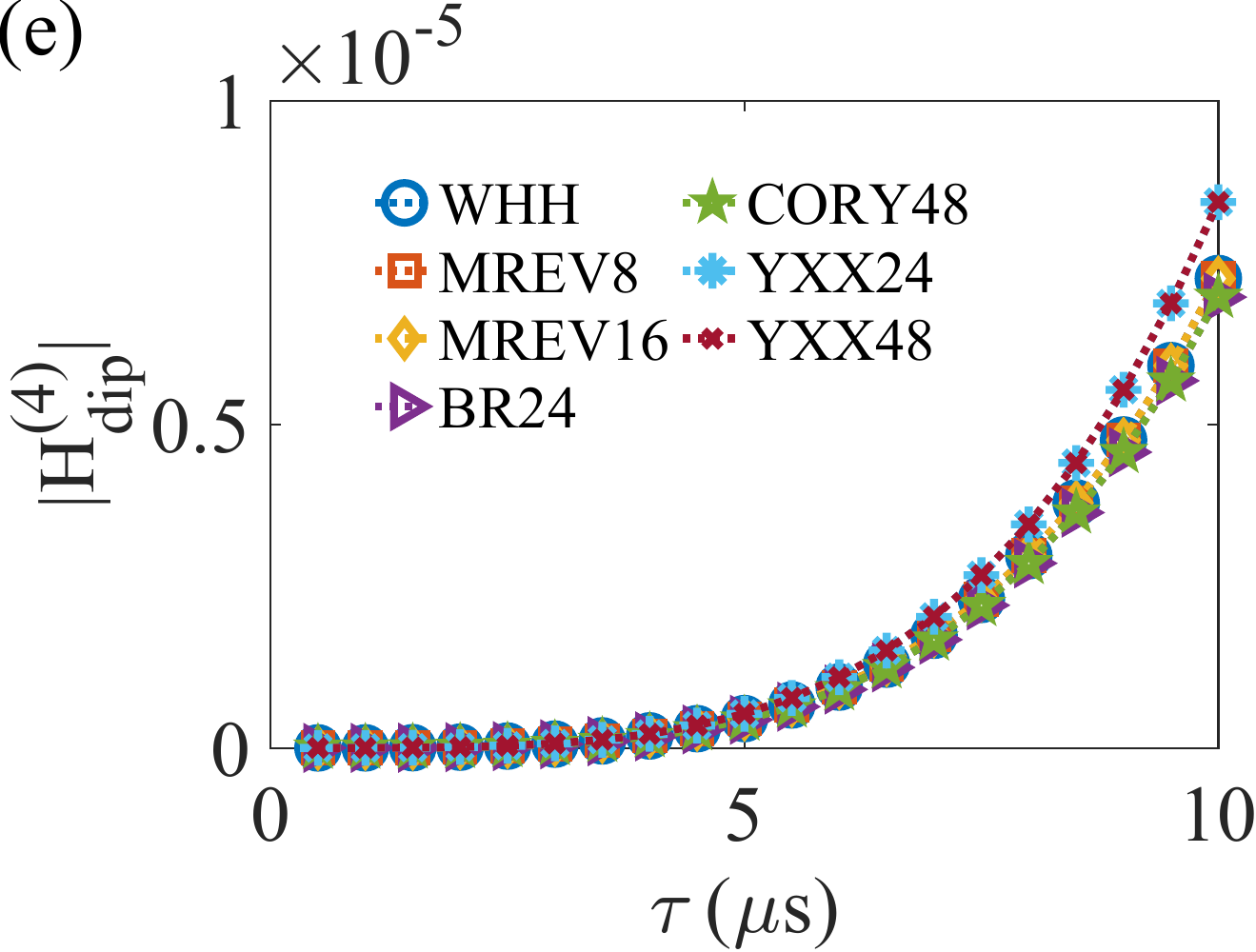} 
        \hspace{0.3cm}
    \includegraphics[width=0.22\textwidth]{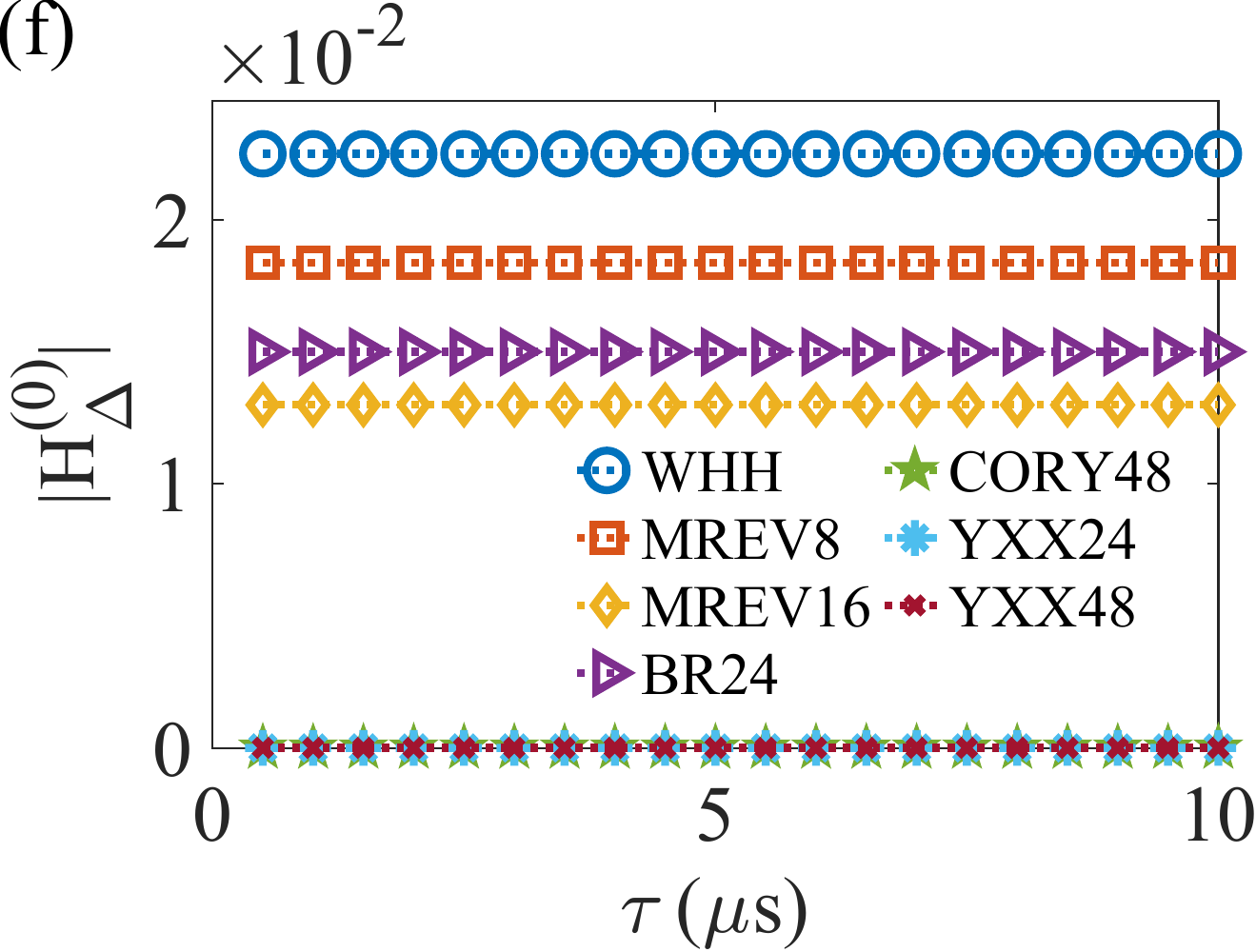} 
        \hspace{0.3cm}
    \includegraphics[width=0.22\textwidth]{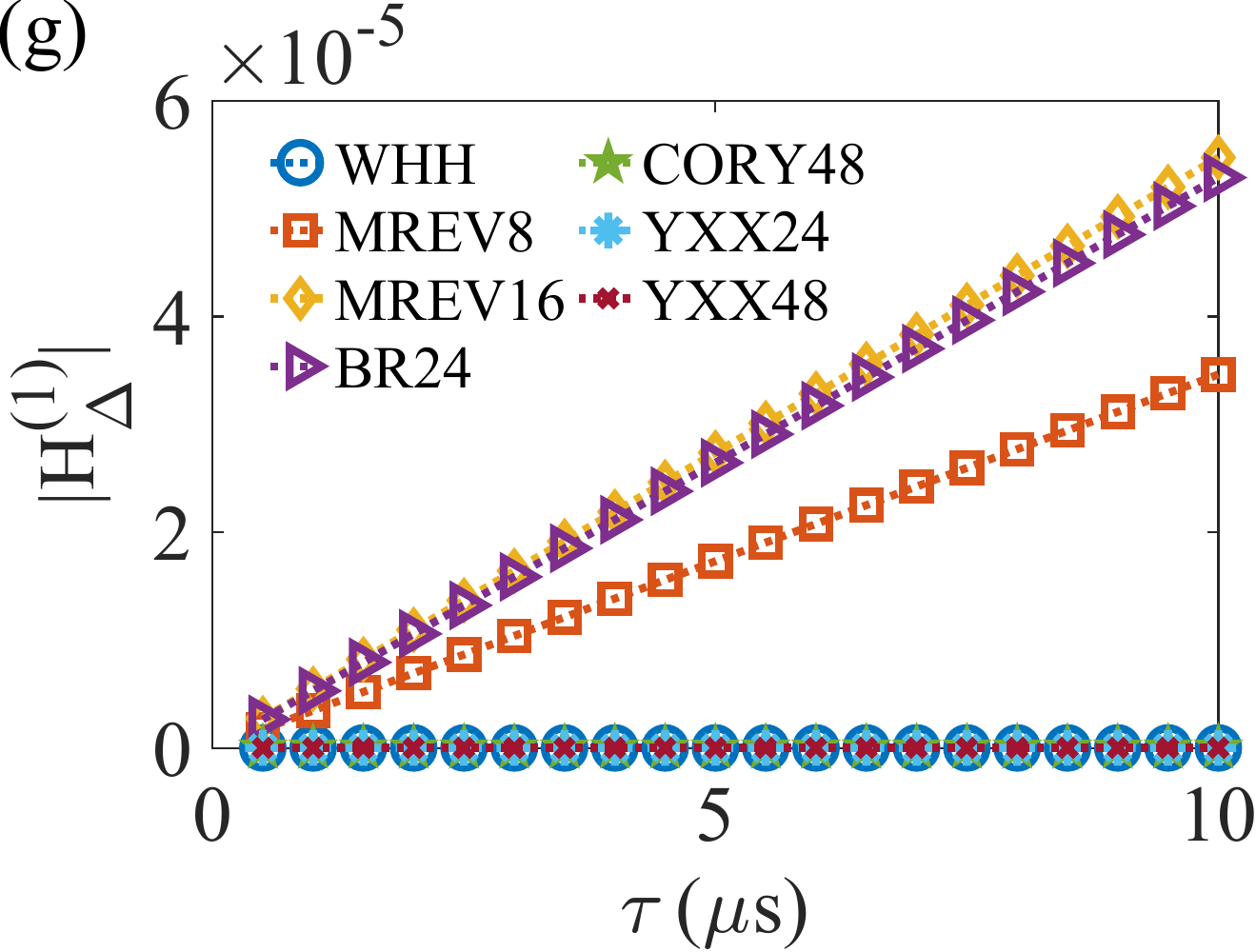} 
        \hspace{0.3cm}
    \includegraphics[width=0.22\textwidth]{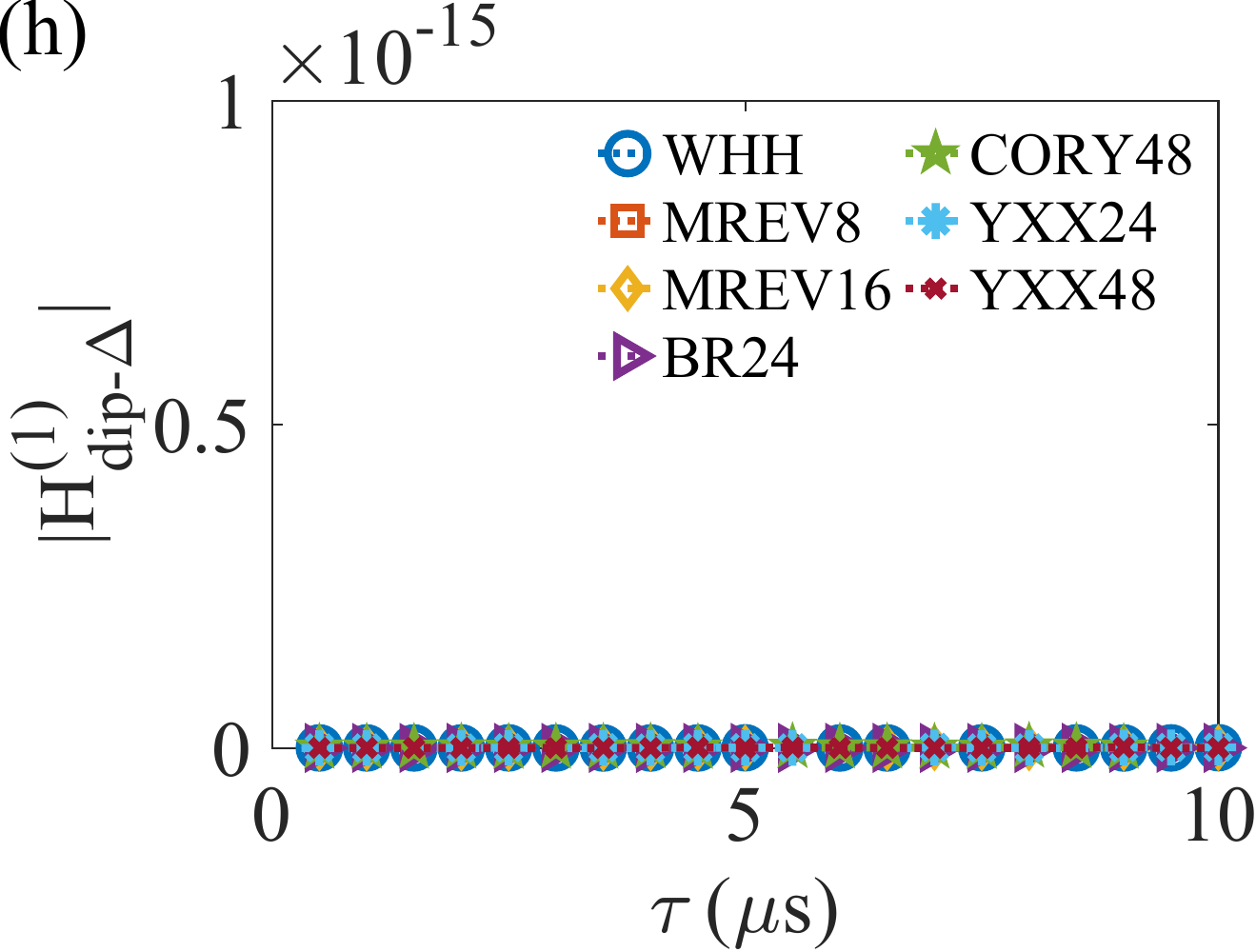} 
    \caption{Numerical estimates of the magnitude of the Magnus expansion terms for a system of four spins. Dipolar coupling strength and global resonance offset are set to 420 Hz and 30 Hz respectively. The Size of a Hamiltonian term is estimated as $|H^{(n)}|=(\Tr[{H^{(n)}H^{(n)}}])^{1/2}$/$(\Tr[{H_{\text{dip}}H_{\text{dip}}}])^{1/2}$. Figures show the estimated magnitudes of the effective Hamiltonian terms for (a) zeroth to fourth order for the dipolar Hamiltonian, (f) - (g) zeroth and first order for the resonance offset Hamiltonian, and (h) first order dipolar-offset cross term. Note that $|H^{(0)}_{\text{dip}}|$, $|H^{(1)}_{\text{dip}}|$, and $|H^{(1)}_{\text{dip}-\Delta}|$ are negligible for all sequences considered here.} 
	\label{fig_supp:magnus_terms} 
\end{centering}
\end{figure*}

\begin{figure*}[]
\begin{centering}
    \includegraphics[width=0.3\textwidth]{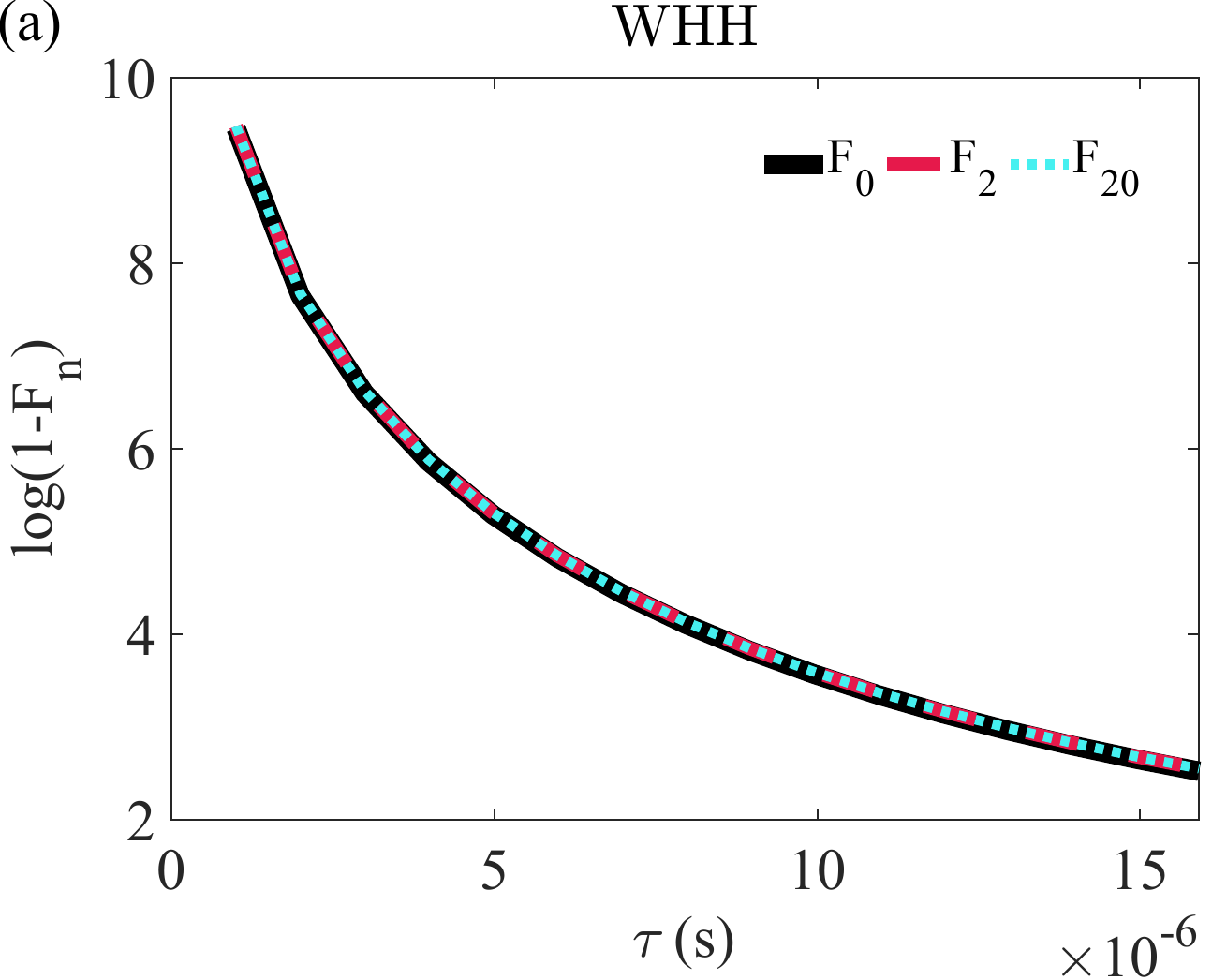} 
        \hspace{0.3cm}
    \includegraphics[width=0.3\textwidth]{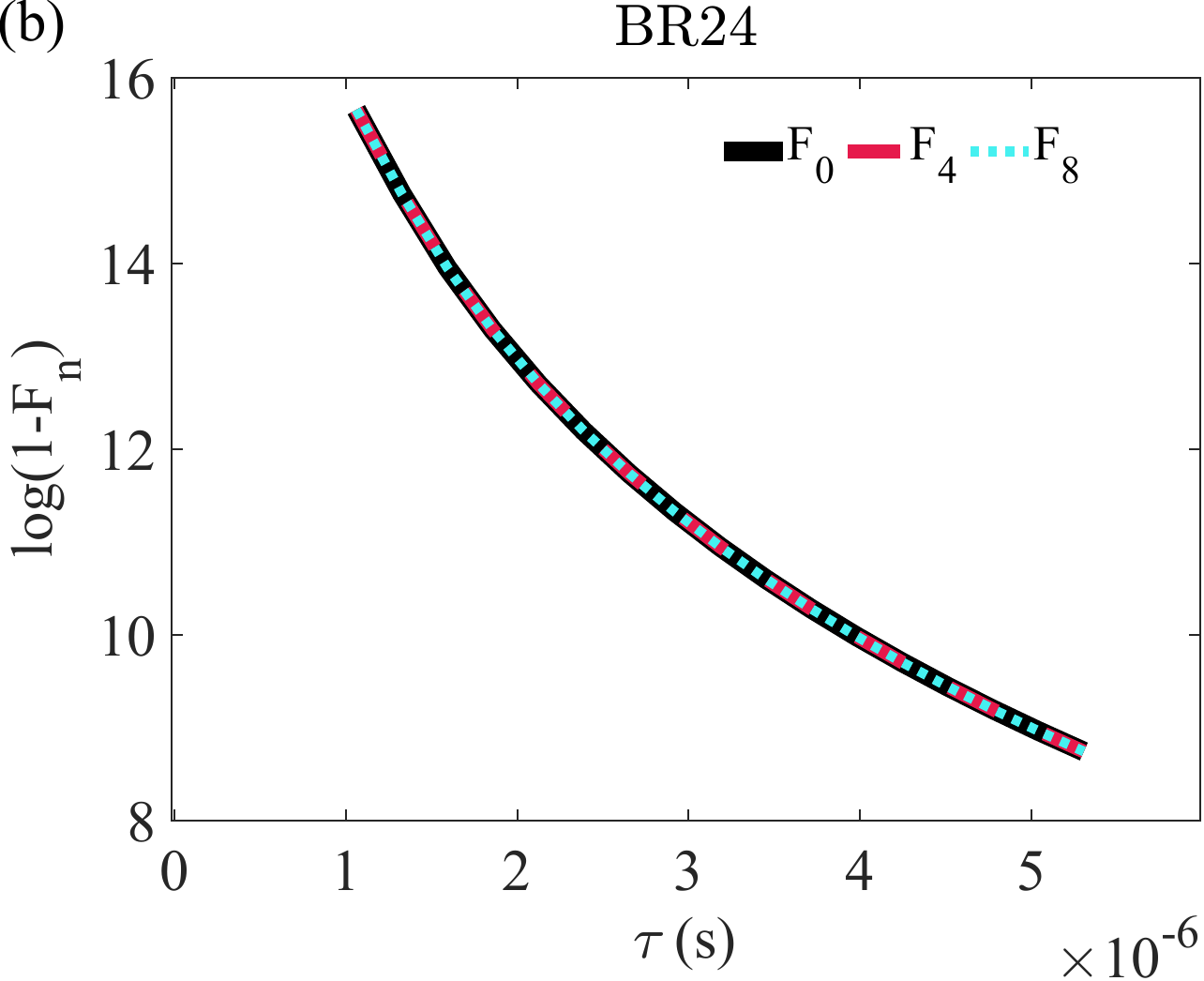} 
        \hspace{0.3cm}
    \includegraphics[width=0.3\textwidth]{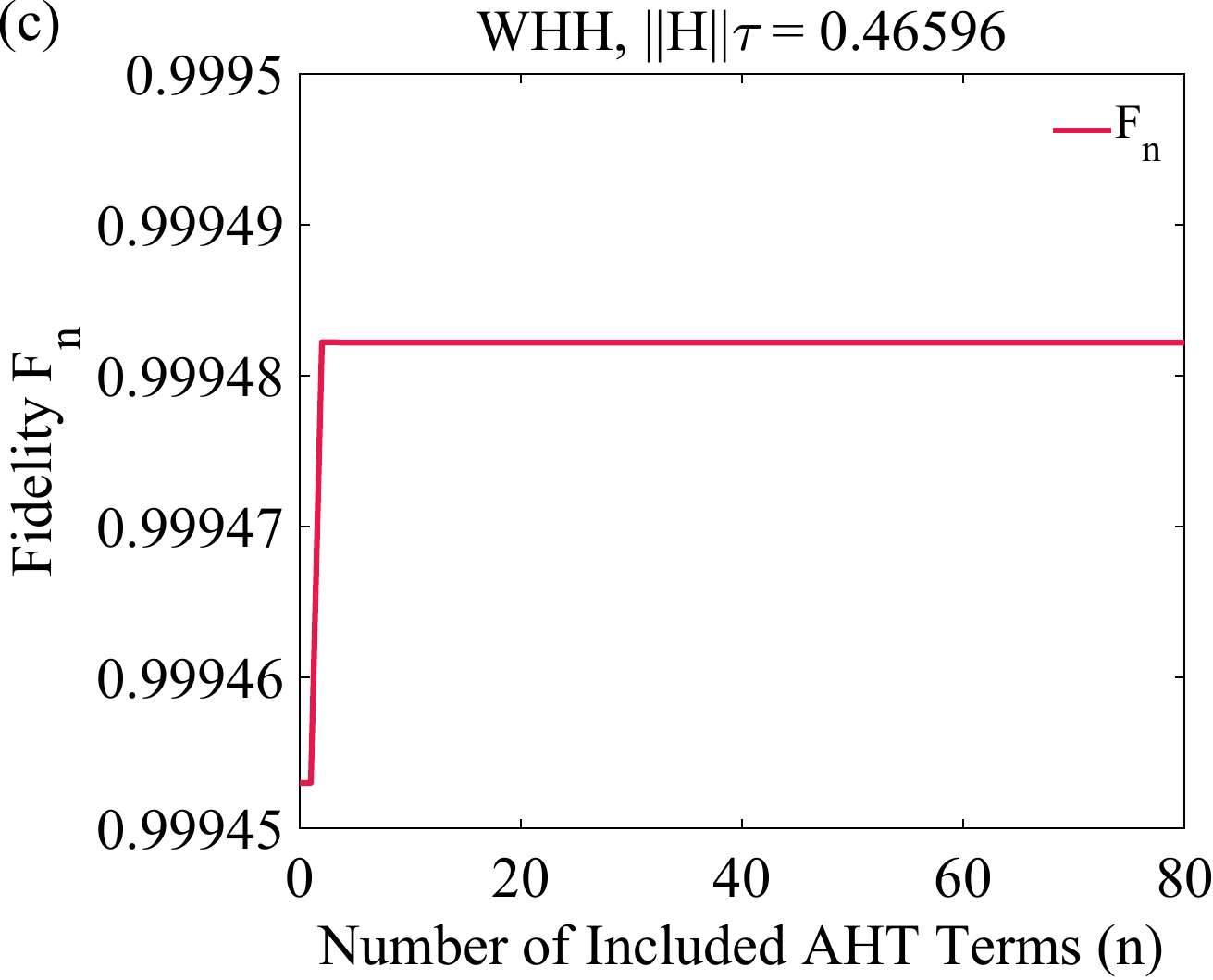} 
        
    \caption{Effect of considering higher order average Hamiltonian terms. Dipolar coupling strength is set to 5000 Hz, resonance offset is set to zero, and pulses are treated instantaneously. (a) Log-Infidelity versus $\tau$ for WHH with a 6-spin system. Note that if $\Hbar{2}$ was a large source of error, we would expect the $F_2$ curve to lie somewhere above the black curve; surprisingly, the curves overlap instead. (b) Log-Infidelity versus $\tau$ for BR-24 with a 4-spin system. (c) $F_n$  versus $n$ for WHH for a 4-spin system, with $||H||\tau=0.466$. Since we know $F_n\to1$ for $n\to\infty$, it is surprising that even considering the first 70 terms of the average Hamiltonian is not sufficient to avoid errors due to higher-order terms.}
	\label{fig_supp:magnus_terms_fidelity} 
\end{centering}
\end{figure*}

In this section, we show our numerical results on the magnitudes of higher-order Magnus expansion terms and their effects on the fidelity of the dipolar decoupling sequences. Both spectroscopic and time-suspension sequences are designed to cancel dipolar terms in the Magnus expansion up to a certain order, as shown in Table \ref{table:sequence_comparison}. Thus, even in the absence of pulse errors and in the $\delta$-pulse limit, there will still be some decoupling error caused by higher-order terms in the average Hamiltonian that are not canceled out. In general, the terms scale as $\left(||H||\tau\right)^n$, so provided that $\tau$ is not too long, most higher-order terms should be negligibly small. Recently, there has been work investigating the regimes of convergence and validity of AHT specifically in the context of applications in quantum sensing and Hamiltonian engineering~\cite{oon_beyond_2024}.

Figure \ref{fig_supp:magnus_terms} shows numerical estimates of the size of various Magnus expansion terms, $H^{(n)}$, given by $(\Tr[{H^{(n)}H^{(n)}}])^{1/2}$ and normalized by the size of the system Hamiltonian, $|H_{\text{dip}}|$ for a system of 4 spins, with a dipolar coupling strength of 420 Hz and a resonance offset of 30 Hz. Results are given up to order 4 for $H_{\text{dip}}$ and up to order 1 for $H_{\Delta}$. Numerical values $<10^{-15}$ can be considered negligible. These results agree with Table \ref{table:sequence_comparison} and with the scaling of sequence infidelity with various errors, discussed in detail in Sections \ref{sec:Comparisons} and \ref{sec:Errors}. 

We can further numerically assess the size of errors due to higher order terms by introducing the $n$th-order fidelity:
\begin{gather}
    F_n=\Tr(U_{\text{th},n}^{\dag}U_{\text{exp}})^{1/M}\\
    U_{\rm{th},n}=\exp\left\{it_c\sum_{j=0}^n\Hbar{n}\right\}
\end{gather}
This is defined similarly to the fidelity metric in Section \ref{sec:sims}, but $U_{\rm{th},n}$ now refers to the unitary generated by considering up to the $n$-th term in the average Hamiltonian. For this section, we consider instantaneous pulses with no errors when constructing $U_{\rm{exp}}$. Thus, if a pulse sequence decouples dipolar interactions up to order $k$, then $U_{\rm{th},n}=\openone$ if $n\leq k$. We can then compare $U_{\rm{th},k}$ with $U_{\rm{th},k+1}$ to find the error resulting from the lowest order term that is not decoupled. A large difference in fidelity indicates a large error. The Magnus series is known to converge if $\int_0^{t_c}dtH(t)<\pi$ \cite{casas_sufficient_2007}. Thus, if this condition is met, then we must have $F_n\to1$ for $n\to\infty$. Furthermore, we expect this to converge quickly, since if $\left(||H||\tau\right)$ is small then the sizes of the terms in the series decrease exponentially. 

Figure \ref{fig_supp:magnus_terms_fidelity}(a)--(b) show the log-infidelities, $-\log(1-F_n)$, for $n=0$, $n=k+1$ (with $k$ the decoupling order), and a larger value of $n$ for WHH and BR24. We would expect the fidelity $F_{k+1}$ to be noticeably larger than $F_0$, and then to see a smaller increase in fidelity when considering further terms beyond the lowest nonzero term. Rather surprisingly, the three curves for both sequences are almost identical, suggesting that the error due to the lowest few non-decoupled terms in the Magnus expansion is still very small. 

To assess how quickly the average Hamiltonian converges for these sequences, we compute higher order terms using the recursive method first presented by Burum \cite{burum_magnus_1981-1, klarsfeld_recursive_1989}. Starting from the Dyson series:
\begin{equation}
    \label{eqn:dyson}
    P_n=\int_0^{t_c}dt_1\int_0^{t_1}d t_2... \int_0^{t_{n-1}}d t_n H(t_1)H(t_2)...H(t_n)
\end{equation}
we recognize that $P_1=\bar H^{(0)}$. The terms $\bar H^{(n)}=\frac{i}{t_c}\Omega_{n-1}$ can then be computed recursively by:
\begin{gather}
\Omega_n=P_n-\sum_{k=2}^{n}Q_n^{(k)}\\
Q_n^{(k)}=\sum_{m=1}^{n-k+1}Q_m^{(1)}Q_{n-m}^{(k-1)}\\
Q_n^{(1)}=\Omega_n\hspace{0.5em},\hspace{0.5em}Q_n^{(n)}=\Omega_1^n 
\end{gather}   

Using this method, we can numerically compute terms up to $\Hbar{8}$ for the 24-pulse sequences, and as high as $\Hbar{70}$ for WHH. 

Results for WHH for 70 terms are shown in Figure \ref{fig_supp:magnus_terms_fidelity}(c), with $||H||\tau=0.47$. This is still within the regime where the Magnus series is known to converge, so we must have $F_n\to1$ for $n\to\infty$. However, computing the terms in the Magnus series as high as $\Hbar{70}$ is not sufficient to see a fidelity close to $1$; in fact, we find $F_{70}$ to be almost identical to $F_4$ and only nominally higher than $F_0$. This likely indicates that the Magnus series converges quite slowly in this regime. If this is the case, then AHT methods may not be ideal for constructing dipolar-decoupling pulse sequences for systems with particularly strong dipolar couplings. We believe that further invesitgations are required in this direction for conclusive results. 

\section{Experimental Data}
\label{sec_supp:Expts}

\subsection{Raw data and analysis}
\label{sec_supp:raw_data}
\begin{figure*}[ht]
\begin{centering}
    \includegraphics[width=0.3\textwidth]{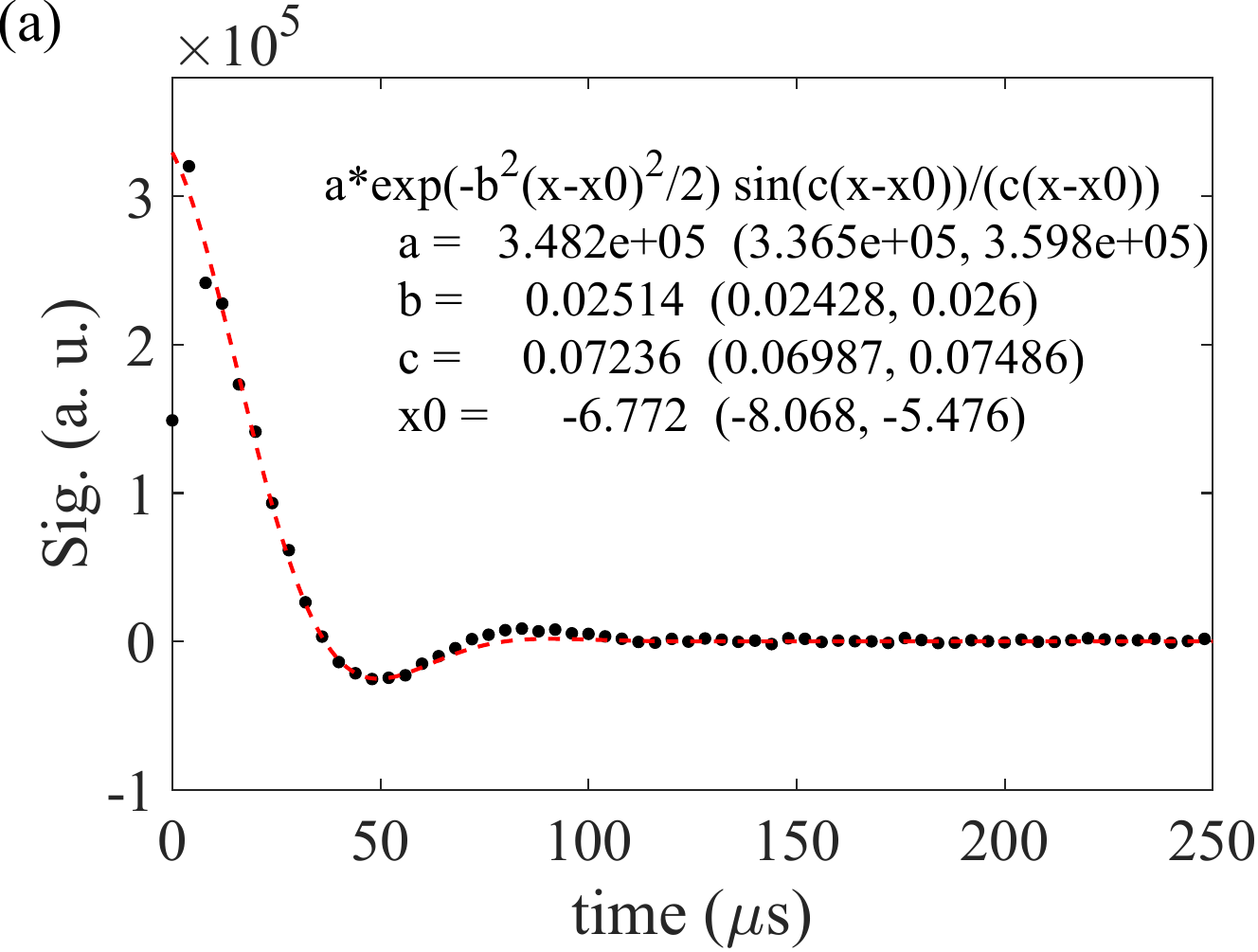} 
    \hspace{0.3cm}  
    \includegraphics[width=0.3\textwidth]{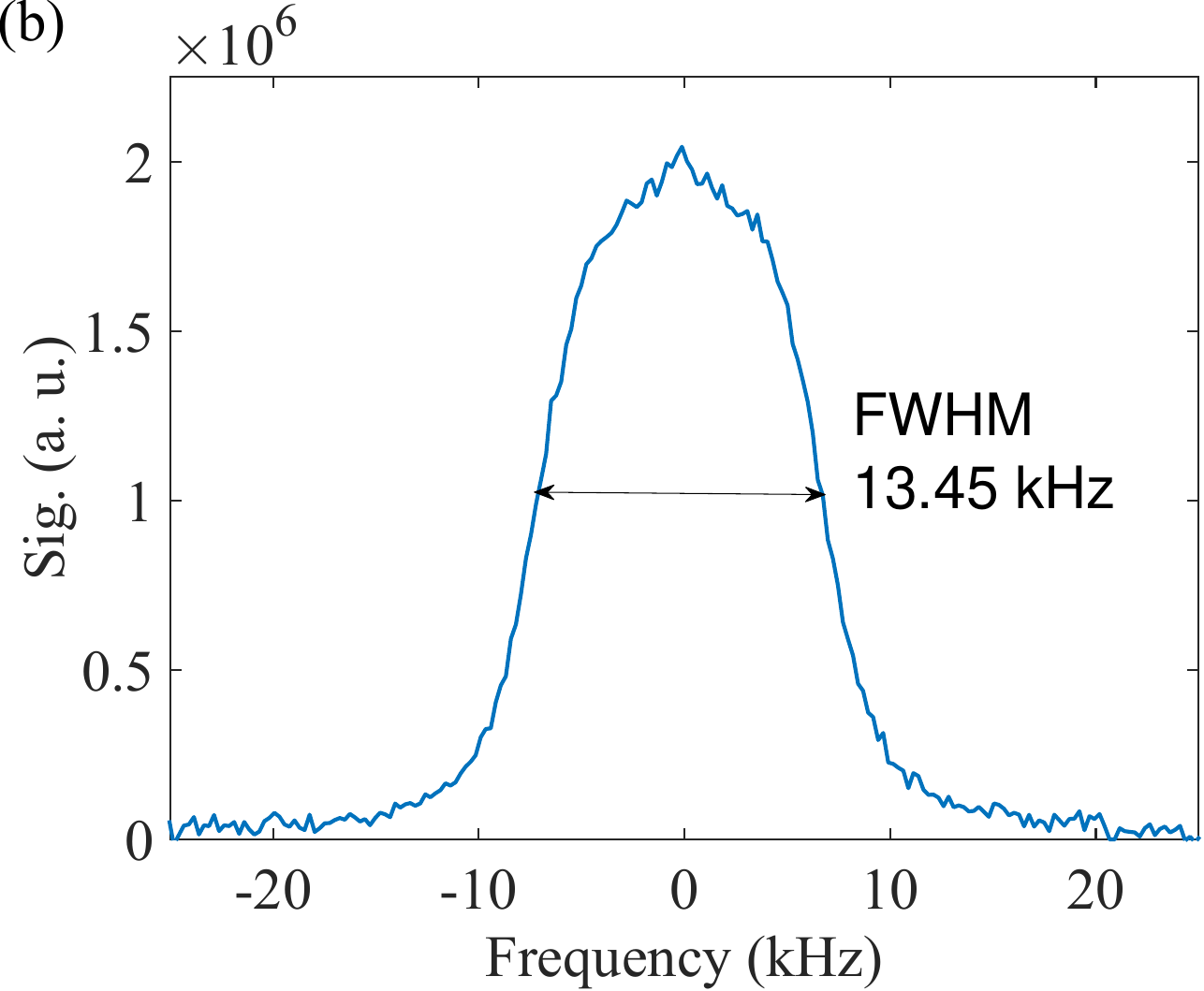} 
    \hspace{0.3cm} 
    \includegraphics[width=0.3\textwidth]{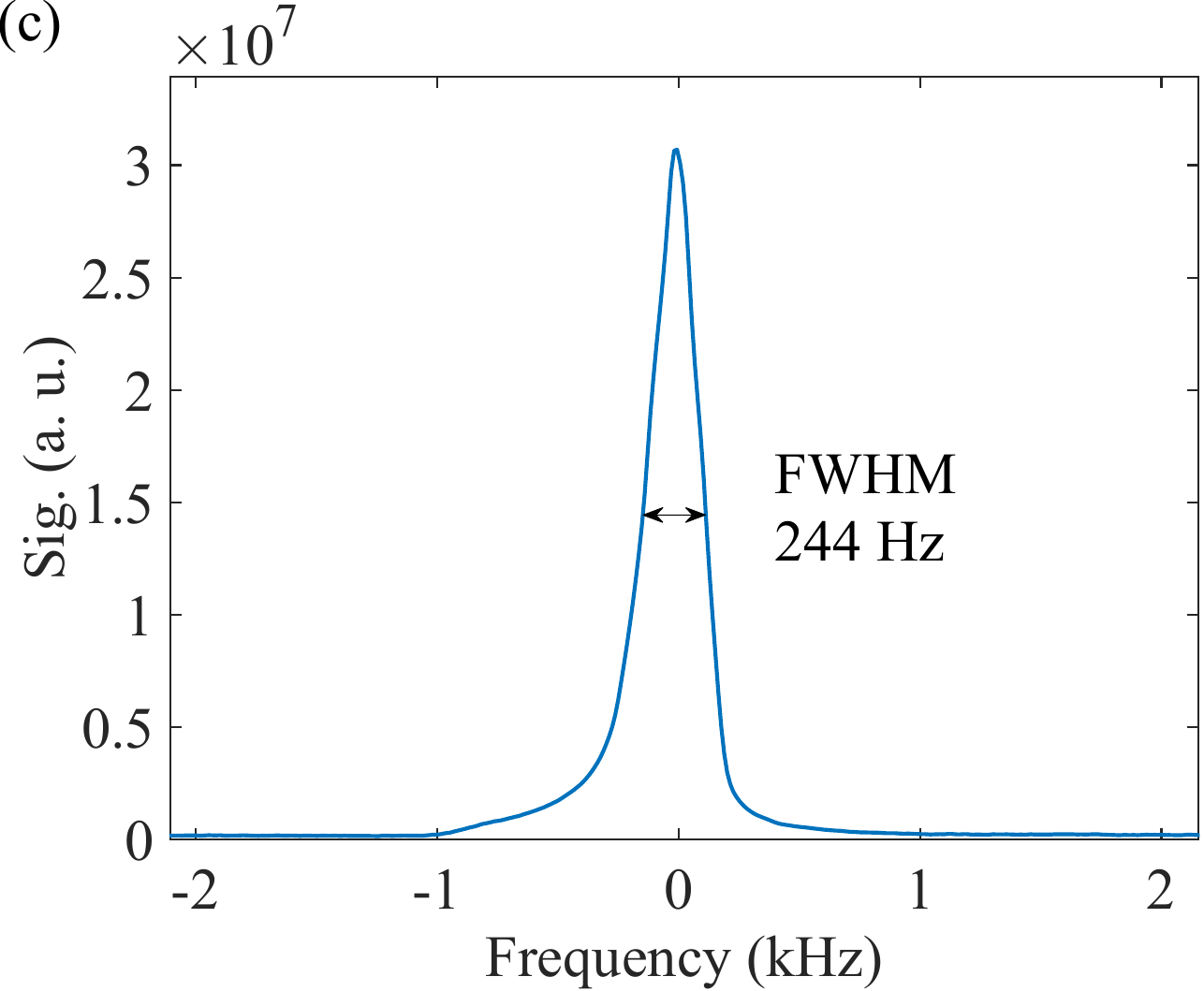}
    \caption{NMR experimental signal details. (a) The on-resonance FID signal of the adamantane sample at a $\pi/2$ pulse length of 1.22 $\mu$s. (b) The NMR spectrum of adamantane --- the real part of the Fourier transformed and phase-corrected signal from (a). (c) The spectrum of the water sample used for calibration at the same pulse length.}
	\label{fig_supp:signal_details} 
\end{centering}
\end{figure*}

Figure \ref{fig_supp:signal_details} shows details of the raw NMR experimental data. The on-resonance free induction decay (FID) signal fits the empirical function described by Abragam~\cite{abragam_principles_1983}. The FID is Fourier transformed into the frequency domain to construct the spectrum. The signal at a given time is obtained by averaging over the entire peak of the corresponding spectrum. The error bar for the data point corresponds to the standard deviation of the last 50 data points in the FID, where the signal has died down to zero. The figure also shows a spectrum of the water sample used for calibration. The width of this signal provides an estimate of the magnet's longitudinal ($B_{0}$) field inhomogeneity.

\subsection{Normalization}
\label{sec_supp:normalization}
We normalize the $X$, $Y$, and $Z$ autocorrelation decays using signals from experiments with the same structure and delays as the decoupling experiments, only omitting the sequences themselves. 

\subsection{Decay curve fits}
\label{sec_supp:fits}
\begin{figure}[htb]
\begin{centering}                  
    \includegraphics[width=0.23\textwidth]{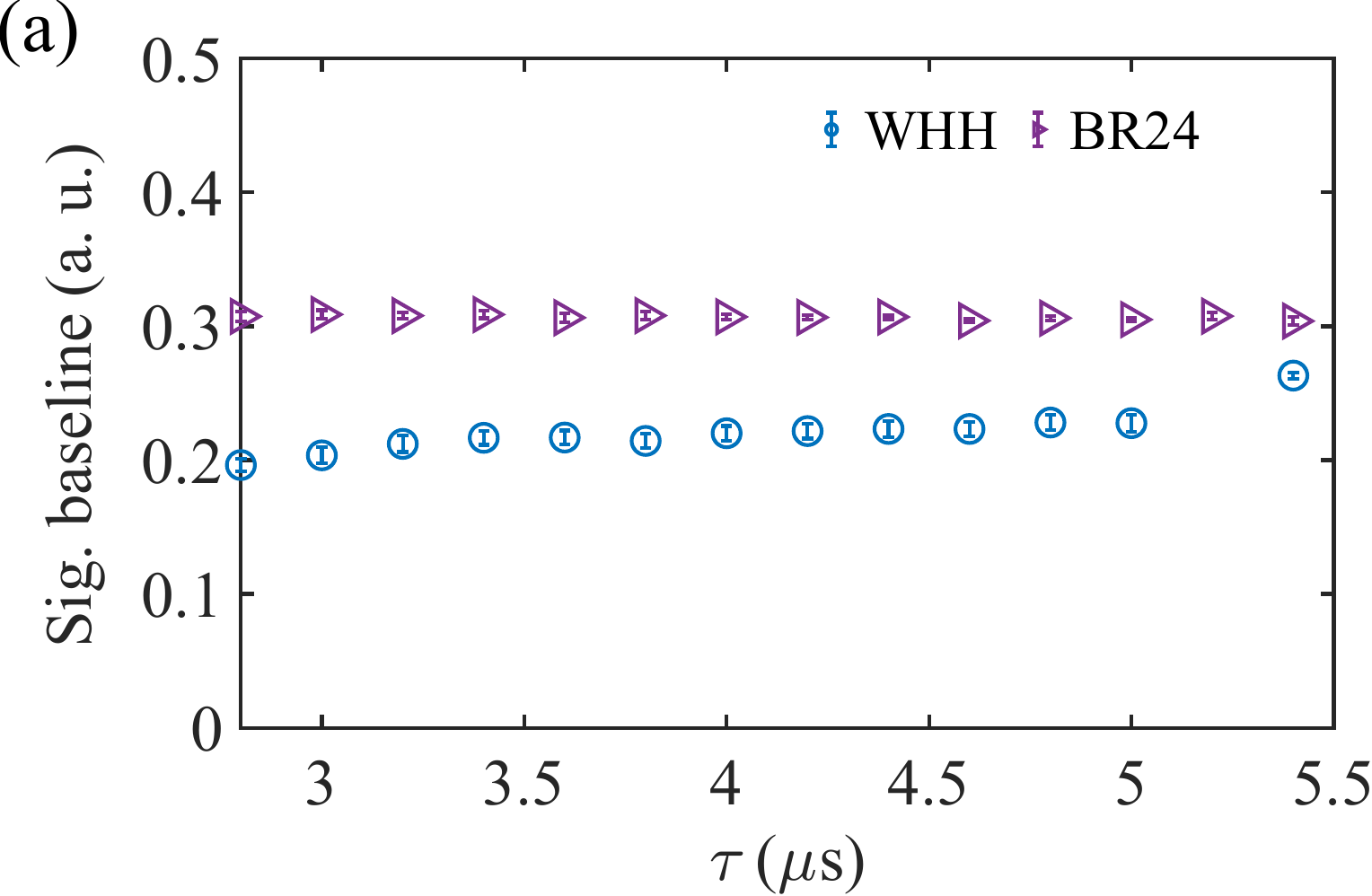} 
    \hspace{0.2cm} \includegraphics[width=0.22\textwidth]{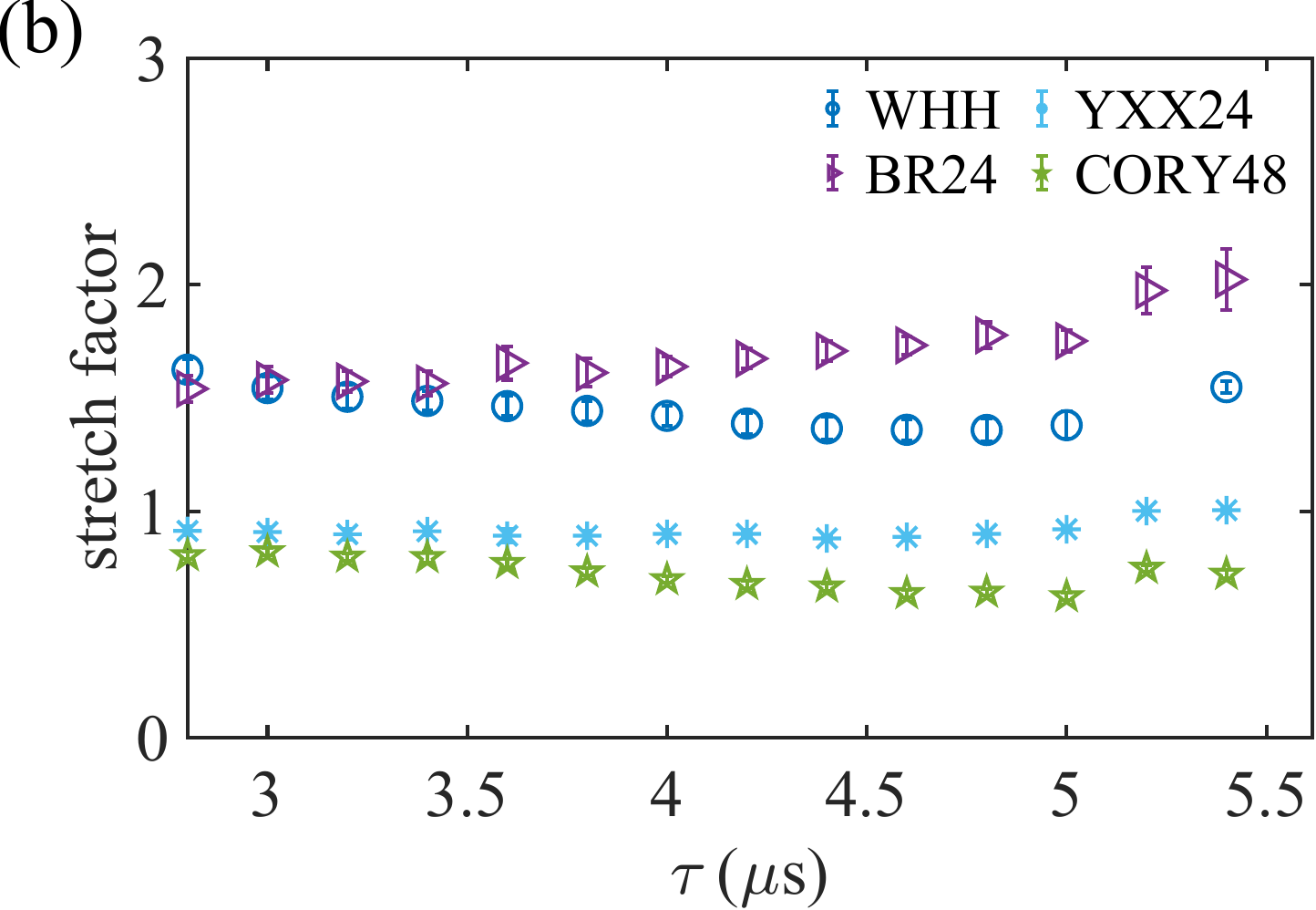} 
    \hspace{0.3cm}
    \includegraphics[width=0.22\textwidth]{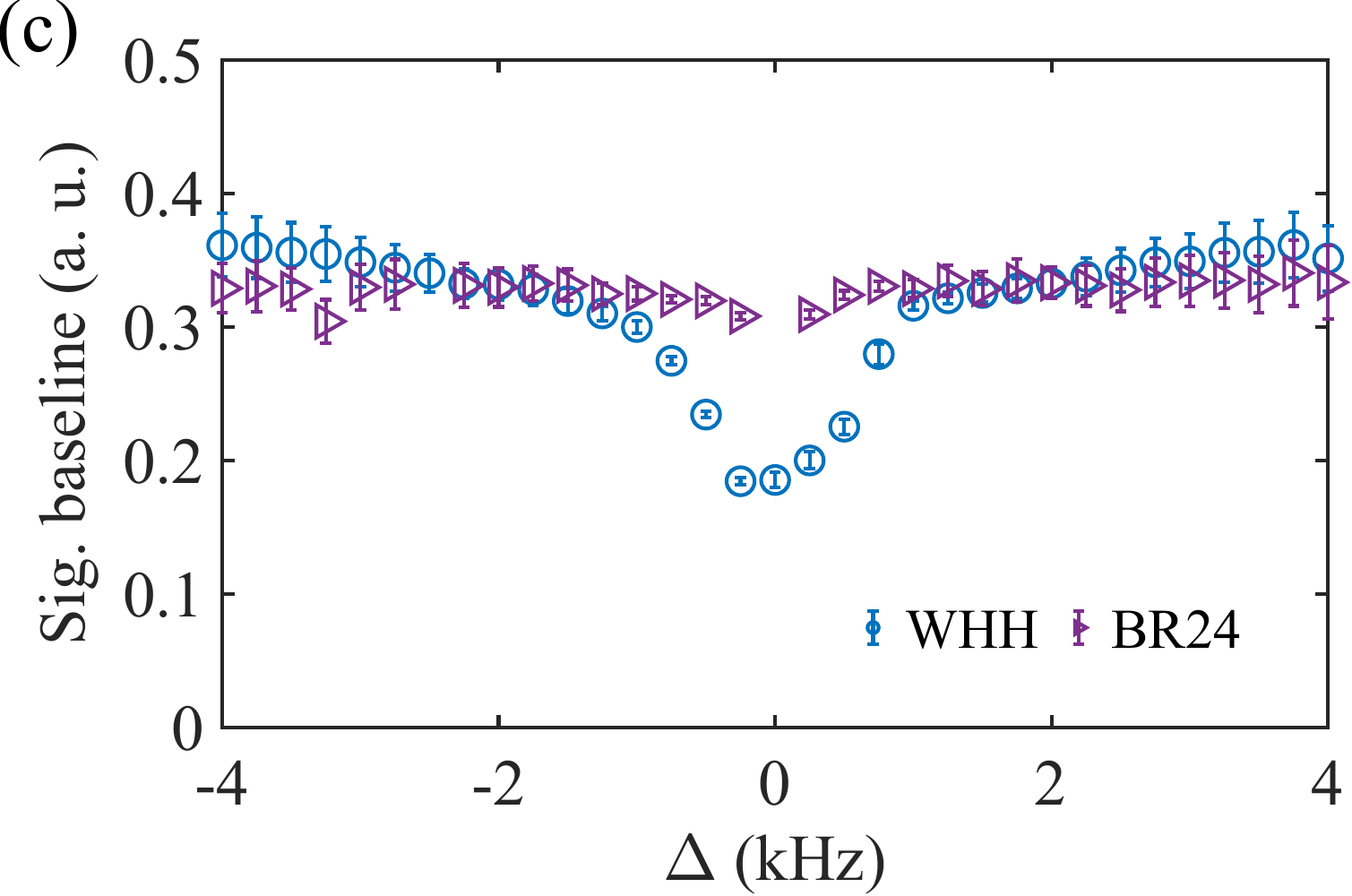}
    \hspace{0.3cm} \includegraphics[width=0.22\textwidth]{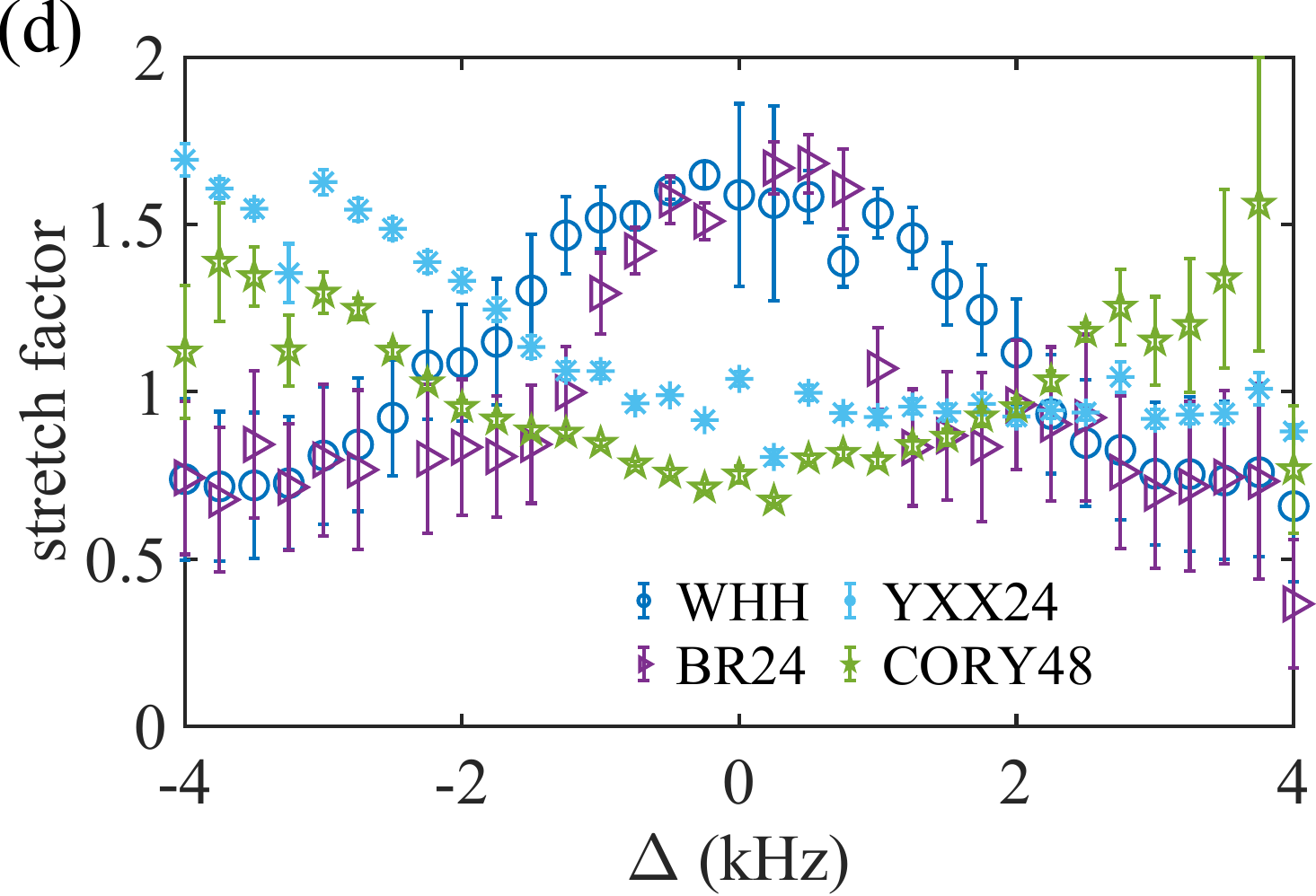}
    \caption{Details of the fit to $C_{\text{avg}}$ decay curves. (a) and (b) show baselines and stretch factors for the $\tau$ dependence data. (c) and (d) show baseline and stretch factor values for the off-resonance data.}
	\label{fig_supp:fit_details} 
\end{centering}
\end{figure}
Figure \ref{fig_supp:fit_details} shows the details of the fit discussed in Section \ref{sec:expts}. The normalized decay curves are fit to a stretched exponential of the form $C_{\text{avg, ts}}=C_0 e^{-t^{g}/T_{2,\text{eff}}}$ for the time-suspension sequences. For spectroscopic sequences, we fit the data to $C_{\text{avg,spectro}}=C_0\cos(2\pi ft)+C_1$. Here, we show the stretch factors ($g$) and baselines ($C_{1}$) for all the fits for the inter-pulse delay dependence (Figure \ref{fig_supp:fit_details}(a) and (b), $T_{2, \text{eff}}$ data shown in Figure \ref{fig:tau_data}) and the off-resonance dependence (Figure \ref{fig_supp:fit_details}(c) and (d), $T_{2, \text{eff}}$ data shown in Figure \ref{fig:delta_data}) of the $C_{\text{avg}}$ decay data. For the stretch factor, we see a grouping of the time suspension vs spectroscopic sequences in the inter-pulse delay data, especially at lower $\tau$ and a larger deviation at higher $\tau$. The grouping of spectroscopic and time suspension sequences also holds for the off-resonance data, although there is a larger variation in the values with $\Delta$. It is also interesting that the trends of this variation are opposite for the two groups of sequences. For the signal baseline data, since WHH and BR24 are both sequences that have an effective field composed of components in $X$, $Y$, and $Z$ directions (see Table \ref{table:sequence_comparison}), the decay curves show conserved magnetization baselines for all three autocorrelation experiments, and so do their geometric mean, as captured here in sub-figures (a) and (c). 

\section{Pulse Sequence Details}
\label{sec_supp:Sequences}
Pulse phases and delay timings that make up each of the seven sequences studied in this work are given in Table \ref{table_supp:pulse_sequences}. 
\subsection{F-matrix representation}
\label{sec_supp:f_matrix}
\begin{figure*}[]
\begin{centering}
    \includegraphics[width=0.95\textwidth]{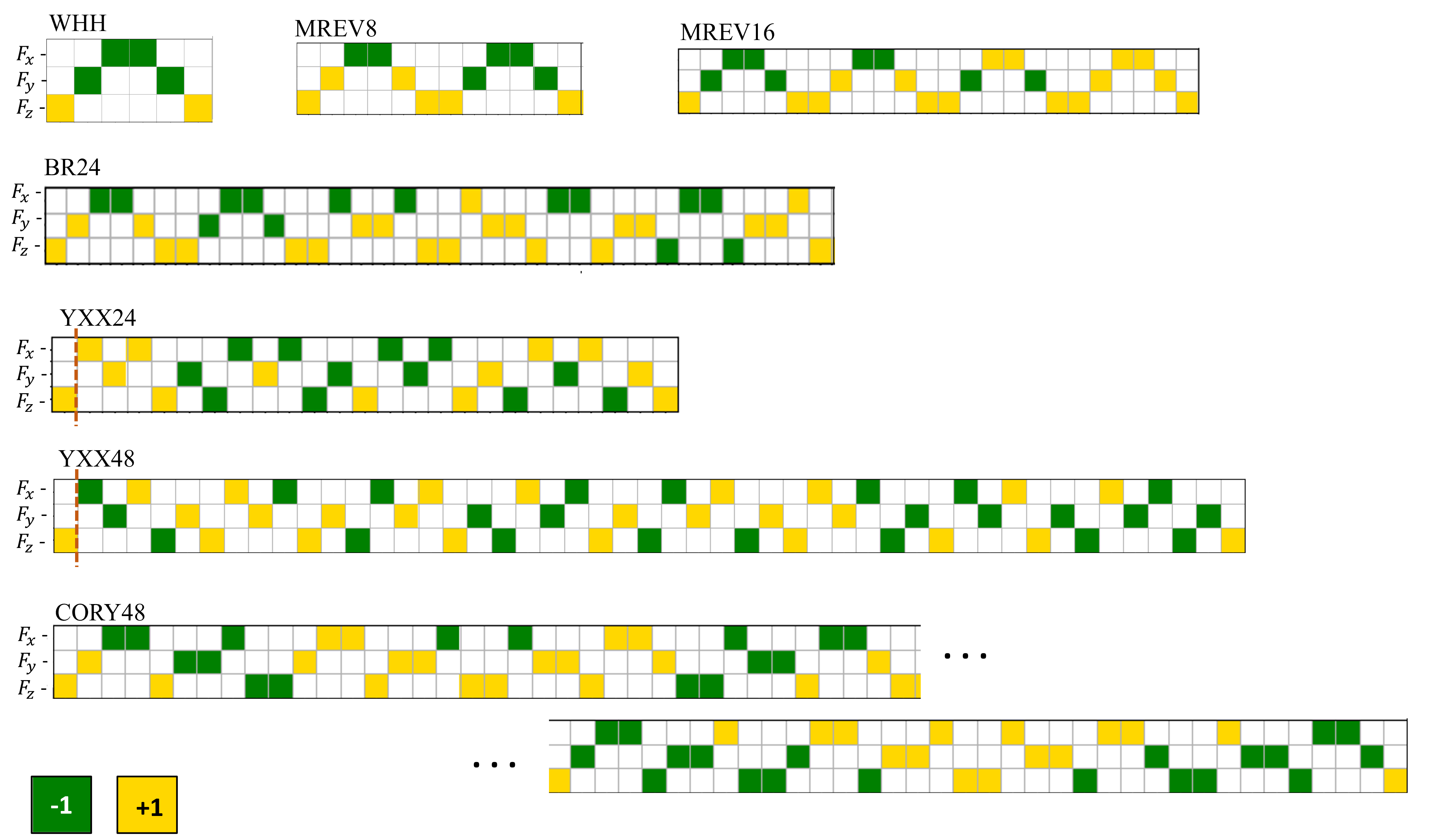} 
    \caption{F-matrix representation for the sequences. The x-axis represents time and each column is of uniform width $\tau$. Finite pulse widths are not considered. Each row corresponds to one of the three possible orientations of the toggling frame in any given time window $\tau$, green and yellow colors correspond to negative or positive orientations along a given axis, respectively. The red dotted lines after the first column in YXX24 and YXX48 indicate that these sequences start with a pulse and do not have a $\tau$ delay at the start unlike the other five sequences. In these cases, the initial (yellow) $+Z$-oriented frame precedes the first pulse and do not factor into the analysis of the cycle.}
	\label{fig_supp:Fmatrix} 
		\end{centering}
\end{figure*}

In Figure \ref{fig_supp:Fmatrix}, we give the frame matrix representations of the seven sequences, following the formalism developed in ~\cite{choi_robust_2020}. This pictorial representation of the toggling frame orientation during a full cycle of a given sequence allows us to understand underlying patterns and conveniently analyze sequence characteristics. For example, interactions $\propto S_{z}$ are averaged to zero only by sequences whose ``weighted row sums" equal zero, i.e., they have an equal number of positive and negative squares in each row. Further `rules' and decoupling conditions have been derived for various interaction types and pulse imperfections ~\cite{zhou_quantum_2020,zhou_robust_2023,zhou_robust_2023-1,tyler_higher-order_2023-1}.  

{\renewcommand{\arraystretch}{1.5}
\begin{table*}
  \centering
\begin{tabular}{|c|c|} 
 \hline
\textbf{Sequence} &  \textbf{Pulses and Delays} \\
 \hline
WHH &  $\tau-\bar{x}-\tau-y-2\tau-\bar{y}-\tau-x-\tau$\\
\hline
 MREV8 & $\tau-x-\tau-y-2\tau-\bar{y}-\tau-\bar{x}-2\tau-\bar{x}-\tau-y-2\tau-\bar{y}-\tau-x-\tau$ \\  
 \hline
 MREV16 & $\tau-\bar{x}-\tau-y-2\tau-\bar{y}-\tau-x-2\tau-x-\tau-y-2\tau-\bar{y}-\tau-\bar{x}-2\tau-$\\
  & $\bar{x}-\tau-\bar{y}-2\tau-y-\tau-x-2\tau-x-\tau-\bar{y}-2\tau-y-\tau-\bar{x}-\tau$ \\  
 \hline
 BR24 &  $\tau-\bar{x}-\tau-y-2\tau-\bar{y}-\tau-\bar{x}-2\tau-\bar{x}-\tau-y-2\tau-\bar{y}-\tau-x-2\tau-y-\tau-x-2\tau-\bar{x}-\tau-\bar{y}-2\tau-$ \\
 & $\bar{y}-\tau-x-2\tau-y-\tau-x-2\tau-\bar{x}-\tau-\bar{y}-2\tau-\bar{y}-\tau-x-2\tau-\bar{x}-\tau-y-2\tau-\bar{x}-\tau-y-\tau$ \\
 \hline
 YXX24 &  $\bar{y}-\tau-x-\tau-\bar{x}-\tau-y-\tau-\bar{x}-\tau-\bar{x}-\tau-y-\tau-\bar{x}-\tau-x-\tau-\bar{y}-\tau-x-\tau-x-\tau-$\\
 & $y-\tau-\bar{x}-\tau-x-\tau-\bar{y}-\tau-x-\tau-x-\tau-\bar{y}-\tau-x-\tau-\bar{x}-\tau-y-\tau-\bar{x}-\tau-\bar{x}-\tau$ \\  
 \hline
 YXX48 & $y-\tau-\bar{x}-\tau-\bar{x}-\tau-y-\tau-\bar{x}-\tau-\bar{x}-\tau-\bar{y}-\tau-x-\tau-x-\tau-y-\tau-\bar{x}-\tau-\bar{x}-\tau-$ \\
 & $\bar{y}-\tau-x-\tau-x-\tau-\bar{y}-\tau-x-\tau-x-\tau-y-\tau-\bar{x}-\tau-\bar{x}-\tau-y-\tau-\bar{x}-\tau-\bar{x}-\tau-$ \\ 
  & $\bar{y}-\tau-x-\tau-x-\tau-y-\tau-\bar{x}-\tau-\bar{x}-\tau-\bar{y}-\tau-x-\tau-x-\tau-\bar{y}-\tau-x-\tau-x-\tau-$ \\
  & $y-\tau-\bar{x}-\tau-\bar{x}-\tau-\bar{y}-\tau-x-\tau-x-\tau-y-\tau-\bar{x}-\tau-\bar{x}-\tau-\bar{y}-\tau-x-\tau-x-\tau$ \\ 
 \hline
CORY48 & $\tau-x-\tau-y-2\tau-\bar{x}-\tau-y-2\tau-x-\tau-y-2\tau-x-\tau-y-2\tau-$
   $x-\tau-\bar{y}-2\tau-x-\tau-y-2\tau-$\\
   & $\bar{y}-\tau-\bar{x}-2\tau-y-\tau-\bar{x}-2\tau-\bar{y}-\tau-\bar{x}-2\tau-\bar{y}-\tau-\bar{x}-2\tau-\bar{y}-\tau-x-2\tau-\bar{y}-\tau-\bar{x}-2\tau-$ \\
  & $\bar{x}-\tau-y-2\tau-\bar{x}-\tau-\bar{y}-2\tau-\bar{x}-\tau-y-2\tau-x-\tau-\bar{y}-2\tau-\bar{x}-\tau-\bar{y}-2\tau-x-\tau-\bar{-y}-2\tau-$\\
  & $y-\tau-\bar{x}-2\tau-y-\tau-x-2\tau-y-\tau-\bar{x}-2\tau-\bar{y}-\tau-x-2\tau-y-\tau-x-2\tau-\bar{y}-\tau-x-\tau$ \\ 
 \hline
\end{tabular}
\caption{Pulses and delays of the dipolar decoupling sequences studied in the manuscript.}
\label{table_supp:pulse_sequences}  
\end{table*}}

\clearpage

\bibliography{Main.bib}

\end{document}